\documentclass[a4paper,11pt]{article}
\usepackage{jcappub} 
\usepackage{lineno}
\usepackage{aas_macros}

\usepackage{upgreek}
\usepackage{amsmath,amssymb}

\usepackage{mathtools}

\usepackage{hyperref}
\usepackage{xcolor}

\title{\boldmath Pixelization effects in cosmic shear angular power spectra}

\author[a]{Alex Hall}
\author[b]{and Nicolas Tessore}

\affiliation[a]{Institute for Astronomy, University of Edinburgh,\\Royal Observatory, Blackford Hill, Edinburgh EH9 3HJ, UK}
\affiliation[b]{Department of Physics and Astronomy, University College London,\\Gower Street, London WC1E 6BT, UK}

\emailAdd{ahall@roe.ac.uk}

\abstract{We conduct a comprehensive study into the impact of pixelization on cosmic shear, uncovering several sources of bias in standard pseudo-$C_\ell$ estimators based on discrete catalogues. We derive models that can bring residual biases to the percent level on small scales. We elucidate the impact of aliasing and the varying shape of HEALPix pixels on power spectra and show how the HEALPix pixel window function approximation is made in the discrete spin-2 setting. We propose several improvements to the standard estimator and its modelling, based on the principle that source positions and weights are to be considered fixed. We show how empty pixels can be accounted for either by modifying the mixing matrices or applying correction factors that we derive. We introduce an approximate interlacing scheme for the HEALPix grid and show that it can mitigate the effects of aliasing. We introduce bespoke pixel window functions adapted to the survey footprint and show that, for band-limited spectra, biases from using an isotropic window function can be effectively reduced to zero. This work partly intends to serve as a useful reference for pixel-related effects in angular power spectra, which are of relevance for ongoing and forthcoming lensing and clustering surveys.}

\begin{document}
\maketitle
\flushbottom

\section{Introduction}
\label{sec:intro}

The statistical constraining power of galaxy imaging surveys will soon increase significantly, with the Euclid survey~\citep{2024arXiv240513491E} well underway and the Rubin Observatory Legacy Survey of Space and Time (LSST; \citep{2019ApJ...873..111I}) to begin soon. Amongst the main science drivers of these surveys are precision tests of the consensus cosmological model $\Lambda$CDM and percent level constraints on the Dark Energy equation of state.

The headline constraints from these surveys will be made from two-point statistics measured from catalogues of galaxy shapes and positions. Common estimators used for weak lensing analyses include the real-space $\xi_\pm$ correlation functions~\citep{1998MNRAS.301.1064K, 2002ApJ...568...20C, 2002A&A...396....1S, 2021A&A...645A.104A, 2022PhRvD.105b3515S, 2022PhRvD.105b3514A, 2023PhRvD.108l3518L}, harmonic space angular power spectra~\citep{2003MNRAS.341..100B, 2011MNRAS.412...65H, 2018MNRAS.479..454A, 2019MNRAS.484.4127A, 2021JCAP...03..067N, 2023PhRvD.108l3519D}, and hybrid estimators such as COSEBIs (Complete Orthogonal Sets of E-/B-Integrals, \cite{2010A&A...520A.116S, 2012A&A...542A.122A, 2019A&A...624A.134A}). Each of these methods has its own advantages and disadvantages, and it is now common practice for analysis teams to check the consistency of parameter constraints by repeating their analysis with a different choice of two-point statistic~\citep{2021A&A...645A.104A, 2021MNRAS.503.3796D}.

In this work, we investigate biases inherent to the standard pseudo-$C_\ell$ estimator. This statistic is fast to compute~\citep{2019MNRAS.484.4127A, PKWL}, allows a fairly clean separation of linear and non-linear modes, and can return reasonably uncorrelated measurements if mode-mixing due to the survey mask is properly accounted for~\citep{2004MNRAS.349..603E}. Furthermore, the scale separation allows a Gaussian likelihood approximation to be used with high accuracy for parameter inference~\citep{2022PhRvD.105l3527H}. Although not statistically optimal, information loss is mild, in particular for the $E$-mode spectrum on small scales~\citep{2023MNRAS.520.4836M}. The estimator proceeds by first constructing a map from the discrete set of points, taking the angular power spectrum, and then interpreting the result using mixing matrices constructed from weight maps that trace the survey footprint and other observational inhomogeneities in the data.

Much of the formalism for the measurement and interpretation of pseudo-$C_\ell$ has been derived from its use on Cosmic Microwave Background (CMB) data (see Ref.~\cite{2024arXiv241012951S} for a recent review of map making and power spectrum estimation in the CMB context). However, the application to weak lensing presents several additional challenges~\citep{2019MNRAS.484.4127A, 2021JCAP...03..067N}. These include the fact that the input data set is usually in the form of catalogues, i.e.~shear measurements at discrete samples across the survey footprint, rather than pixelized maps or time-ordered data. Secondly, there is no analogue of the beam in CMB experiments and hence the only small-scale filtering is due to the discrete sampling of the galaxy position field. Furthermore, weak lensing and galaxy clustering spectra have significant power on small scales, unlike in the case of the primordial CMB where diffusion damping suppresses small-scale power. This means that angular power spectra constructed from shear catalogues are more susceptible to aliasing and other pixelization effects. This is exacerbated by the fact that most of the information on cosmological parameters comes from small angular scales which have low statistical uncertainty, so particular care must be taken to ensure that no biases arise due to insufficient understanding of the measurement process. Surveys such as Euclid aim to make precise measurements down to very small scales corresponding to a multipole, $\ell$, of roughly 5000~\citep{2024arXiv240513491E}. While pixel-related effects can be mitigated by using maps at higher resolution and truncating to scales far above the pixel scale, this increases the run-time of the estimator and can lead to empty pixels that must be accounted for in the interpretation of the power spectra.

In recognition of these challenges, biases due to pixelization-related effects have been studied in several recent works that push the pseudo-$C_\ell$ method to small scales in weak lensing. Ref.~\citep{2021JCAP...03..067N} identified two limiting regimes depending on whether pixels are densely populated or sparsely populated. In the former case, the effect of pixelization is close to that of a convolution of the shear field with the pixel window function, while in the latter case it is closer to a point sampling. Since small-scale lensing analyses lie between these two regimes, Ref.~\citep{2024arXiv240313794G} defined an effective pixel window function based on simulations and found that although the current generation surveys are not sensitive to these effects, near-future surveys will have to carefully account for pixel-related effects in order to avoid large biases in parameters. Ref.~\citep{2023arXiv231212285B} presented a method to bypass the map making step altogether by computing the power spectra via direct summation, and recent advances in developing fast spherical harmonic transforms over arbitrary pixelizations~\citep{2024arXiv240614542B} offer highly promising routes to avoid many of the pitfalls associated with making shear maps.

This paper has several objectives. Firstly, we bring together many of the discussions and derivations of pixelization effects in the standard pseudo-$C_\ell$ method and provide a pedagogical resource for understanding how effects such as window functions, empty pixels, shot noise, aliasing, varying and anisotropic pixel shapes, shear weight variations, and survey masks interact with each other, and how they should be accounted for in a precise analysis. Secondly, we elucidate the approximations that go into the HEALPix pixel window function when applied to shear fields, which is a common method for dealing with pixelization smoothing. Finally, we use analytic modelling and simulations to test a variety of improvements to the standard pseudo-$C_\ell$ method. Our paper is partly a companion paper to Ref.~\citep{PKWL}, where many of the ideas presented here are implemented and rigorously tested in the context of Euclid.

This paper is structured as follows. In section~\ref{sec:pCl}, we introduce the standard pseudo-$C_\ell$ estimator. In section~\ref{sec:aliasing}, we present models for the impact of pixelization on shear power spectra and derive the HEALPix pixel window function approximation for spin-2 fields. This section also presents derivations of power spectrum biases due to finite pixel occupancy and stochasticity of the underlying source galaxy number counts and their shear weights. In section~\ref{sec:simulations}, we test our expressions against simulated shear catalogues. In section~\ref{sec:nn}, we explore an alternative to the standard estimator that does not normalize by the total weight in each pixel, and in section~\ref{sec:fixedpos} we present an alternative measurement and testing procedure that considers the source galaxies and weights as fixed in the analysis. In section~\ref{sec:interlacing}, we present a new method for estimating power spectra that approximately interlaces HEALPix grids. In section~\ref{sec:masked}, we present an alternative approach to constructing pixel window functions that restricts to the survey footprint, and test it against simulations. We present our main conclusions in section~\ref{sec:finalconc}. In a series of appendices, we derive a hitherto overlooked bias due to neglect of a phase factor when constructing shear maps (appendix~\ref{sec:pt}), present a model for the impact of varying pixel shapes that applies on large scales for any pixelization scheme (appendix~\ref{subsec:superpix_pert}), and present auxiliary results on the conditional statistics of density fields (appendix~\ref{app:conddensstats}).

\section{The standard pseudo-$C_\ell$ estimator}
\label{sec:pCl}

We will start with an introductory discussion of the standard pseudo-$C_\ell$ approach to measuring shear power spectra. Fundamentally, this involves taking the harmonic transform of the observed shear field. To make use of the fast spherical harmonic transforms provided by the HEALPix library~\citep{2005ApJ...622..759G}, one first pixelizes the catalogue to make a shear map. This may be done, for example, by taking a weighted average of galaxy ellipticity measurements in HEALpix pixels~\citep{2021JCAP...03..067N, 2021JCAP...10..030G, 2021arXiv211006947L, 2022arXiv220307128D}. This map is given by
\begin{equation}
\label{eq:wrongeq0}
    \hat{\gamma}_p = \frac{\sum_{i \in p} w_i \, \hat{\gamma}_i}{\sum_{i\in p}w_i},
\end{equation}
where $\gamma = \gamma_1 + {\mathrm{i}}\gamma_2$ is the complex shear, $
\hat{\gamma}_i$ is the measured galaxy ellipticity (or any proxy for the local shear), $w_i$ are galaxy weights, and the sums run over galaxies within pixel $p$. Equation~\eqref{eq:wrongeq0} neglects a phase factor required to consistently sum the spin-2 ellipticity at different points on the sphere. This factor is briefly discussed and quantified in Ref.~\cite{PKWL}, where it is found to be negligible for Euclid. We investigate this term in further detail in appendix~\ref{sec:pt}.

Once a shear map has been formed, the next step is to extract its spherical multipole coefficients. The standard procedure is to form the $E$ and $B$ pseudo-multipoles given by
\begin{equation}
    \tilde{E}_{\ell m} \pm \mathrm{i} \tilde{B}_{\ell m} =  \Omega\sum_{p=1}^{N_{\mathrm{pix}}}  A_p \, \omega_p \, (\hat{\gamma}_{1,p} \pm  \mathrm{i}\hat{\gamma}_{2,p}) \, {}_{\pm 2}Y_{\ell m}^*(\hat{n}_p),
    \label{eq:map2alm}
\end{equation}
where $\omega_p$ are pixel weights, $A_p$ is the survey mask and $\Omega$ is the pixel area (constant for HEALPix at a given resolution). $N_{\rm pix}$ is the total number of pixels on the sphere, related to the HEALPix resolution parameter $N_{\rm side}$ by $N_{\rm pix} = 12 N_{\rm side}^2$. The pixel weights $\omega_p$ can be freely chosen to optimise the estimator\footnote{For clarity, we will omit from this discussion the per-pixel HEALPix quadrature weights that should be included when going from maps to multipoles from our expressions. The weights are very close to unity for almost all pixels and are included in our numerical results.}. The rule of thumb, as proposed in Ref.~\citep{2004MNRAS.349..603E}, is to choose uniform weights in the signal dominated regime and inverse variance weights in the noise dominated regime\footnote{Formally, for Gaussian fields the optimal weighting derives from the Quadratic Maximum Likelihood (QML) estimator, which effectively interpolates between these two regimes. For Stage-IV weak lensing surveys like Euclid, it has been shown that the statistical precision of the pseudo-$C_\ell$ estimator is close to that of the optimal QML estimator on small scales~\citep{2023MNRAS.520.4836M}.}. In this work, we will consider two choices for the weights; uniform weights with $\omega_p = 1$, and an approximate inverse variance weighting with $\omega_p = \sum_{i \in p}w_i$, which is equivalent to not dividing by the local pixel weight when forming the shear map.

Finally, spectra are constructed from the pseudo-multipoles by cross-correlating and averaging over the azimuthal index. For a single map, we have
\begin{equation}
    \tilde{C}^{EE}_{\ell} = \frac{1}{2\ell+1}\sum_{m=-\ell}^\ell \tilde{E}_{\ell m} \tilde{E}_{\ell m}^*,
\end{equation}
with analogous expressions for the $BB$ and $EB$ spectra.

The quadrature scheme in Equation~\eqref{eq:map2alm} is known to be inexact for the HEALPix sampling of the sphere, in the sense that the true pseudo-multipoles are not exactly recovered (for alternative sampling schemes, see Ref.~\cite{2011ITSP...59.5876M}). Nevertheless, for band-limited shear fields with spectra having no support beyond $\ell_{\mathrm{Ny}} \equiv 3N_{\mathrm{side}} -1$, the quadrature scheme can be very accurate when using the pixel weights provided with HEALPix, or by applying an iterative scheme. The reverse transform then recovers the (weighted, masked) signal with high precision:
\begin{equation}
    A_p \, \omega_p \, (\hat{\gamma}_{1,p} \pm  \mathrm{i}\hat{\gamma}_{2,p}) \approx \sum_{\ell=2}^{\ell_{\mathrm{Ny}}} \sum_{m=-\ell}^\ell(\tilde{E}_{\ell m} \pm  \mathrm{i} \tilde{B}_{\ell m}) \, {}_{\pm 2}Y_{\ell m}(\hat{n}_p).
    \label{eq:alm2map}
\end{equation}
In other words, the (spin-weighted) spherical harmonics evaluated at the HEALPix pixel centres can be considered a complete set of orthogonal (with respect to the quadrature weights) functions that describe band-limited functions on the sphere. The convergence of the iterative scheme can however be very slow at $\ell$ close to the Nyquist frequency $\ell_{\mathrm{Ny}} $, so reliable results are most easily obtained for $\ell \lesssim 2N_{\mathrm{side}}$, as discussed in the HEALPix \texttt{anafast} documentation. In reality of course the true shear field is not band-limited, and the pseudo-multipoles extracted with Equation~\eqref{eq:map2alm} will suffer from aliasing. As we shall see, for sufficiently red power spectra a degree of low-pass filtering is naturally imposed by the averaging of galaxy ellipticities into pixels. This mitigates aliasing due to sub-pixel modes, but there is still a residual impact on the multipoles.

The standard pseudo-$C_\ell$ estimator may be refined in several ways. For example, as introduced in Ref.~\cite{PKWL}, one may average galaxy ellipticities not in HEALPix pixels but in spherical caps centered on the pixels of a HEALPix map, thus making the shear map a point sampling of a true convolution. Alternatively, and as also explored in Ref.~\cite{PKWL}, one may choose not to divide by the sum of weights when forming the shear map in Equation~\eqref{eq:wrongeq0}, instead forward modelling this term, equivalent to the approximate inverse-variance weighting discussed above. We shall investigate the consequences of this approach in section~\ref{sec:aliasing}. Finally, one may try to avoid making shear maps entirely by directly taking the spherical harmonic transform of the catalogue, as explored in Refs.~\citep{2023arXiv231212285B, 2024arXiv240721013W, PKWL}.

\section{Modelling the bias from pixelization}
\label{sec:aliasing}

The shear map produced by the estimator Equation~\eqref{eq:wrongeq0} does not correspond to the true shear field sampled at pixel centres. The averaging procedure amounts to approximately a convolution of the shear with the pixel window (referred to as a pseudo-convolution in Ref.~\cite{PKWL}). Theory predictions for the expectation value of the pseudo-$C_\ell$ need to account for this window function, and often do so by including the square of the HEALPix-provided window function, $W_\ell$, a function of the map resolution $N_{{\rm side}}$, as a multiplicative factor in theory predictions, i.e.
\begin{equation}
    C_\ell \rightarrow W_\ell^2 C_\ell.
\end{equation}
This simple correction does not account for aliasing due to the presence of shear modes with frequency higher than the Nyquist frequency of the HEALPix grid, nor does it account for the fact that the HEALPix pixels vary in shape over the sky. While the discreteness of the galaxy catalogue imposes a degree of low-pass filtering, this is inexact and dependent on the particular realisation of the source galaxy distribution\footnote{ Ref.~\citep{2024arXiv240721013W} discusses the distinction between these effects in their appendix B, finding that the HEALPix pixel window function performs well in the case of band-limited catalogues. In this paper, we are primarily concerned with the more realistic scenario of catalogues which are not band limited. The resulting maps display more sensitivity to pixel scales.}.

\begin{figure}
\centering
    \includegraphics[width=0.45\columnwidth]
    {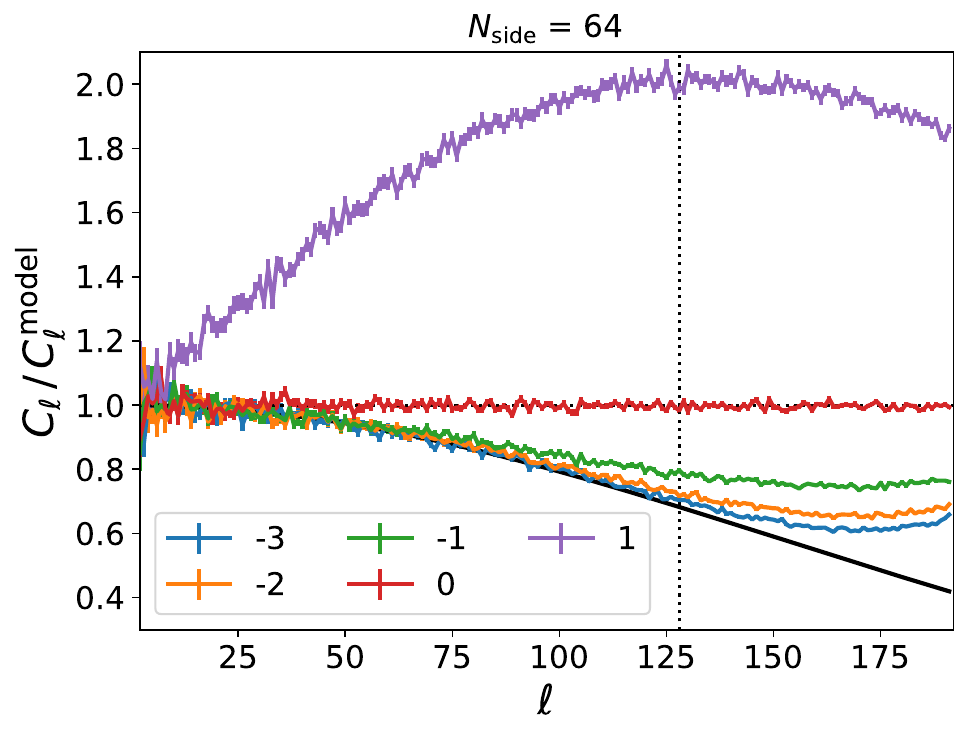}
    \includegraphics[width=0.45\columnwidth]
    {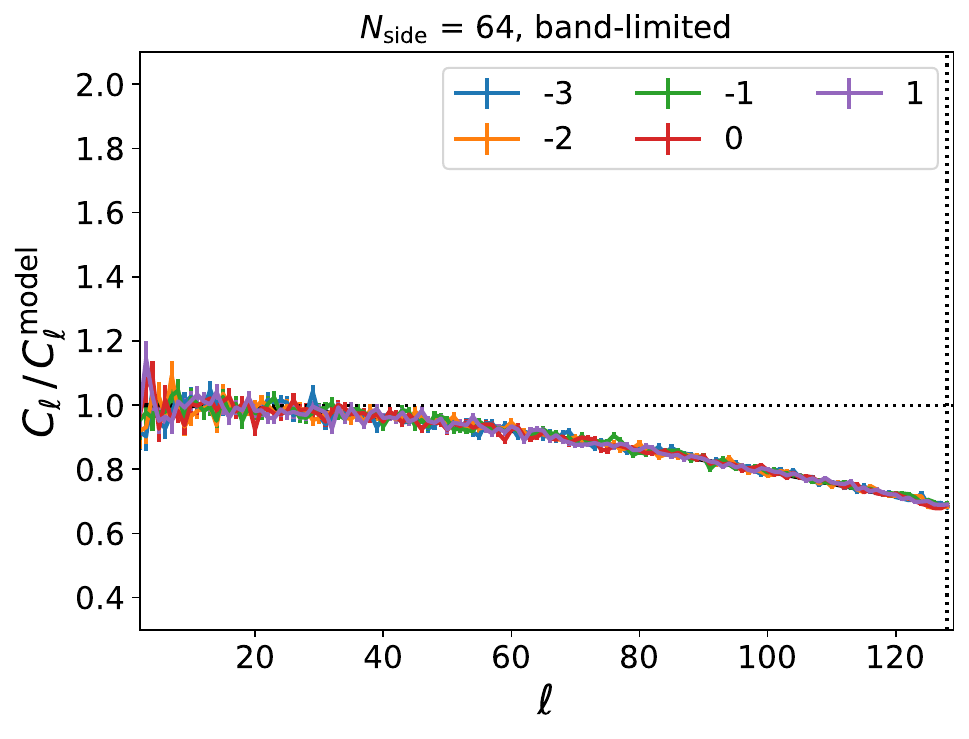}
    \caption{\emph{Left}: Ratio of recovered to input power spectra for power law $C_\ell$ having exponents indicated by the different coloured lines. Error bars are derived from 100 simulations of $N_{{\rm side}}=512$ maps averaged in $N_{{\rm side}}=64$ pixels. The black curve is the HEALPix pixel window function. The vertical dotted line indicates $\ell = 2 N_{{\rm side}}$. \emph{Right}: same as left panel for spectra having the indicated power-law slopes but set to zero above $2N_{\mathrm{side}}$ for $N_{{\rm side}}=64$.} 
    \label{fig:pows}
\end{figure}

The residual biases due to these effects are illustrated in the left panel of Figure~\ref{fig:pows}. In this example, aliasing is by far the main contributor to the bias. We make high-resolution full-sky spin-0 maps drawn from power spectra having a power law form $C_\ell \propto \ell^n$, with $n$ varied between $-3$ and $1$. Realistic cosmic shear spectra have $n$ between $-1.5$ and $-2$ at $\ell \approx 5000$. We normalize all spectra to have the same variance, and draw Gaussian realizations on an $N_{{\rm side}}=512$ grid. We then average the maps in pixels of an $N_{{\rm side}}=64$ grid, and compute the power spectrum. Figure~\ref{fig:pows} shows the ratio of the resulting spectra to the input model, averaged over 100 maps. The prediction based on the HEALPix pixel window function for $N_{{\rm side}}=64$ is shown as the black curve. All spectra tend to the input at low $\ell$, but large biases are seen at higher $\ell$. These biases become more severe the bluer the input spectrum. For $\Lambda$CDM-like exponents $n=-2$ and $n=-1$, the HEALPix pixel window function provides a reasonably good fit, with residuals of several percent at $\ell = N_{{\rm side}}$ but growing rapidly at higher $\ell$ due to aliasing. The blue spectrum $n=1$ is extremely biased by aliasing for all but the lowest $\ell$. An important feature to note is that although the white noise spectrum $n=0$ is also very biased, it remains white and therefore easily predictable. This is why noise terms are understood to be exempt from the pixel window correction, but the figure demonstrates that the issue is really to do with the power on pixel scales. The right panel of Figure~\ref{fig:pows} shows the ratio of the recovered power spectrum to the model after the input spectra have been set to zero above $\ell = 2 N_{{\rm side}}$. These maps are hence band limited, and are immune from aliasing. In this case, the HEALPix pixel window function provides an excellent fit for all input spectra, demonstrating that the biases seen in the left panel of Figure~\ref{fig:pows} are due to aliasing rather than a breakdown of the HEALPix pixel window function approximation. Later on we will explore situations where aliasing is sub-dominant and corrections to the pixel window function to account for anisotropic pixels must be made.

A derivation of the HEALPix pixel window function for spin-0 fields can be found in the HEALPix documentation\footnote{Specifically, \url{https://healpix.jpl.nasa.gov/html/intronode14.htm}.}, and implementations for spin-0 fields (e.g.~CMB temperature) and spin-2 fields (e.g.~CMB polarization) are both available through HEALPix. However, it is not clear how the approximation is derived for spin-2 fields that are discretely sampled, which is the situation in cosmic shear. The absence of beam smoothing in this case also demands that inaccuracies in the pixel window function and sensitivity to aliasing should be studied carefully. In this section, we provide that study, testing our modelling against simple simulations, and show how the HEALPix pixel window function approximation is derived for discrete spin-2 catalogues.

\subsection{Exact expressions for pixel window functions}
\label{subsec:superpix_exact}

To decouple the effects of cosmic shear fluctuations and source galaxy number density fluctuations, we will write the shear map (including the correct phase factor, $\beta_p$, derived in appendix~\ref{sec:pt}) as
\begin{equation}
    (\hat{\gamma}_1 \pm  \mathrm{i}\hat{\gamma}_2)_p = \frac{1}{N_p w_{(1)}(p)} \int \mathrm{d}^2\hat{n} \, W_p(\hat{n})\, n_w(\hat{n})\,(\gamma_1 \pm  \mathrm{i}\gamma_2)(\hat{n}) \,e^{-2 \mathrm{i}\beta_p(\hat{n})},
    \label{eq:hatgammaallscales}
\end{equation}
%
%
%
%
%
where $w_{(1)}(p) \equiv \sum_{i \in p} w_i/N_p$, for non-empty pixels, is the mean weight in pixel $p$, $N_p$ is the number of galaxies in pixel $p$, the pixel window function is defined as
\begin{equation}
    W_p(\hat{n}) = \left\{\begin{array}{lr}
        1, & \hat{n} \in p \\
        0, & \text{otherwise,}
        \end{array}\right. 
\end{equation}
and the weighted source density field is
\begin{equation}
    n_w(\hat{n}) \equiv \sum_{i=1}^{N_{\mathrm{tot}}} w_i \, \delta^D(\hat{n} - \hat{n}_i).
    \label{eq:nwdef}
\end{equation}
Importantly, in the case of a pixel having no source galaxies inside it, the shear map is set to \emph{zero} in that pixel. Inserting the spin-weighted spherical harmonic expansion of the shear field gives $(\hat{\gamma}_1 \pm i\hat{\gamma}_2)(\hat{n}_p) = \sum_{\ell m}(E_{\ell m} \pm iB_{\ell m}) \, {}_{\pm2}\Upsilon_{\ell m}^{p}$ where
\begin{equation}
    {}_{\pm2}\Upsilon_{\ell m}^{p} = \frac{1}{N_p w_{(1)}(p)} \int \mathrm{d}^2\hat{n} \, W_p(\hat{n}) \, n_w(\hat{n}) \, {}_{\pm2}Y_{\ell m}(\hat{n}) \, e^{-2 \mathrm{i}\beta_p(\hat{n})},
    \label{eq:ups}
\end{equation}
are the pixelized spin-weighted spherical harmonics.

%
%

%
%

The HEALPix pixel window function approximation is based on the idea that the pixels are axisymmetric `on average'. Axisymmetry of the pixels must be defined in a frame aligned with the centroid of the pixel. Therefore, consider transforming the integral in Equation~\eqref{eq:ups} to a frame having its $z$-axis aligned with the pixel centre $\hat{n}_p$. Let the required rotation from the original frame to this new frame be defined by the Euler angles $\{\phi_p, \theta_p, 0\}$, where $\hat{n}_p = (\theta_p, \phi_p)$ in the global (original) coordinate system. Denote this rotation by $R$. The image of the $x$-axis under this rotation, $x'$, is now parallel to the $\mathbf{e}_\theta$ basis vector at $\hat{n}_p$, so an arbitrary point $\hat{n}$ has a $\phi'$ coordinate in the rotated frame of $\pi + B(\hat{n}, \hat{n}_p)$, where the angle $B$ is defined in Figure~\ref{fig:sphere}. The new $(\mathbf{e}_{\theta'}, \mathbf{e}_{\phi'})$ basis at $\hat{n}$ is rotated by an angle $B(\hat{n}, \hat{n}_p)$ in a left-handed sense about $\hat{n}$ compared to $(\mathbf{e}_{\theta}, \mathbf{e}_{\phi})$. As shown in Ref.~\citep{2000PhRvD..62l3002C}, the tensor spherical harmonics in the new frame are related to those in the original frame by
\begin{equation}
    Y_{\ell m, ab}' = \sum_{m'} D^{\ell}_{m'm}(\phi_p, \theta_p, 0) Y_{\ell m', ab},
\end{equation}
where $D^\ell(R)$ is the Wigner $D$-matrix representation of $R$. Since the spin-weighted spherical harmonics are ${}_{\pm 2}Y_{\ell m}(\hat{n}) \propto e_{\pm}^a e_{\pm}^b Y_{\ell m,ab}$, we can relate the spin-weighted harmonics in the original frame to those in the new frame by
\begin{equation}
    {}_{\pm 2}Y_{\ell m}(\hat{n}) = e^{\mp 2 \mathrm{i}A(\hat{n}, \hat{n}_p)}  \sum_{m'} D^{\ell *}_{mm'}(\phi_p, \theta_p, 0) {}_{\pm 2}Y_{\ell m'}'(\hat{n}),
\end{equation}
where the angle $A$ is defined in Figure~\ref{fig:sphere}. Substituting this into Equation~\eqref{eq:ups}, using standard relations between the spin-weighted spherical harmonics and the Wigner $D$-matrix elements, and noting that $W_p(\hat{n}) = W_p'(\hat{n}')$ and similarly for the weighted number density, we can write 
%
%
%
%
%
%
%
\begin{equation}
    {}_{\pm2}\Upsilon_{\ell m}^{p} = \sum_{|m'|\leq \ell} {}_{m'}Y_{\ell m}(\hat{n}_p) \, {}_{m'}W_{\ell,p}^{\pm}
    \label{eq:conv_theorem}
\end{equation}
where we have introduced the spin $s=\pm2-m$ quantities
\begin{equation}
    {}_{m}W_{\ell,p}^{\pm} \equiv \frac{1}{N_p w_{(1)}(p)} \int \mathrm{d}^2 \hat{n} \, W_p(\hat{n}) \, n_w(\hat{n})\, d_{m \pm2}^\ell(\theta)\, e^{(\pm2 - m)\mathrm{i}\phi},
    \label{eq:mWpm}
\end{equation}
where $|m| \leq \ell$, and the integral is performed in a coordinate system having its $z$-axis aligned with the pixel centre. Note that ${}_2 W_{\ell,p}^+ =  {}_{-2} W_{\ell,p}^-$.

Equation~\eqref{eq:conv_theorem} shows that pixelization cannot generally be accounted for with a simple factor multiplying the harmonics. Instead, the harmonic-space pixel window functions couple to spherical harmonics of higher spin and depend on the pixel shape. Furthermore, the window functions depend on the source density realization. However, as we shall see, for \emph{axisymmetric} pixels the harmonic-space window functions simplify substantially when averaged over source positions. The azimuthal integral in Equation~\eqref{eq:mWpm} is then trivial and equals $2\pi \delta_{m \mp 2}$. This highlights the advantage of transforming the convolution integral to the pixel-aligned frame.
\footnote{
In appendix~\ref{subsec:superpix_pert} we give a perturbative expression for ${}_{\pm2}\Upsilon_{\ell m}^{p}$ averaged over galaxy positions valid on scales far above the pixel scale. To show that Equation~\eqref{eq:conv_theorem} agrees with this expression, first average Equation~\eqref{eq:conv_theorem} over source positions. We then have
\begin{equation}
    \langle {}_{\pm2}\Upsilon_{\ell m}^{p} \rangle = \frac{1}{\Omega}\sum_{|m'|\leq \ell} {}_{-m'}Y_{\ell m}(\hat{n}_p) \int \mathrm{d}^2\hat{n}' \, W_p'(\hat{n}') \, d^\ell_{-m' \pm 2}(\theta') \, e^{ \mathrm{i}(m' \pm 2)\phi'}. \nonumber
\end{equation}

On scales far above the pixel scale we can Taylor expand the Wigner $d$-function around $\theta'=0$ using the expressions in Ref.~\citep{1988qtam.book.....V}. To first order this gives
\begin{equation}
    \langle {}_{\pm2}\Upsilon_{\ell m}^{p} \rangle \approx {}_{\pm 2}Y_{\ell m}(\hat{n}_p) - \frac{1}{2} \eth^{\mp} {}_{\pm 2}Y_{\ell m}(\hat{n}_p) \langle {}_{\pm 1} d \rangle_p - \frac{1}{2}\eth^{\pm} {}_{\pm 2}Y_{\ell m}(\hat{n}_p) \langle {}_{\mp 1} d \rangle_p + \dots \nonumber
\end{equation}
where $\eth^{+} \equiv  \eth$ and $\eth^{-} \equiv \bar{\eth}$, and ${}_{\pm 1} d$ are the spin $\pm1$ displacements on the global frame, as in appendix~\ref{subsec:superpix_pert}. This agrees with the leading-order term in Equation~\eqref{eq:pixbias}. Note that $\langle {}_{\pm 1} d \rangle_p = 0$ since the pixel centres are chosen to be the geometric centre of the pixel.}

We can now form the correlation functions for a given pair of pixels. Averaged over realizations of the shear field these are
\begin{equation}
    \xi_{\pm}(p,p') = \sum_{\ell} \frac{(2\ell+1)}{4\pi} (C_{\ell}^E \pm C_{\ell}^B) \sum_{m,m'} {}_m W_{\ell,p}^+ \, {}_{m'} W_{\ell,p'}^{\pm \, *} \, d^{\ell}_{mm'}(r_{pp'}).
    \label{eq:xipm_unaveraged}
\end{equation}
Similarly, taking the expression for the pseudo-multipoles Equation~\eqref{eq:map2alm}, we can write the angular power pseudo-spectra averaged over shears as
\begin{equation}
\tilde{C}_L^E \pm \tilde{C}_L^B = \frac{\Omega^2}{4\pi}\sum_p \sum_q  A_p \, A_q \, \omega_p \, \omega_q \, \xi_{\pm}(p,q) \, d^L_{2\pm2}(r_{pq}).
\label{eq:pcl_pure}
\end{equation}

These expressions are analytically intractable, but represent mathematically what pixelization actually does to the underlying shear field. The fact that a major simplification of Equation~\eqref{eq:mWpm} is possible if the pixels are axisymmetric and the source number density is smooth pushes us to consider the effects of pixelization after averaging over the ensemble of possible source galaxy positions.


%
%
%
%

\subsection{The impact of pixelization after averaging over unclustered sources}
\label{subsec:unclustered}

The simplest source galaxy distribution we can consider is a Poisson process. First, consider taking the expectation value of the pixel window functions over unclustered galaxy positions at fixed pixel occupancy $N_p>0$. Using the conditional moments of the weighted density field given in appendix~\ref{app:conddensstats}, we have
\begin{equation}
    \langle {}_{m}W_{\ell,p}^{\pm} \rangle = \frac{1}{\Omega} \int \mathrm{d}^2 \hat{n} \, W_p(\hat{n})\, d_{m \pm2}^\ell(\theta)\, e^{(\pm2 - m) \mathrm{i}\phi},
\end{equation}
and therefore axisymmetric pixels enforce $m = \pm 2$. Note that on small scales we have $d^{\ell}_{22}(\theta) \approx d^{\ell}_{00}(\theta) \approx J_0(\ell \theta)$, so the spin-2 window function asymptotes to the spin-0 window function. 

Using Equation~\eqref{eq:wncov}, the second moments of the window functions averaged over unclustered source positions are
\begin{align}
    \langle {}_m W_{\ell,p}^{+} \, {}_{m'}W_{\ell,p'}^{\pm \, *} \rangle =& \left[1 - \frac{\delta_{pp'}}{N_p^{\mathrm{eff}}}\right] \langle {}_m W_{\ell,p}^{+} \rangle \langle  {}_{m'}W_{\ell,p'}^{\pm \, *} \rangle \nonumber \\
    &+ \frac{\delta_{pp'}}{N_p^{\mathrm{eff}}}\int \frac{\mathrm{d}^2\hat{n}}{\Omega} \, W_p(\hat{n}) \, d^\ell_{m2}(\theta) \, d^{\ell}_{m'\pm2}(\theta) \, e^{(2\mp2 +m' - m) \mathrm{i}\phi},
\end{align}
where we have defined the effective number of galaxies in each pixel as
\begin{equation}
    N_p^{\mathrm{eff}} \equiv N_p \, \frac{w_{(1)}^2(p)}{w_{(2)}(p)},
\end{equation}
with $w_{(2)}(p) \equiv \sum_{i \in p} w_i^2/N_p$ for non-empty pixels.

Using these expressions, we can write down the correlation functions averaged over the shear field and unclustered source positions at fixed pixel occupancy,
\begin{equation}
\xi_+(p,p') =
\begin{dcases}
\sum_{\ell} \frac{(2\ell+1)}{4\pi} (C_{\ell}^E + C_{\ell}^B) \sum_{m,m'} \langle {}_m W_{\ell,p}^+ \rangle \langle {}_{m'} W_{\ell,p'}^{+ \, *} \rangle  d^{\ell}_{mm'}(r_{pp'}) \quad \quad (p \neq p')  \\
\frac{\sigma^2_\gamma}{N_p^{\mathrm{eff}}}  + \left(1-\frac{1}{N_p^{\mathrm{eff}}}\right) \sum_{\ell} \frac{(2\ell+1)}{4\pi} (C_{\ell}^E + C_{\ell}^B) \sum_m \big \lvert \langle {}_m W_{\ell,p}^+ \rangle \big \rvert^2   \quad (p=p') 
\end{dcases}
\label{eq:xip_map}
\end{equation}
and
\begin{equation}
\xi_-(p,p') =
\begin{dcases}
\sum_{\ell} \frac{(2\ell+1)}{4\pi} (C_{\ell}^E - C_{\ell}^B) \sum_{m,m'} \langle {}_m W_{\ell,p}^+ \rangle \langle {}_{m'} W_{\ell,p'}^{- \, *} \rangle  d^{\ell}_{mm'}(r_{pp'}) \quad \quad (p \neq p')  \\
\left(1-\frac{1}{N_p^{\mathrm{eff}}}\right)  \sum_{\ell} \frac{(2\ell+1)}{4\pi} (C_{\ell}^E - C_{\ell}^B) \sum_m \langle {}_m W_{\ell,p}^+ \rangle \langle {}_{m} W_{\ell,p}^{- \, *} \rangle  \quad (p=p'),
\end{dcases}
\label{eq:xim_map}
\end{equation}
where $\sigma^2_\gamma$ is the ellipticity (shear plus intrinsic shape) variance at a point. In the case that only a single galaxy is present in the survey, $\xi_+(p,p') = \delta_{p,p'} \sigma^2_\gamma$ and $\xi_-(p,p') = 0$, i.e.~pixelization has no effect due to isotropy. For axisymmetric pixels $\xi_-(p,p) = 0$ for any $N_p$, as in the unpixelized case.

\sloppy As well as the pixel smoothing exhibited in the mode sums in Equations~\eqref{eq:xip_map} and \eqref{eq:xim_map}, stochasticity in the source galaxy field adds a shot noise term with amplitude $\sigma_{\gamma}^2/N_p^{\mathrm{eff}}$. The map variance therefore consists of a genuine shot noise term depending on the variance at a point, plus the shear correlation function smoothed within a pixel. Which of these is more important depends on the small-scale behaviour of the shear power spectrum, but it is the first term which is usually termed `shape noise' in weak lensing analyses.


%
%
%
%
%


Using Equations~\eqref{eq:xip_map} and \eqref{eq:xim_map} we can derive the impact of pixelization on the angular power spectrum, which is the focus of this paper. Substituting these expressions into Equation~\eqref{eq:pcl_pure} and assuming zero intrinsic $B$-mode gives
%
%
%
\begin{align}
\tilde{C}_L^{E/B} &= \frac{\Omega^2}{8\pi}\sum_{p, q}  \tilde{A}_p \tilde{A}_q  \sum_{\ell,m,m'} \frac{(2\ell+1)}{4\pi} C_{\ell}^E   \langle {}_m W_{\ell,p}^+ \rangle d^{\ell}_{mm'}(r_{pq}) \nonumber \\
&\times\left[\langle {}_{m'} W_{\ell,q}^{+ *} \rangle   d^L_{22}(r_{pq}) \pm \langle {}_{m'} W_{\ell,q}^{- *} \rangle\, d^L_{2-2}(r_{pq})\right]\nonumber \\
&+ \frac{\Omega^2}{8\pi}\sum_{p} \frac{\tilde{A}_p^2}{N_p^{\mathrm{eff}}} \left[\sigma^2_\gamma -\sum_{\ell} \frac{(2\ell+1)}{4\pi} C_{\ell}^E \sum_m \big \lvert \langle {}_m W_{\ell,p}^+ \rangle \big \rvert^2   \right],  \label{eq:EEBB_full}
\end{align}
where we have defined the quantity $\tilde{A}_p \equiv A_p \omega_p$, the product of the survey mask and the pixel weights. In the pure noise case and with non-overlapping pixels we have
\begin{equation}
\tilde{N}_L^E = \tilde{N}_L^B = \frac{\Omega^2}{8\pi}\sum_{p} \frac{\tilde{A}_p^2}{N_p^{\mathrm{eff}}} \sigma_e^2,
\label{eq:shapenoisevar}
\end{equation}
where $\sigma_e^2$ is the total intrinsic shape variance.

Equation~\eqref{eq:EEBB_full} looks cumbersome, but its interpretation is clear. The first two lines express the pseudo-power at a given multipole in terms of the shear power smoothed on the pixel scale by pixel-dependent window functions, and then coupled with the survey window. The third line is additional white noise power coming from the mean-square ellipticity variance within each pixel, which scales inversely with the number of galaxies in each pixel. The shear variance which contributes effectively uses the \emph{complement} of the pixel window function(s), i.e. only \emph{sub-pixel modes} contribute to this white noise term. Note that there is non-zero $B$-mode power generated by pixelization even on the full sky. This comes from both the white noise term on the last line of Equation~\eqref{eq:EEBB_full}, but also from the aliasing and the shape dependence of the HEALPix pixels in the smoothing terms on the first two lines of Equation~\eqref{eq:EEBB_full}.

\subsection{Axisymmetry and the HEALPix window approximation}
\label{subsubsec:healpix}

In section~\ref{subsec:superpix_exact} we saw that the impact of pixelization could be substantially simplified assuming axisymmetric pixels and averaging over source positions. In section~\ref{subsec:unclustered} we derived the pixelized correlation functions and angular power spectra after source averaging. This made it clear how pixelization impacts shot noise, but we are now in a position to simplify these expressions further by considering axisymmetric pixels.

First, consider the standard HEALPix window functions. These are built under the assumption that the pixelized harmonics averaged over source positions can be written as
\begin{equation}
    \langle {}_{\pm2}\Upsilon_{\ell m}^{p} \rangle \approx   {}_{\pm 2}Y_{\ell m}(\hat{n}_p) \, W_{\ell}^p ,
    \label{eq:hpx}
\end{equation}
for some spin-0 window function $ W_{\ell}^p  $. A sky-averaged window is then built as
\begin{equation}
    W_{\ell} ^2 \equiv \frac{1}{N_{\mathrm{pix}}} \sum_{p} (W_{\ell}^p)^2 = \frac{\Omega}{4\pi}\sum_{p} (W_{\ell}^p)^2.
\end{equation}
As we have seen, the source-averaged window functions satisfy Equation~\eqref{eq:hpx} only when we enforce axisymmetry in the pixels. In this case, we have (we will not yet assume that all the pixels are the same size)
\begin{equation}
     W_{\ell}^p  = \langle {}_{2}W_{\ell,p}^{+} \rangle = \frac{2\pi}{\Omega} \int_0^\pi \mathrm{d}\theta \, \sin\theta \, W_p(\theta)\, d_{22}^\ell(\theta).
\end{equation}
The correlation functions in this case become
\begin{equation}
\xi_+(p,p') =
\begin{dcases}
\sum_{\ell} \frac{(2\ell+1)}{4\pi} (C_{\ell}^E + C_{\ell}^B) W_{\ell}^p \,  W_{\ell}^{p'}  d^{\ell}_{22}(r_{pp'}) \quad \quad (p \neq p') \nonumber \\
\frac{\sigma^2_\gamma}{N_p^{\mathrm{eff}}} + \left(1-\frac{1}{N_p^{\mathrm{eff}}}\right) \sum_{\ell} \frac{(2\ell+1)}{4\pi} (C_{\ell}^E + C_{\ell}^B)   (W_{\ell}^p)^2   \quad (p=p') 
\end{dcases}
\label{eq:xip_map_asym}
\end{equation}
and
\begin{equation}
\xi_-(p,p') =
\begin{dcases}
\sum_{\ell} \frac{(2\ell+1)}{4\pi} (C_{\ell}^E - C_{\ell}^B)   W_{\ell}^p \, W_{\ell}^{p'}  d^{\ell}_{2-2}(r_{pp'}) \quad \quad \quad \quad \quad \quad(p \neq p') \nonumber \\
0 \quad (p=p') .
\end{dcases}
\label{eq:xim_map_asym}
\end{equation}
Note that the approximation of axisymmetric pixels means $\xi_-(p,p')$ now vanishes for $p=p'$. The pseudo-spectra, assuming zero intrinsic $B$-mode, become
\begin{align}
\tilde{C}_L^{E/B} =& \frac{\Omega^2}{8\pi}\sum_{p, q} \tilde{A}_p \tilde{A}_q  \, \sum_{\ell} \frac{(2\ell+1)}{4\pi} C_{\ell}^E   W_{\ell}^p  \, W_{\ell}^{q}  \left[d^{\ell}_{22}(r_{pq}) \, d^L_{22}(r_{pq}) \pm d^{\ell}_{2-2}(r_{pq})\, d^L_{2-2}(r_{pq})\right]\nonumber \\
&+ \frac{\Omega^2}{8\pi}\sum_{p} \frac{\tilde{A}_p^2}{N_p^{\mathrm{eff}}} \left[\sigma^2_\gamma -\sum_{\ell} \frac{(2\ell+1)}{4\pi} C_{\ell}^E (W_{\ell}^{p})^2   \right].  \label{eq:EEBB_full_asym}
\end{align}

Equation~\eqref{eq:EEBB_full_asym} suggests that it is difficult to control the accuracy of the axisymmetry approximation, due to the coupling of small-scale modes to the power at a given $L$ due to both the survey mask and the size variation of the pixels across the survey. Both may be mitigated by restricting to sufficiently low $L$; although this does not control the mode coupling in the white noise terms in the second line of Equation~\eqref{eq:EEBB_full_asym}, these terms can be estimated from the data directly and subtracted from the power spectrum estimator (note that in the presence of shape noise, $\sigma_\gamma^2$ should be replaced with $\sigma_\gamma^2 + \sigma_e^2$).

\begin{figure}
\centering
    \includegraphics[width=0.8\columnwidth]{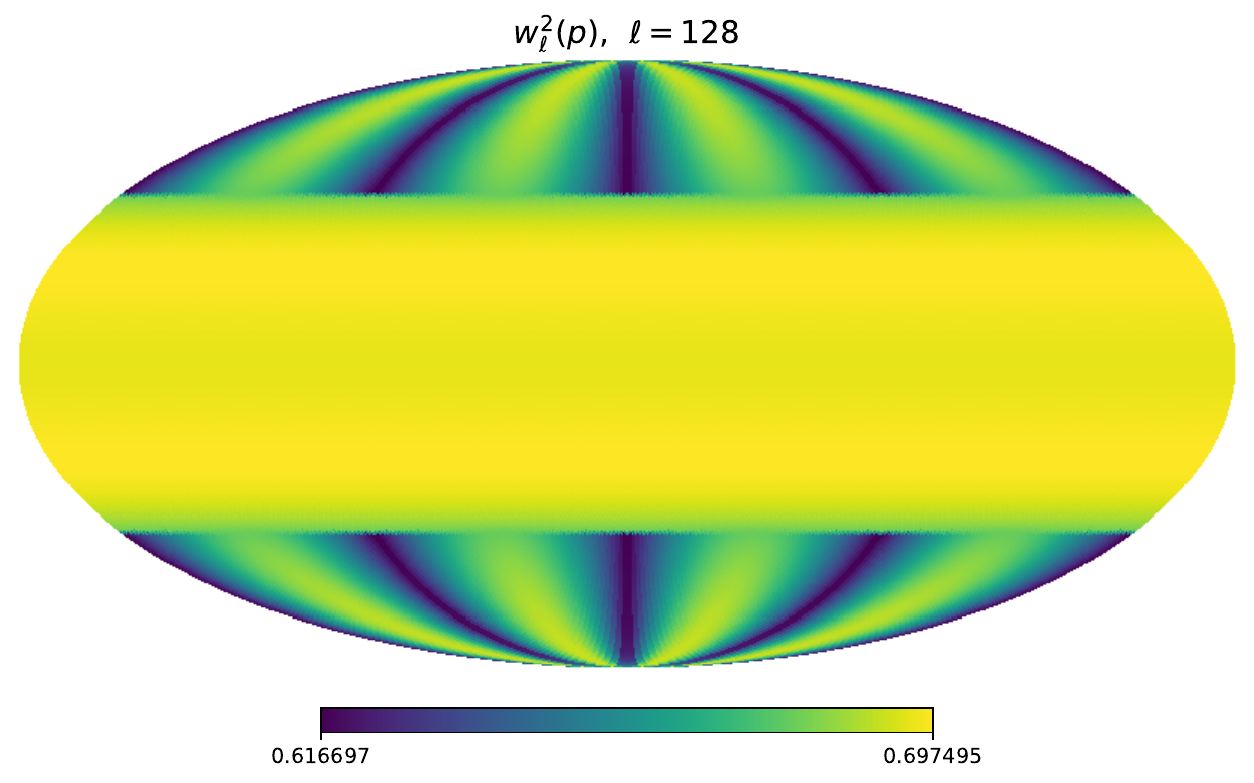}
    \caption{Square of the spherically averaged per-pixel window function, for $\ell=128$. The average of this quantity over the map gives the nominal HEALPix window function for this $\ell$.} 
    \label{fig:wlp}
\end{figure}

In the axisymmetric approximation the source-averaged pixel window functions are azimuthally symmetric, but their size still varies over the sky. The next step in the HEALPix approximation is to assume that the angular scale of this variation is much smaller than the modes of interest. Again, this is difficult to control for, particularly for the white noise from sub-pixel scales. In Figure~\ref{fig:wlp} we plot $ (W_\ell^p)^2$ for $\ell=128$. The large-scale spatial structure is roughly independent of $\ell$, and is very similar to the pattern of pixel inertia shown in Figure~\ref{fig:varMoI}. Variations are rather modest, with a maximal spread of less than 10\% of the median. The equatorial region is particularly stable. The assumption of constant shape therefore seems reasonably accurate, and can be made by replacing $(W_\ell^p)^2$ by its average over all pixels, $W_\ell^2$. This gives
\begin{align}
\tilde{C}_L^{E/B} =& \frac{\Omega^2}{8\pi} \sum_{\ell} \frac{(2\ell+1)}{4\pi} C_{\ell}^E   W_{\ell}^2 \sum_{p, q}  \tilde{A}_p \tilde{A}_q  \left[d^{\ell}_{22}(r_{pq}) \, d^L_{22}(r_{pq}) \pm d^{\ell}_{2-2}(r_{pq})\, d^L_{2-2}(r_{pq})\right]\nonumber \\
&+ \frac{\Omega^2}{8\pi}\left[\sigma^2_\gamma -\sum_{\ell} \frac{(2\ell+1)}{4\pi} C_{\ell}^E W_{\ell}^2   \right] \sum_{p} \frac{ \tilde{A}_p^2}{N_p^{\mathrm{eff}}}  \label{eq:EEBB_full_hp}.
\end{align}
From these expressions we see that the white noise term produced by pixelization remains in the HEALPix approximation, and depends purely on sub-pixel shear modes. We remind the reader that in the presence of shape noise, $\sigma_\gamma^2$ should be replaced with $\sigma_\gamma^2 + \sigma_e^2$.

Equation~\eqref{eq:EEBB_full_hp} gives the expectation value of the shear spectra, conditioned on a fixed set of (weighted) pixel occupancies, in the HEALPix pixel window function approximation. It has assumed unclustered sources, but is otherwise valid for general survey footprints and pixel weights, and is valid on the curved sky. The derivation of this expression has shown, to our knowledge for the first time, how the HEALPix pixel window function approximation works for discrete sources and spin-2 fields on the curved sky. The resulting isotropic window function $W_\ell$ can be applied to the underlying theory model as a simple multiplicative factor. Equation~\eqref{eq:EEBB_full_hp} also highlights that mode coupling comes from survey inhomogeneities ($\tilde{A}_p \neq 1$) and aliasing (loss of orthogonality between the Wigner $d$-functions even for $\tilde{A}_p = 1$ when $\ell$ exceeds the Nyquist frequency). In the next section, we will further average this expression over pixel occupancy. 

As is apparent from Equation~\eqref{eq:EEBB_full_hp}, the power spectrum at each $L$ is an average over many pixels; on scales far above the pixel scale, we can expect the actual pixel window function to be close to its source ensemble-averaged form. In other words, our analysis of the power spectrum does not have to explicitly average or marginalize over source positions for the HEALPix pixel window function approximation to provide an accurate model on large scales. We will demonstrate this quantitatively in section~\ref{sec:fixedpos}.

We close this subsection by noting that a better approximation for the HEALPix window function would use an average over the pixels in the \emph{observed} region.
%
%
%
We will explore this idea further in section~\ref{sec:masked}. The accuracy of the window function approximation can in any case be easily checked against mock data. A validation can be made, for example, by repeating the shear analysis with a rotated map.

\subsection{Bias due to finite pixel occupancy, occupancy stochasticity, and galaxy weights}
\label{sec:weights_and_Np}

So far, we have assumed two things about the shear weights: that they are uncorrelated with the shear, and independent of the source density. Both of these are untrue in reality, the first because the weights depend on the PSF-convolved galaxy ellipticity which is correlated with the shear, and the second because both the photometry and the shear measurement quality varies with galaxy position, for example due to the effects of blending in crowded regions. There are also effects from the correlation of magnification (affecting size and flux) with the shear that we have neglected. In other words, we are assuming that the weights are uncorrelated with any mass inhomogeneities along our past light cone.

Within these assumptions, we now wish to average the power spectra and correlation functions derived in the previous section over the distribution of the shear weights. This is made complicated by the fact that, for uniform pixel weighting $\omega_p=1$, we divide by the sum of weights in each pixel when forming the shear map, making it impossible to write down an exact expression for the average of $w_{(2)}(p)/w_{(1)}^2(p)$. We can avoid this by considering the weights to be fixed and make no attempt to average over them or by not dividing by the summed weights, equivalent to an approximate inverse-variance pixel weighting. We will consider this latter approach in section~\ref{sec:nn} and \emph{assume uniform pixel weighting for the remainder of this section}. We note that uniform pixel weighting could avoid incurring biases due to source-lens clustering and other related effects, as shown in Ref.~\cite{2024arXiv241115063D}.

In the uniform pixel weighting scenario, we need to average over the weight terms in both the numerator and denominator of the shear map. We will make the approximation that
\begin{equation}
\left \langle \frac{\sum_{i\in p} w_i^2}{\left(\sum_{i \in p}w_i\right)^2} \right \rangle \approx \frac{\left \langle \sum_{i\in p} w_i^2 \right \rangle }{\left \langle \left(\sum_{i \in p}w_i\right)^2 \right \rangle} = \frac{N_p\langle w^2 \rangle}{N_p(N_p-1)\langle w \rangle^2 + N_p\langle w^2 \rangle}.
\label{eq:wratio}
\end{equation}
This approximation has been made assuming $N_p \gg 1$ but it is technically exact for any $N_p$ when the weight `field' has no variation over the pixel. Note that we continue to assume that $N_p>0$. Shear weight terms only appear in the white noise terms in Equation~\eqref{eq:EEBB_full}, and these terms now take the form
\begin{equation}
   \Omega \sum_p \Theta(N_p) A_p^2 \left \langle \frac{1}{N_p^{{\rm eff}}} \right \rangle \approx \sum_p\Theta(N_p) \frac{\Omega^2 A_p^2 \, \langle w^2 \rangle}{(N_p-1)\langle w \rangle^2 + \langle w^2 \rangle} ,
    \label{eq:general_weights}
\end{equation}
where $\Theta(N_p)$ is a step function whose value is unity if $N_p>0$ and zero otherwise. Note that step functions imposing $N_p>0$ appear in all the sums over pixels in our expressions so far, but have been omitted to ease the notation.

\subsubsection{Poisson sources}
\label{subsec:pixpoiss}

Consistent with the assumption throughout this section that the sources are unclustered, we now average over pixel occupancies by Poisson sampling $N_p$. This requires evaluating the sum
\begin{equation}
    \left \langle \frac{\Theta(N_p)}{N_p+\alpha} \right \rangle = \sum_{N=1}^{\infty} \frac{\lambda_p^N e^{-\lambda_p}}{(N +\alpha)N!}=
    \begin{cases}
    e^{-\lambda_p}\left[ (-\lambda_p)^{-\alpha} \gamma(\alpha, -\lambda_p) - \alpha^{-1}\right] & \alpha>0 \\
    e^{-\lambda_p}[\mathrm{Ei}(\lambda_p) - \log \lambda_p - \upgamma] & \alpha=0,
    \end{cases}
    \label{eq:ei}
\end{equation}
where $\alpha \equiv \mathrm{Var}(w)/\langle w \rangle^2 \geq 0$, $\lambda_p \equiv \bar{n}\Omega$ is the expected pixel occupancy for a source number density $\bar{n}$, $\mathrm{Ei}$ is the exponential integral, $\upgamma$ is the Euler-Mascheroni constant, and $\gamma(\alpha, x)$ is the lower incomplete Gamma function. Note that $(-\lambda_p)^{-\alpha} \gamma(\alpha, -\lambda_p)$ is single valued and real for all $\alpha>0$.

Similarly, we need to average over occupancies in the smoothed-signal terms in the first lines of Equation~\eqref{eq:EEBB_full}, which uses the results
\begin{align}
    \langle \Theta(N_p) \rangle &= \sum_{N=1}^{\infty} \frac{\lambda_p^N e^{-\lambda_p}}{N!} = 1 - e^{-\lambda_p},\nonumber \\
    \langle \Theta(N_p)\Theta(N_q) \rangle &= (1 - e^{-\lambda_p})(1 - e^{-\lambda_q}) + \delta_{pq}(1 - e^{-\lambda_p})e^{-\lambda_p}.
    \label{eq:meannp}
\end{align}
The HEALPix pixels are of equal area, so we can drop the pixel subscript on $\lambda$.


The suppression factors in Equation~\eqref{eq:ei} and \eqref{eq:meannp} are effectively corrections to the sky fraction that account for the fraction of the map with non-zero occupancy in a typical realization of the source counts. These should be compared with the $f_{\mathrm{sky}}$ factors that are the leading-order effect on the power spectra of the survey mask.

\subsubsection{Full sky, unit shear weights}
\label{subsec:fs_uw}

At this point, it simplifies the discussion to consider full sky coverage, since the assumption of isotropy in the pixels combined with source averaging has removed the coupling between the survey footprint and the pixel shapes. It is also instructive to consider the case of constant shear weights, in which case $\alpha=0$. We then have $A_p=1$, and choosing $L$ small enough such that the effects of aliasing and pixel shapes can be neglected, we have, for zero intrinsic shear $B$-mode,
\begin{align}
    \tilde{C}_L^E &\approx h(\lambda)C_L^E W_L^2 + \frac{g(\lambda)}{2\bar{n}}  \sum_{\ell} \frac{(2\ell+1)}{4\pi} C_{\ell}^E   W_{\ell}^2 + \frac{f(\lambda)}{2\bar{n}}\left[\sigma^2_\gamma -\sum_{\ell} \frac{(2\ell+1)}{4\pi} C_{\ell}^E W_{\ell}^2   \right] \nonumber \\
    \tilde{C}_L^B &\approx \frac{g(\lambda)}{2\bar{n}} \sum_{\ell} \frac{(2\ell+1)}{4\pi} C_{\ell}^E  W_{\ell}^2 + \frac{f(\lambda)}{2\bar{n}}\left[\sigma^2_\gamma -\sum_{\ell} \frac{(2\ell+1)}{4\pi} C_{\ell}^E W_{\ell}^2   \right] , 
    \label{eq:fullsky_EEBB}
\end{align}
where
\begin{align}
    f(\lambda) &=\lambda e^{-\lambda}[\mathrm{Ei}(\lambda) - \log \lambda - \upgamma] \nonumber \\
    g(\lambda) &= \lambda e^{-\lambda}(1-e^{-\lambda}) \nonumber \\
    h(\lambda) &= (1 - e^{-\lambda})^2.
    \label{eq:fgh}
\end{align}
Note that $f$ and $h$ tend to unity when $\lambda \gg 1$, while $g$ tends to zero. In the opposite limit, $\lambda \ll 1$, all three functions go to zero as $\lambda^2$.

%

\begin{figure}
\centering
    \includegraphics[width=0.48\columnwidth]{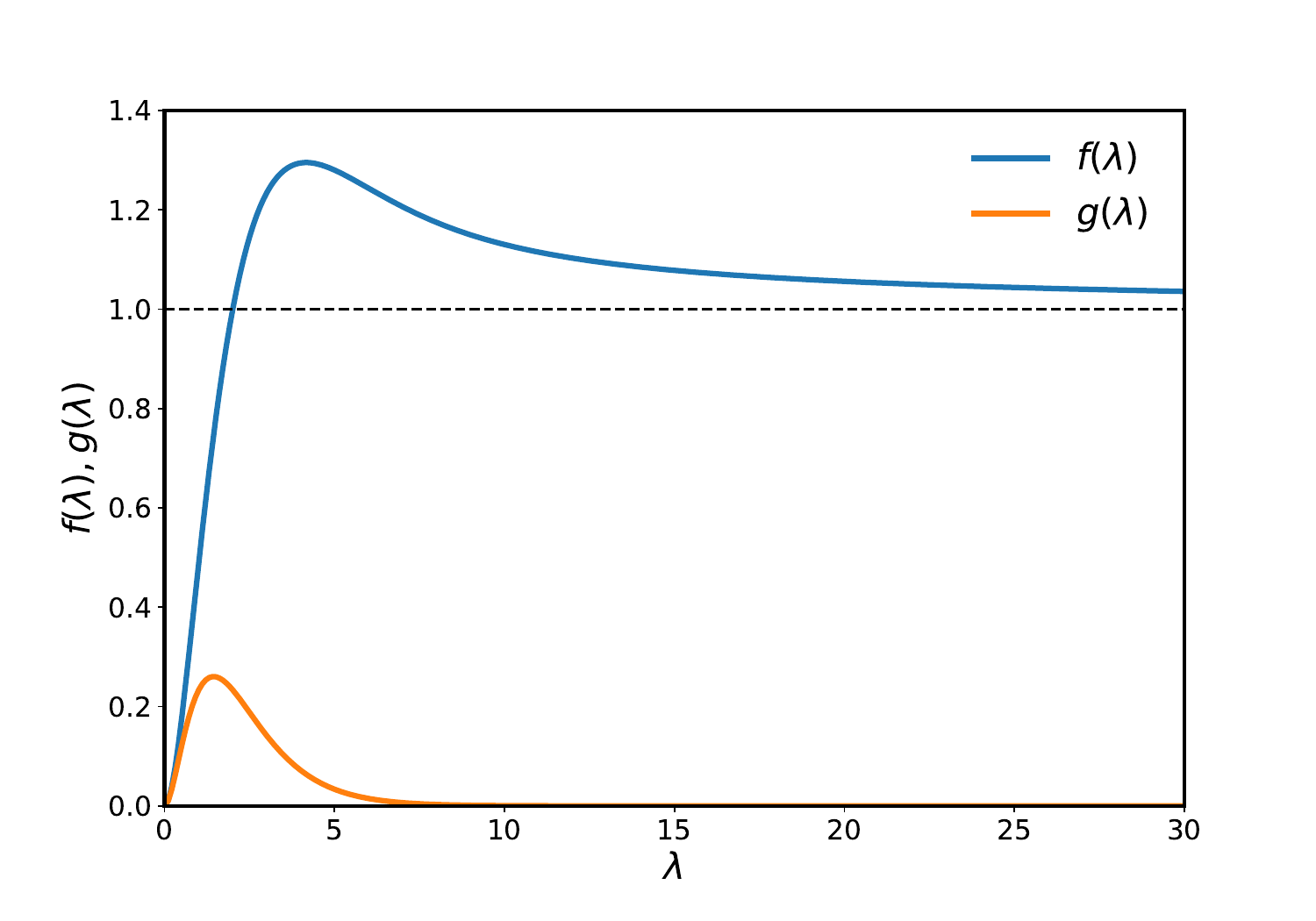}
    \includegraphics[width=0.48\columnwidth]{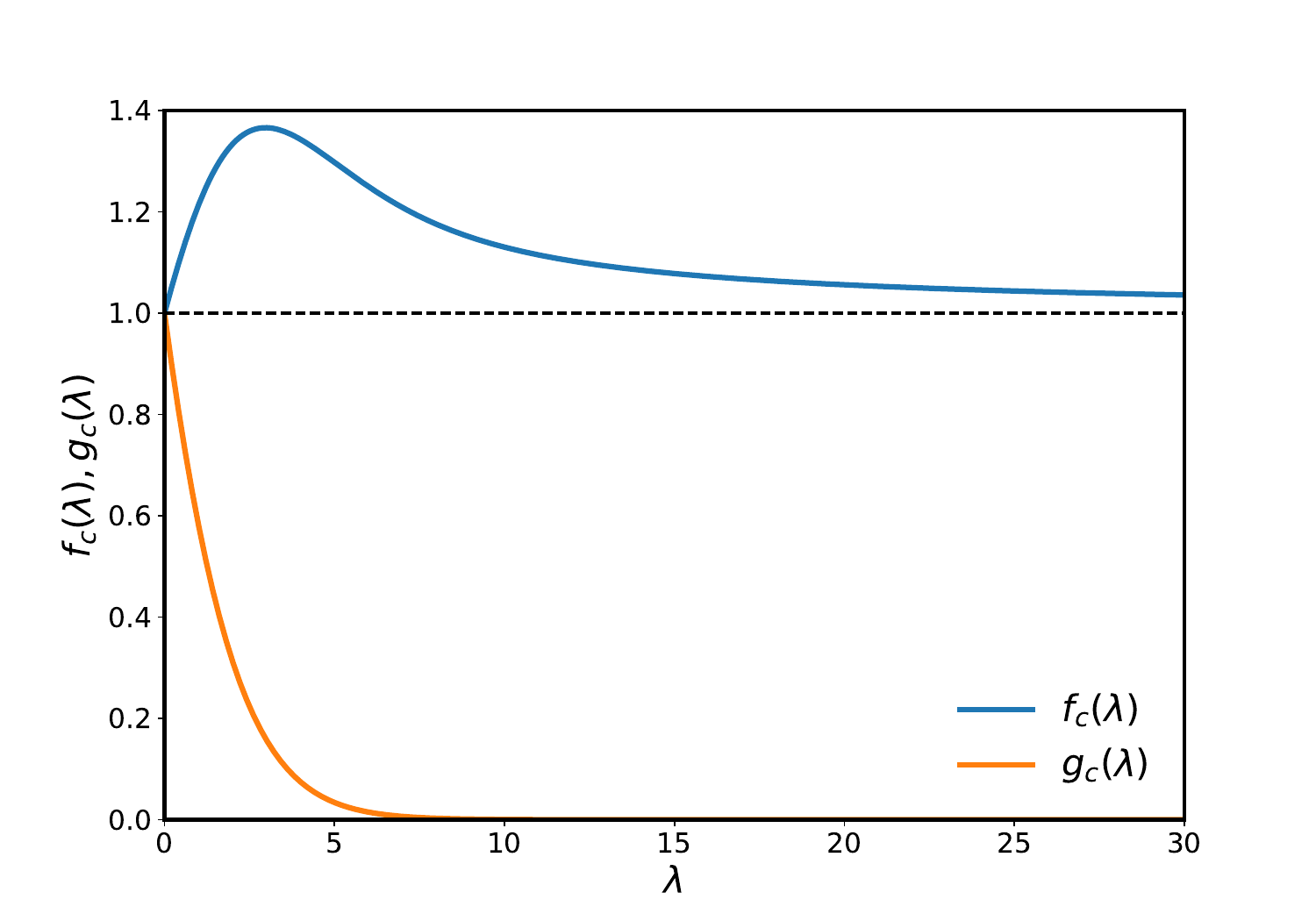}
    \caption{\emph{Left}: The function $f(\lambda)$ defined in Equation~\eqref{eq:fgh}, controlling the amplitude of shape noise in the power spectrum (blue, upper curve), and the function $g(\lambda)$ controlling the shot noise terms in Equation~\eqref{eq:fullsky_EEBB} (orange, lower curve), assuming constant shear weights. $\lambda$ is the expected occupancy of pixels in the shear map. \emph{Right}: The function $f_c(\lambda)$ (blue, upper curve) and $g_c(\lambda)$ (orange, lower curve) defined in Equation~\eqref{eq:fcgc}, controlling the amplitude of shape noise in the power spectrum after correcting for the primary bias to the signal power spectrum.} 
    \label{fig:fglam}
\end{figure}

We can now see that the power spectrum estimators are severely biased when $\lambda \ll 1$. In the limit that $\lambda \rightarrow 0$ both estimators are zero, as expected, with the leading order expression given by
\begin{align}
    \tilde{C}_L^E\big \rvert_{\lambda \ll 1}  &\rightarrow  \lambda^2\left[C_L^E W_L^2 + \frac{\sigma^2_\gamma}{2\bar{n}}\right]\label{eq:lowlamEE}\\
    \tilde{C}_L^B\big \rvert_{\lambda \ll 1}  &\rightarrow  \lambda^2\frac{\sigma^2_\gamma}{2\bar{n}}. \label{eq:lowlamBB}
\end{align}
Intuitively, this result makes sense. In the limit of zero expected pixel occupancy, the shot noise is given by the unsmoothed shear variance, since there is negligible averaging of multiple shears with each pixel. The overall amplitude is strongly suppressed essentially because $f_{\mathrm{sky}} \ll 1$ in each realization. In the opposite limit, $\lambda \gg 1$, we have the expected results that
\begin{align}
    \tilde{C}_L^E\big \rvert_{\lambda \gg 1}  &\rightarrow  C_L^E W_L^2 + \frac{1}{2\bar{n}}\left[\sigma^2_\gamma -\sum_{\ell} \frac{(2\ell+1)}{4\pi} C_{\ell}^E W_{\ell}^2   \right] \label{eq:simpleEE}\\
    \tilde{C}_L^B\big \rvert_{\lambda \gg 1}  &\rightarrow  \frac{1}{2\bar{n}}\left[\sigma^2_\gamma -\sum_{\ell} \frac{(2\ell+1)}{4\pi} C_{\ell}^E W_{\ell}^2   \right]. \label{eq:simpleBB}
\end{align}
In the left panel of Figure~\ref{fig:fglam} we plot the functions $f(\lambda)$ and $g(\lambda)$, the latter of which controls amplitude of the white noise terms, i.e.~the second terms in Equation~\eqref{eq:fullsky_EEBB}. Features of note in $f(\lambda)$ include the quadratic rise from zero, the peak at $\lambda \approx 4$, and the slow asymptote to unity at large $\lambda$. The amplitude of $g(\lambda)$ is generally less significant than that of $f(\lambda)$, but is comparable at low $\lambda$ and both tend to unity as $\lambda$ approached zero. This behaviour ensures that only the variance-at-a-point terms contribute in each pixel when the probability for a pixel to host more than one galaxy is very small.

To get an idea of typical values of $\lambda$, we can use typical numbers for a Euclid-like survey, giving
\begin{equation}
    \lambda \approx 2.21 \, \left(\frac{\bar{n}_{\mathrm{tot}}}{30 \, \mathrm{arcmin}^{-2}}\right) \left(\frac{10}{N_z}\right) \left(\frac{4096}{N_{\mathrm{side}}}\right)^2,
\end{equation}
where $N_z$ is the number of tomographic redshift bins, and $\bar{n}_{\mathrm{tot}}$ is the survey number density. At this nominal map resolution, we have $f(\lambda) \approx 1$ and $g(\lambda) \approx 0.2$. The pre-factor in front of the signal term is $0.8$, indicating that the power spectra are severely biased at this low resolution. This can be remedied by simply dividing the estimates by the bias factor, analogous to deconvolving by the survey mask mixing matrix; in our case, the `mask' is an average mask induced by empty pixels, and the mixing matrix is a constant multiplicative factor. In section~\ref{sec:fixedpos} we will make this correspondence more precise. After division by this factor, the white noise terms are given by
\begin{align}
    &g_c(\lambda) \frac{1}{2\bar{n}}  \sum_{\ell} \frac{(2\ell+1)}{4\pi} C_{\ell}^E  W_{\ell}^2 + f_c(\lambda) \frac{1}{2\bar{n}}\left[\sigma^2_\gamma -\sum_{\ell} \frac{(2\ell+1)}{4\pi} C_{\ell}^E W_{\ell}^2   \right] \nonumber \\
    &=  \frac{1}{2\bar{n}}  \sum_{\ell} \frac{(2\ell+1)}{4\pi} C_{\ell}^E \left\{f_c(\lambda) + W_\ell^2 [g_c(\lambda) - f_c(\lambda)]\right\},
    \label{eq:corrected_nbias}
\end{align}
where the corrected functions $f_c$ and $g_c$ are defined as
\begin{align}
    f_c(\lambda) &\equiv \frac{\lambda e^{-\lambda}[\mathrm{Ei}(\lambda) - \log \lambda - \upgamma]}{(1-e^{-\lambda})^2} \\
    g_c(\lambda) &\equiv \frac{\lambda e^{-\lambda}}{(1-e^{-\lambda})}. \label{eq:fcgc}
\end{align}
These functions are plotted in the right panel of Figure~\ref{fig:fglam}. At low expected occupancy, $g_c(\lambda) - f_c(\lambda) \rightarrow 0$ and $f_c(\lambda) \rightarrow 1$, leaving us with a pure shot noise term expected from point-sampling the shear field, given by $\sigma_\gamma^2/2\bar{n}$. In the limit of high occupancy, $g_c(\lambda) \rightarrow 0$ and $f_c(\lambda) \rightarrow 1$, leaving us with a modified shape noise term with $\sigma_\gamma^2$ replaced by the shear variance smoothed within a pixel.


\subsection{Including shape noise}
\label{subsec:results_with_sn}

Let us now extend the discussion to include shape noise in the intrinsic source galaxy ellipticity distribution. In the presence of shape noise, the power spectra gain white noise terms given by Equation~\eqref{eq:shapenoisevar}. In the scenario of full sky coverage, unit weights, and averaging over source realizations, this becomes
\begin{equation}
    \langle \tilde{N}^{E/B}_{\ell} \rangle = f(\lambda) \frac{\sigma_e^2}{2\bar{n}},
\end{equation}
with $f(\lambda)$ defined in Equation~\eqref{eq:fgh}, which simply modifies Equation~\eqref{eq:EEBB_full_hp} by replacing $\sigma_\gamma^2 \rightarrow \sigma_\gamma^2 + \sigma_e^2$.

Rather than include the shape noise term in the model, we can try to remove it from the data vector by estimating the noise power. There are several estimators we could use for this, all having the correct mean. Two examples are
\begin{equation}
    \hat{N}_\ell^{E/B, (1)} = \frac{\Omega^2}{8\pi} \sum_p |\hat{\gamma}_p|^2
    \label{eq:nest1}
\end{equation}
and
\begin{equation}
    \hat{N}_\ell^{E/B, (2)} = \frac{\Omega^2}{8\pi} \sum_p \frac{\sum_{i \in p}w_i^2 |\hat{\gamma}_i|^2}{\left(\sum_{i \in p} w_i\right)^2},
    \label{eq:nest2}
\end{equation}
where we remind the reader that in this section we only consider uniform pixel weighting. Both these estimators have the correct mean to remove shape noise, but Equation~\eqref{eq:nest1} uses the variance of the pixelized shears over the map while Equation~\eqref{eq:nest2} uses the average of the per-galaxy ellipticity variance over the map. The expectation value of Equation~\eqref{eq:nest1} in the presence of both shape noise and the spatially correlated shear field is
\begin{align}
    \langle \hat{N}_L^{E/B, (1)} \rangle &= \frac{\Omega^2}{8\pi} \sum_\ell \frac{(2\ell+1)}{4\pi} C_\ell^E \sum_p \sum_m |{}_m W_{\ell,p}^+ |^2 + \frac{\Omega^2}{8\pi}\sigma_e^2 \sum_p \frac{\Theta(N_p)}{N_p^{{\rm eff}}} \nonumber \\
    &\approx \frac{\Omega^2}{8\pi} \sum_p \left\{\frac{\Theta(N_p)}{N_p^{{\rm eff}}}\left[\sigma_\gamma^2 + \sigma_e^2 - \sum_\ell \frac{(2\ell+1)}{4\pi}C_\ell^E (W_\ell^p)^2 \right] \right.\nonumber \\
    &\left. + \Theta(N_p) \sum_\ell \frac{(2\ell+1)}{4\pi}C_\ell^E (W_\ell^p)^2\right\}\nonumber \\
    &\approx \frac{f(\lambda)}{2\bar{n}}\left(\sigma_\gamma^2 + \sigma_e^2 - \sum_\ell \frac{(2\ell+1)}{4\pi}C_\ell^E W_\ell^2 \right) +  \frac{g(\lambda)}{2\bar{n}} \sum_\ell \frac{(2\ell+1)}{4\pi}C_\ell^E W_\ell^2,
    \label{eq:nhat1}
\end{align}
where in the second line we made the approximation of unclustered sources and axisymmetric pixels, and in the third line we assumed unit shear weights, full sky coverage, and averaged over the Poisson source pixel occupancy. Note that partial sky coverage would introduce an overall factor of $f_{{\rm sky}}$ in front of this expression. Comparison with Equation~\eqref{eq:EEBB_full_hp} demonstrates that the estimator Equation~\eqref{eq:nest1} is sufficient to remove all of the white noise terms from power spectra, including those due to the shear variance. By contrast, the expectation of Equation~\eqref{eq:nest2} is
\begin{align}
     \langle \hat{N}_L^{E/B, (2)} \rangle &= \frac{\Omega^2}{8\pi} (\sigma_\gamma^2 + \sigma_e^2)\sum_p \frac{\Theta(N_p)}{N_p^{{\rm eff}}}\nonumber \\
     &\approx \frac{f(\lambda)}{2\bar{n}} (\sigma_\gamma^2 + \sigma_e^2),
\end{align}
where in the second line we have made the approximation of unit shear weights, full sky coverage, and averaged over Poisson sources. Therefore, subtracting Equation~\eqref{eq:nest2} from the power spectrum estimator removes the shape noise, but leaves behind white noise due to shear correlations. This may be included in the model by modifying the mixing matrices, as shown in Ref.~\citep{PKWL}.

\subsection{Non-Poissonian sources and stochastic shear weights}
\label{subsec:nonpoiss}

Finally, let us briefly discuss how the modelling we have developed in the previous sections is impacted in the more realistic scenario of clustered sources. In this case, we must modify our expressions in Equation~\eqref{eq:meannp} for the statistics of $\Theta(N_p)$. In particular, two-point statistics of the occupancy gain a term proportional to $\xi_{pq}$, the correlation function of the source density field. This introduces additional $\ell$-dependence in the power spectrum and generates additional $B$-mode power, akin to the effects of mode mixing induced by a survey mask. In section~\ref{sec:fixedpos} we will make this analogy more concrete.

Modelling the effects of source clustering is difficult due to the complexity of the source selection and the unknown galaxy bias. In addition, the source density field is correlated with the lensing of background sources, which brings in additional terms due to source-lens clustering (SLC; \citep{1998A&A...338..375B, 2002A&A...389..729S, 2024arXiv240709810L}). Any discussion of the impact of clustered sources on pixelization effects should account for the SLC effect, and so we postpone further discussion to a forthcoming work. Despite this, the fixed-position estimator and tests discussed in section~\ref{sec:fixedpos} offer a way of accounting naturally for source clustering; effectively the source positions are deconvolved from the estimator and hence their impact on the signal is suppressed (although not completely removed).

Deconvolving the source density field also partially accounts for stochasticity of the shear weights, as discussed in section~\ref{sec:fixedpos}. What still remains is the possible impact of shear-dependence in the weights. Although one can try to model this in a similar way to SLC modelling, the impact is best quantified through image simulations with realistic shear and clustering, not least because it is dependent on the specific shear measurement algorithm and shear calibration process.

\section{Tests of the standard estimator against simulations}
\label{sec:simulations}

Having built theoretical models for the bias due to pixelization, finite pixel occupancy and, in appendix~\ref{sec:pt}, due to neglecting parallel transport of shear, we will now test our models against simulations. We produce a suite of mock galaxy catalogues, construct shear maps from them, and produce angular power spectra. The input power spectrum is a fiducial $\Lambda$CDM weak lensing spectrum with an \texttt{HMcode-2020} non-linear correction~\citep{2021MNRAS.502.1401M}, with no intrinsic alignments. We assume a single source redshift bin peaked at $z\approx 0.8$ corresponding to the fifth bin of a 10-bin Euclid-like survey, following the specifications of Ref.~\cite{2020A&A...642A.191E}.

We draw Gaussian realizations of the shear $E$-mode multipoles from this angular power spectrum and create a set of high-resolution full-sky shear maps with $N_{\mathrm{side}}=4096$, which support modes out to $\ell \approx 13000$. We then sample galaxies in a Poisson process with a uniform number density across the sky, such that the expected number of galaxies in a pixel on an $N_{\mathrm{side}}=512$ grid is 32, i.e.~a number density of $\bar{n} \approx 0.68 \, \mathrm{arcmin}^{-2}$. This is more than a factor 4 less dense than the nominal Euclid Wide Survey~\citep{2022A&A...662A.112E} for such a redshift bin, but keeps the catalogues to a manageable size. The density of galaxies was chosen such that we can sample a good range mean pixel occupancies in three grids of lower resolution, while ensuring that there are always shear modes on scales smaller than a pixel.

Once galaxy positions have been drawn we assign them shears based on the shear of the parent pixel in the underlying high-resolution $N_{\mathrm{side}}=4096$ grid. We generate $N_{\mathrm{cat}}=1000$ realizations, resulting in $N_{\mathrm{cat}}$ catalogues.

Once the baseline catalogues have been built, we construct shear maps on grids of varying resolution corresponding to $N_{\mathrm{side}}=512$, $N_{\mathrm{side}}=1024$, and $N_{\mathrm{side}}=2048$. The expected number of galaxies in a pixel on these grids is $\lambda = 32$, $8$, and $2$ respectively. The finest grid has an expected occupancy comparable to that expected in the Euclid Wide Survey at the target map resolution~\citep{PKWL}. Our shear maps incorporate the correct phase factors discussed in appendix~\ref{sec:pt}. For each map we compute the pseudo power spectra. This results in a baseline set of $N_{\mathrm{cat}}$ spectra sampling galaxy number, galaxy positions, and shears. 

Note that our simulations do not include shape noise; we have shown how the noise bias from this can be removed exactly in section~\ref{subsec:results_with_sn}. Shape noise will add extra variance to the power spectrum of course, which affects the implications of any pixel-related biases, as will be discussed below.

\subsection{Results: full sky coverage, unit weights, Poisson sources, zero shape noise}

\begin{figure}
\centering
    \includegraphics[width=0.48\columnwidth]{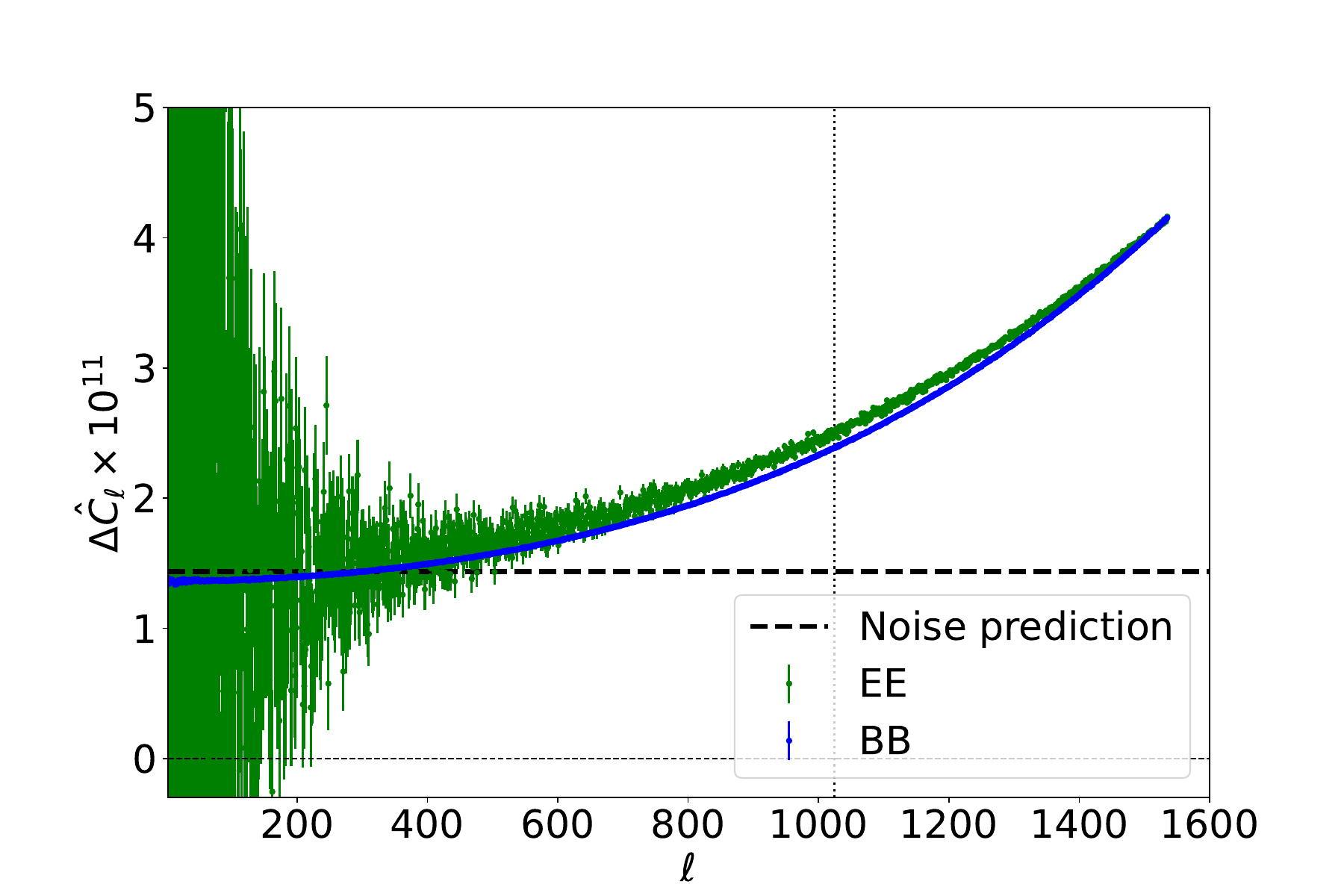}
    \includegraphics[width=0.48\columnwidth]{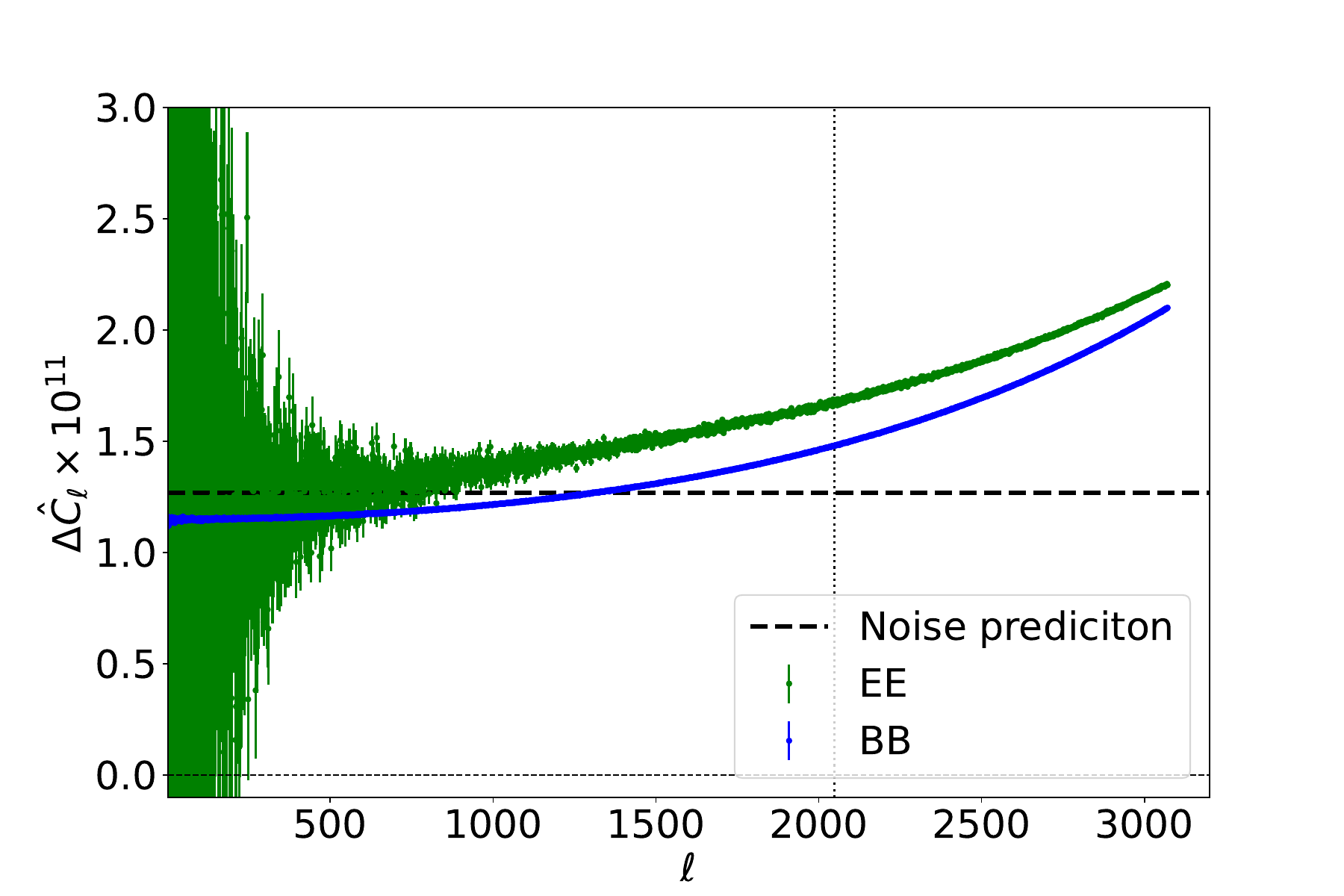}
    \includegraphics[width=0.48\columnwidth]{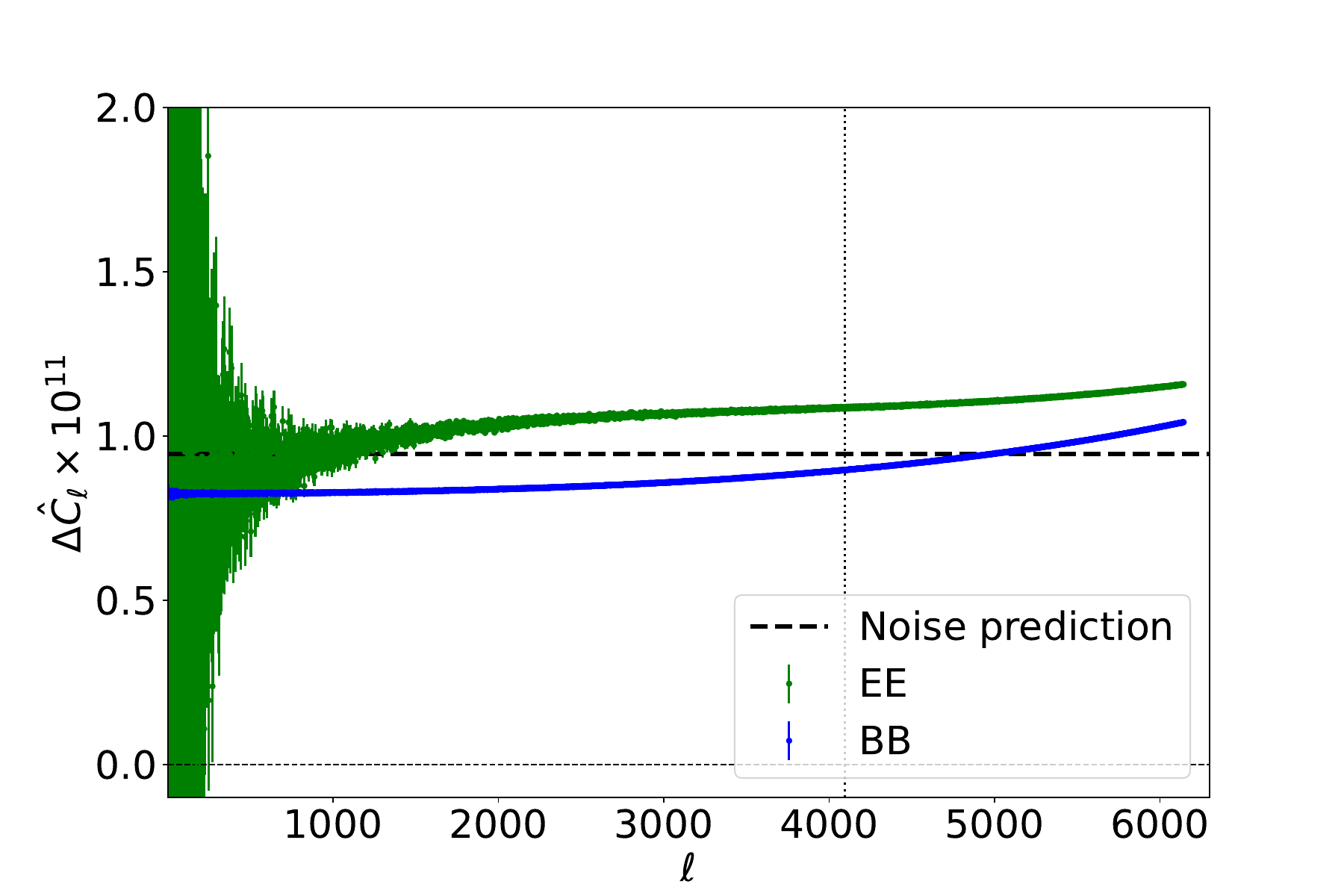}
    \caption{Residuals between measured $E$ (green) and $B$ (blue) power spectra and their input smoothed by the HEALPix window function and corrected by the pixel occupation factor (first term of Equation~\eqref{eq:fullsky_EEBB} for $EE$, zero for $BB$), for $\lambda=32$ (top left panel), $\lambda=8$ (top right panel) and $\lambda=2$ (bottom panel), where $\lambda$ is the mean pixel occupancy. The resolution of the input shear field is $N_{\mathrm{side}} = 4096$ (with no pixel smoothing), while the resolution of the measured shear map is $N_{\mathrm{side}} = 512$ (top left panel), $N_{\mathrm{side}} = 1024$ (top right panel), $N_{\mathrm{side}} = 2048$ (bottom panel). No intrinsic shape noise has been added, and the input power spectrum is a fiducial $\Lambda$CDM spectrum. The thick dashed horizontal line shows the white noise prediction of Equations~\eqref{eq:simpleEE} and \eqref{eq:simpleBB}, and the vertical dashed line indicates a multipole of $2N_{\mathrm{side}}$ for the resolution of the constructed shear map. Error bars are show the empirical scatter based on $N_\mathrm{cat}=1000$ simulations. On scales well above the pixel scale, the residuals are mostly due to white noise and can be correctly captured by our model. Additional corrections, comparable in size to the white noise bias, appear around the pixel scale due to aliasing. These effects are detected with high significance in our simulations. }
    \label{fig:residuals}
\end{figure}

In the top-left panel of Figure~\ref{fig:residuals} we plot the difference between the mean of the $E$-mode spectra and the input theory smoothed with the HEALPix window function and corrected for finite pixel occupation number, i.e.~the first term in Equation~\eqref{eq:simpleEE}, for $\lambda=32$. The $B$-mode spectrum mean is also plotted. Errors on these quantities are estimated from the empirical scatter across the mocks. The value $\lambda=32$ corresponds to a shear map constructed with a $N_{\mathrm{side}}=512$ grid. The residual amplitudes of $E$ and $B$ power are comparable, at around the $10^{-11}$ level, corresponding to a difference of between 0.1\% and 1\% from the input theory. Most of the discrepancy is captured by our model for white noise, indicated by the horizontal dashed line in Figure~\ref{fig:residuals}, although there are still residual differences at the $10^{-12}$ level (roughly a 0.1\% difference from theory) on large scales and roughly $10^{-11}$ differences on smaller scales. These residual differences come mostly from aliasing, in a similar way to the test results shown in Figure~\ref{fig:pows}.


To gain sensitivity to smaller scales in this catalogue we must use a finer grid to produce the shear map, which entails lower expected pixel occupancy. In the top-right panel of Figure~\ref{fig:residuals} we plot mean residuals for a grid with $N_{\mathrm{side}}=1024$, corresponding to an expected occupancy of $\lambda=8$. Again we see differences of around $10^{-11}$ in the power compared with the smoothed theory, with most of this accounted for by the white noise model. Residual differences are at the $10^{-12}$ level. The large-scale $B$-mode power is clearly over-predicted, whereas the large-scale $E$-mode power is in good agreement, although the errors are large. The mismatch between $E$ and $B$ power across most scales is suggestive of aliasing and sensitivity to the variation of pixel shape across the sky, corresponding to the second term in square brackets on the first line of Equation~\eqref{eq:EEBB_full_asym}. 


An occupancy of $\lambda=8$ is still fairly large, such that the correction terms in our white noise model are in the $\lambda \gg 1$ regime. To probe a realistic pixel occupancy we construct a shear map with a resolution of $N_{\mathrm{side}} = 2048$, corresponding to an expected pixel occupancy of $\lambda=2$. The residuals are shown in the bottom panel of Figure~\ref{fig:residuals}. As for the lower resolution grids, most of the difference from the smoothed signal is captured by our white noise model, with residuals of around $10^{-12}$ across all scales. The residual is roughly the same amplitude as the difference between $E$ and $B$ power, again suggestive of mode mixing induced by aliasing and the variation of pixel shapes across the maps. If this could be modelled we could expect residuals of a few times $10^{-13}$, corresponding to a fractional difference of at most $10^{-3}$.

\begin{figure}
\centering
    \includegraphics[width=0.8\columnwidth]{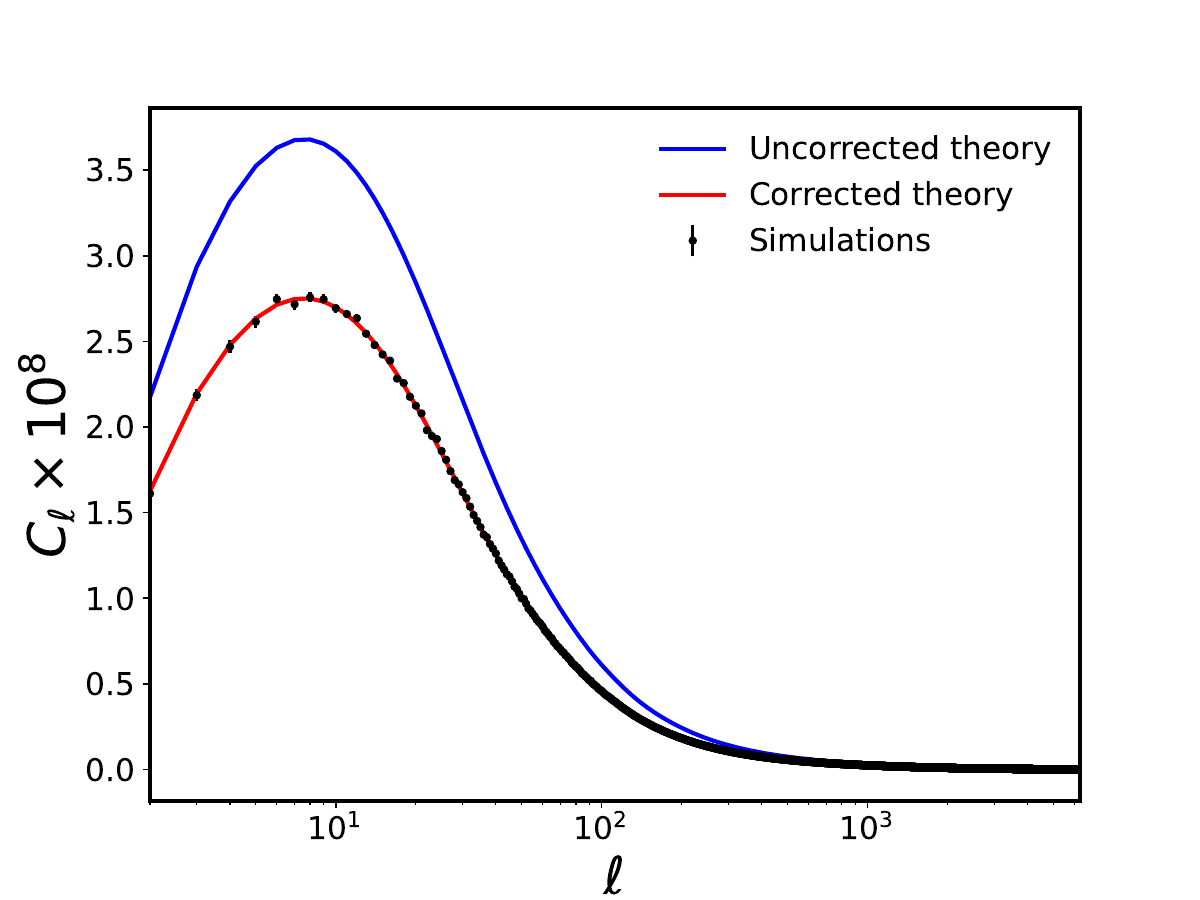}
    \caption{The average of the $E$-mode shear power spectrum from the $N_\mathrm{cat}=1000$ mock catalogues (black point with error bars), compared with the raw input theory (blue solid curve) and the model of Equation~\eqref{eq:simpleEE} with $\lambda=2$ (red solid curve), corresponding to the use of shear maps with $N_{\mathrm{side}} = 2048$ resolution.}
    \label{fig:mean_ns2048}
\end{figure}

It should be appreciated that with $\lambda=2$ all of the terms in the white noise model of Equations~\eqref{eq:simpleEE} and \eqref{eq:simpleBB} are significant. To illustrate this, in Figure~\ref{fig:mean_ns2048} we plot the mean of the measured power spectra compared to the raw input theory, along with our model. The naive prediction is severely biased, at the 75\% level. Most of the difference is captured by the $(1-e^{-\lambda})^2$ correction, with residual differences reduced by an order of magnitude by our white noise model, as shown in the bottom panel of Figure~\ref{fig:residuals}. A caveat to this is that our model is strictly only valid when the approximation of Equation~\eqref{eq:wratio} holds. In the more general case we should avoid dividing the shear map by the weighted occupancy in each pixel if we wish to average over shear weight realizations, as discussed in section~\ref{sec:nn}.


\begin{figure}
\centering
    \includegraphics[width=0.8\columnwidth]{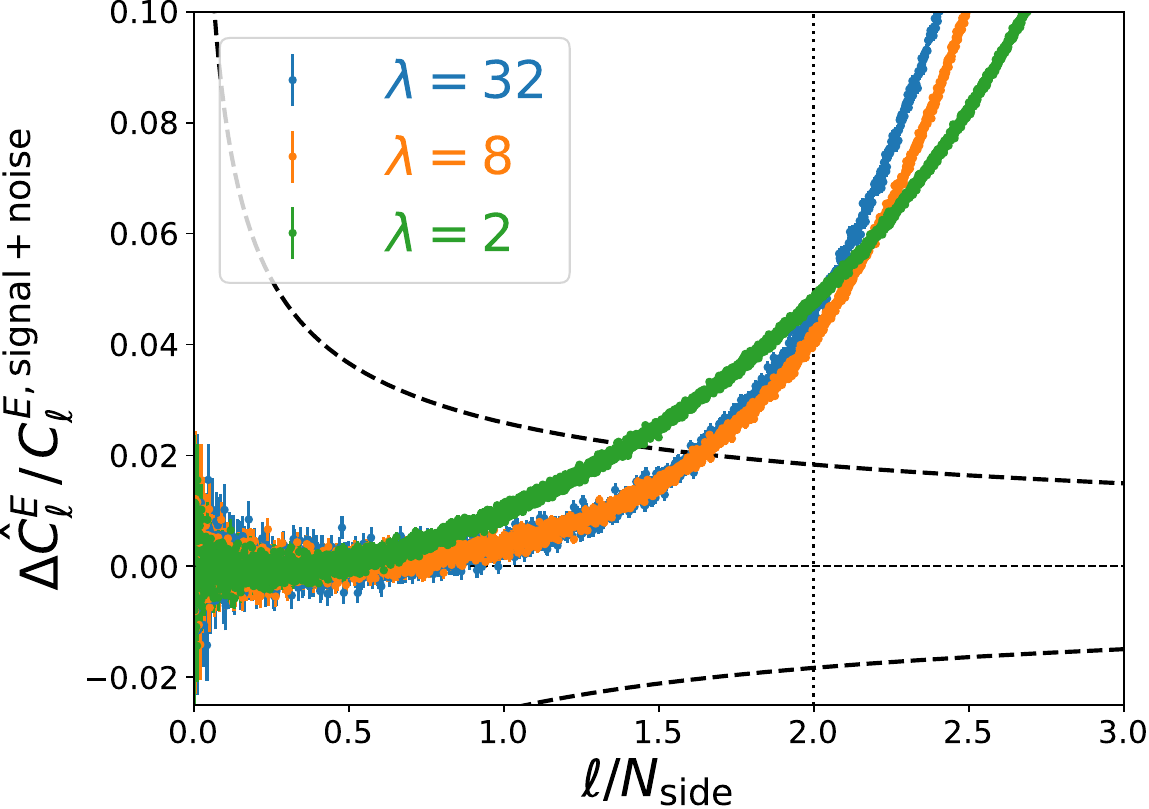}
    \caption{Fractional difference of the measured $E$-mode power spectrum compared to the model of Equation~\eqref{eq:fullsky_EEBB}, with different resolutions of shear map corresponding to mean pixel occupancies of $\lambda=32$ (blue; $N_{{\rm side}}=512$), $\lambda=8$ (orange; $N_{{\rm side}}=1024$) and $\lambda=2$ (green; $N_{{\rm side}}=2048$). The zero line is added for clarity, and the angular scale given by $\ell=2N_{\mathrm{side}}$ is shown by the vertical line. The dashed black curve shows forecast 68\% Gaussian uncertainties in this quantity for a Euclid-like survey covering 15,000 square degrees and using a shear map of resolution $N_{\mathrm{side}}=4096$. Note that the quantity plotted here is the fractional difference of an estimator containing noise bias and a model that includes that noise bias, such that the fractional uncertainty is just that coming from mode counting.} 
    \label{fig:fracdiffs}
\end{figure}

\begin{figure}
\centering
    \includegraphics[width=0.8\columnwidth]{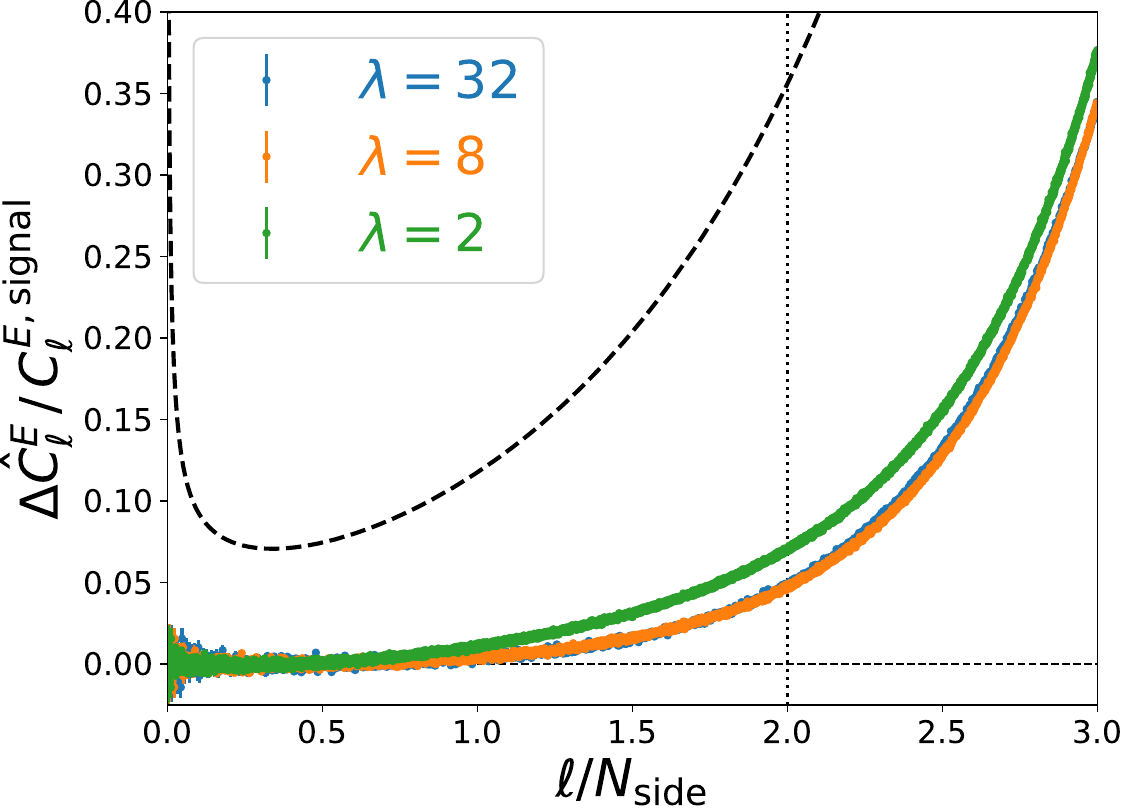}
    \caption{Fractional difference of the measured $E$-mode power spectrum compared to only the signal part of Equation~\eqref{eq:fullsky_EEBB}. The dashed curve shows forecast 68\% Gaussian uncertainties in this quantity for a Euclid-like survey covering 15,000 square degrees and with 30 galaxies per square arcminute, assuming $N_{\mathrm{side}}=4096$. The colour scheme is the same as in Figure~\ref{fig:fracdiffs}, except now the quantity plotted is the fractional difference of an estimator containing noise bias and a model that includes only signal. In this case, shape noise dominates the variance above $\ell \approx 2000$.}
    \label{fig:fracdiffs_EU}
\end{figure}

After making all of the corrections presented in Equation~\eqref{eq:simpleEE}, we find residual differences shown in Figure~\ref{fig:fracdiffs} for each of the three shear map resolutions. Residual biases become visible by eye above a multipole $\ell_* \approx N_{\mathrm{side}}/2$ in all cases. At lower $\ell$, we constrain residual biases to no more than a fraction of a percent. At $\ell \approx N_{{\rm side}}$ biases are still sub-percent, but close to 1\% for the high resolution, low expected occupancy case. At higher $\ell$, biases grow rapidly, reaching several percent at  $\ell \approx 2N_{{\rm side}}$, as expected from our earlier tests with power laws in Figure~\ref{fig:pows}.


 Figure~\ref{fig:fracdiffs} also shows the statistical uncertainty in this fractional difference for a Euclid-like survey covering 35\% of the sky and using a shear map of resolution $N_{{\rm side}}=4096$. Note that because the model that defines the residual bias plotted here includes shape noise, the statistical uncertainty shrinks as the square-root of the number of independent angular modes. Given knowledge of the signal and noise power, a Euclid-like survey will therefore allow a detection of aliasing at this grid resolution with high significance.

More relevant to parameter estimation is the fractional correction to the power spectrum due to aliasing compared with the \emph{signal} power. This is plotted in Figure~\ref{fig:fracdiffs_EU}. The Euclid-like statistical uncertainty in this quantity, roughly the inverse signal-to-noise per $\ell$-mode, now grows on small scales because shape noise dominates over cosmic variance (sourced by the signal power itself, modulated by the pixel window function) above $\ell \approx 2000$. Aliasing is roughly 1\% of the signal for $\ell \gtrsim N_{\mathrm{side}}$, which is an order of magnitude smaller than the statistical uncertainty at $\ell \approx 4000$. Nevertheless, the effect is coherent over a wide range of multipoles where there are many hundreds of independent modes, so the effect on parameter constraints could be significant. This will be mitigated by non-Gaussian contributions to the covariance matrix, and a proper assessment of the impact of aliasing on parameter constraints will likely require a Simulation-Based Inference analysis of the pseudo-$C_\ell$ estimator, such as that in Ref.~\cite{2024arXiv240415402V}.


To summarise this section, the models built in section~\ref{sec:aliasing} are accurate at the sub-percent level for $\ell \lesssim aN_{\mathrm{side}}$, where $a=1-1.5$ depending on the source number density. Residual biases are mainly due to aliasing, which is unavoidable in the HEALPix map making approach. Averaging over source number counts effectively introduces a mixing matrix even on the full sky, and our model for the isotropic part of this matrix (due to shot noise) works with high accuracy across a wide range of source number densities. In the remaining few sections, we will investigate refinements to the standard estimator and test scenarios in which the source counts (and shear weights) are considered fixed.

\section{Modifying the standard estimator: unnormalized and globally normalized shear maps}
\label{sec:nn}

In the previous sections, we have assumed that one wishes to model the power spectrum averaged over the true shear, the shear weights, and the source galaxy distribution, and provided models for the spectra in this scenario that agree reasonably well with simulations. Let us continue to assume that one wishes to average over the source positions and weights in this way, and consider how to overcome some of the technical difficulties in accurately modelling the power spectra.

We can avoid issues encountered when dividing by the noisy total weight in each pixel by considering an alternative shear map given by
\begin{equation}
        \hat{\gamma}_p^{\mathrm{nn}} = \sum_{i \in p} w_i \, \hat{\gamma}_i \, e^{-2 \mathrm{i}\beta_i},
    \label{eq:nn}
\end{equation}
where the superscript `$\mathrm{nn}$' refers to the fact that this map has not been normalized. The properties of this estimator, and two-point statistics derived from it, can be found by setting $\omega_p = N_p w_{(1)}(p) = \sum_{i\in p} w_i$ within summations over pixels. The estimator is equivalent to using approximate inverse-variance pixel weights. The derivation in section~\ref{sec:weights_and_Np} now needs modifying, because we do not have the complications associated with dividing the map by a stochastic quantity. In fact, averages over weights and pixel occupancies can now be written down exactly. In the case of full-sky coverage and unit weights, Equation~\eqref{eq:fullsky_EEBB} still holds, but with the replacements $f(\lambda) = g(\lambda) = h(\lambda) = \lambda^2$, considerably simpler than in the per-pixel normalized map case. The power spectrum bias is now easily accounted for by normalizing by the expected pixel occupancy; the corrected noise bias terms are then given by Equation~\eqref{eq:corrected_nbias} with $f_c(\lambda) = g_c(\lambda) = 1$. This simplification gives the remarkably simple result
\begin{align}
    \tilde{C}_L^{E, \mathrm{nn}}/\lambda^2 &= C_L^E W_L^2 + \frac{\sigma_{\gamma}^2}{2\bar{n}} \nonumber \\
    \tilde{C}_L^{B, \mathrm{nn}}/\lambda^2 &= \frac{\sigma_{\gamma}^2}{2\bar{n}}.
    \label{eq:nn_expected}
\end{align}
Thus, choosing not to normalize the shear map by the total weight in each pixel (or equivalently, applying an inverse variance weight) not only simplifies the statistics of the estimator when averaging over the source population, but makes the resulting expected power spectrum simple in the case of unclustered sources. We have verified that the model of Equation~\eqref{eq:nn_expected} works well, with residual differences coming from aliasing and the failure of $W_\ell$ to capture the effects of varying pixel shapes. 

We caution that these simplifications have assumed uniform shear weights, so should not be taken too literally. Moreover, there are potential disadvantages in not normalizing by the local pixel weight; super-pixel shear modes are no longer faithfully represented when $\lambda \gg 1$, and the sensitivity to source-lens clustering and related biases is potentially greater~\cite{2024arXiv241115063D}.

Jumping ahead to the situation where the galaxy positions and weights are considered fixed (see section~\ref{sec:fixedpos}), the estimator Equation~\eqref{eq:nn} is statistically equivalent to the shear map normalized by the globally averaged total per-pixel shear weight. This estimator is used in Ref.~\citep{PKWL}, and is given by
\begin{equation}
        \hat{\gamma}_p^{\mathrm{gw}} = \frac{\sum_{i \in p} w_i \, \hat{\gamma}_i \, e^{-2 \mathrm{i}\beta_i}}{\bar{w}},
    \label{eq:gamma_pkwl}
\end{equation}
where the superscript `$\mathrm{gw}$' refers to the fact that the map has been normalized by a `global weight', and the quantity $\bar{w}$ is given by
\begin{equation}
    \bar{w} \equiv \frac{N_{{\rm gal}}}{4\pi f_{\mathrm{sky}}} \Omega \langle w_i \rangle
    \label{eq:wbar}
\end{equation}
where $\langle w_i \rangle$ is the mean shear weight over the whole catalogue (for the given tomographic bin). Note that $\bar{w}$ simply normalizes the map in a way comparable to the original map estimator Equation~\eqref{eq:wrongeq0}, but replaces the per-pixel weighted number of galaxies with an average over the survey footprint. In the case of full sky coverage and unit shear weights, we have $\bar{w} \approx \lambda$ to a good approximation. This gives an expected power spectrum equivalent to Equation~\eqref{eq:nn_expected}, with $\tilde{C}_L^{E, \mathrm{gw}} = C_L^E W_L^2 + \sigma_{\gamma}^2/(2\bar{n}) $ and $\tilde{C}_L^{B, \mathrm{gw}} = \sigma_{\gamma}^2/(2\bar{n}) $. Thus, normalizing by the quantity $\bar{w}$ performs the same task as normalizing by $\lambda$, with the advantage that it can be used in the full-complexity situation of arbitrary source positions and weights.

In summary, the statistics of the shear map and the biases that can result from empty pixels and stochasticity of the galaxy weights can much more easily be dealt with by normalizing the shear map by a global shear weight rather than the local weight in each pixel.

\section{Tests with fixed galaxy positions and weights}
\label{sec:fixedpos}

In the tests we have run so far, we have varied both the shear field and the galaxy source positions in each simulation; in the same vein, had we considered realistic shear weights, we would have varied those too. Consequently, the models we have built are intended to model the power spectra averaged over all sources of stochasticity. We have seen how this brings complications when accounting for empty pixels, stochastic shear weights, and correlations between lensing and shear weights. This suggests an alternative; consider the source positions and weights as \emph{fixed} and instead try to model the power spectrum statistics ensemble averaged over realizations of the shear field. This is more aligned with how weak lensing is done in practice; the radial distribution of galaxies is fixed to a (calibrated) distribution of those galaxies that are actually in the sample, rather than some fictitious ensemble-averaged sample.\footnote{Interestingly, this highlights an inconsistency in how shear correlation functions are conventionally modelled; no dependence on the actual galaxy positions is typically included in the model, and instead the theory correlation functions are averaged over angular bins assuming all angular separations are represented in the catalogue (this has been studied in appendix D of Ref.~\cite{2020MNRAS.491...51S}). By contrast, the actual observed pair counts \emph{are} conventionally included in the shape noise part of analytic covariance matrix models, albeit not elsewhere in the standard covariance calculation~\cite{2021A&A...646A.129J, 2021MNRAS.508.3125F}. This has recently been studied in detail in Ref.~\citep{2024arXiv241006962R}, building upon the formalism of Ref.~\citep{2002A&A...396....1S}.} On large scales, we expect this approach to agree well with the situation in which source positions are formally averaged over, since many independent pixels contribute to the estimated power at each $L$. At sufficiently small pair separations we will eventually run out of galaxy pairs, unlike the situation in the source-averaged situation. We therefore expect low-pass filtering to be more effective in the fixed-source case; this potentially mitigates aliasing.

There are some subtle considerations that arise when analyzing weak lensing data assuming a fixed source population. Firstly, the source positions are clustered, and so contain cosmological information. Formally one should therefore compute models for the two-point statistics of the shear field \emph{conditioned} on the positions of the sources. This will have most impact on the cross-correlation of lensing between widely separated redshift bins, effectively due to source-lens clustering. The conventional way of modelling source-lens clustering is to ensemble average over the source positions~\cite{1998A&A...338..375B, 2024A&A...684A.138E}, but if these are instead considered fixed then the calculation changes. This point is partly explored in Ref.~\citep{2015ApJ...803...46Y}, and will be explored thoroughly in a forthcoming work.

Here, we sidestep issues of cosmology dependence in the source population and treat the sources as effectively imposing a cosmology independent `visibility map' onto the survey. The resulting shape noise in the angular power spectrum is still given by Equation~\eqref{eq:shapenoisevar}, but with the effective pixel occupancy given by the moments of the actual weighted source counts in each pixel, rather than a somewhat abstract `source position averaged' version of this quantity. It is not possible however to write down the smoothed-signal terms; we cannot push our analytic expressions further than Equation~\eqref{eq:xipm_unaveraged}. This makes it challenging to develop the HEALPix window function approximation, as we did in section~\ref{sec:aliasing}. Qualitatively, as discussed in Refs.~\citep{2021JCAP...03..067N, 2024arXiv240313794G}, for high-resolution maps where pixels typically have low source galaxy occupancy, the shear map is closer to a point-sampling of the shear field than it is to a convolution. This means that the effective pixel window function transitions from something close to the HEALPix pixel window at high expected occupancy (i.e., for low-resolution grids), and a constant function (with aliasing) at low expected occupancy (i.e., for high-resolution grids).

To gain further intuition, suppose our survey footprint covers the full sky, and consider first the extreme scenario of a low-resolution grid where every pixel is well populated with source galaxies. The shear map should be well approximated by the HEALPix window function, with residuals comparable to those seen in section~\ref{sec:simulations}, i.e.~greatest on angular scales close to the pixel scale (which is much larger than the mean inter-particle separation by construction). In the opposite extreme, the shear map is at such high resolution that each pixel has at most one galaxy in it, and most pixels are empty. In this limit, the shear weights cancel out of Equation~\eqref{eq:wrongeq0}, and we have (with the correct phase factor, see appendix~\ref{sec:pt})
\begin{equation}
    \hat{\gamma}_p \rightarrow \Theta(N_p) \, \hat{\gamma}_{i(p)} \, e^{-2 \mathrm{i}
    \beta_{i(p)}},
\end{equation}
where we remind the reader that $\Theta(N_p)$ is a binary mask that is zero for empty pixels and one elsewhere, and $i(p)$ labels the galaxy in pixel $p$.\footnote{Note that in this limit the shear weights are completely redundant, which would motivate inverse-variance pixel weighting in this regime as discussed in section~\ref{sec:nn}.} Since a given $\ell$-mode averages over many pixels, it should be a reasonable approximation in this limit to model the full-sky power spectrum with a multiplicative HEALPix window function and convolved with the mixing matrix implied by the map $\Theta(N_p)$.

Putting these results together, for fixed source positions and weights, the best model we can recommend for the standard shear map estimator follows from writing, schematically,
\begin{equation}
    \hat{\gamma}_p \approx \Theta(N_p) \tilde{\gamma}_p + n_p ,
    \label{eq:model_fixedpos}
\end{equation}
where $\tilde{\gamma}_p$ is the shear field convolved with the HEALPix pixel window function and evaluated at the centre of the pixel $p$, and $n_p$ is a noise term. If the map is unnormalized as described in section~\ref{sec:nn} we can replace $\Theta(N_p)$ with $N_p^w \equiv \sum_{i \in p}w_i$, i.e.~the shear weight map, or if a globally normalized map is used as in Ref.~\citep{PKWL} we can replace $\Theta(N_p)$ with $N_p^w/\bar{w}$. The model for the shear power spectra then follows from multiplying the model power spectrum by the HEALPix window function, and then applying the mixing matrices that follow from $\Theta(N_p)$ (or $N_p^w$ if not normalizing by the pixel weight), as described in Refs.~\citep{2005MNRAS.360.1262B, 2011MNRAS.412...65H, 2019MNRAS.484.4127A, 2021JCAP...03..067N}. Note that this is still an approximation; in reality the weighted shear in a pixel cannot be factored as a weight map multiplied by a shear map, but this will at least recover the expected behaviour in the extreme cases of low occupancy and high occupancy described above.

\subsection{Unclustered sources on the full sky}
\label{subsec:unclustered_pp}

To understand what form the noise term in Equation~\eqref{eq:model_fixedpos} should take, consider the situation of unclustered sources and uniform pixel weighting. Using the statistics of $\Theta(N_p)$ given in Equation~\eqref{eq:meannp}, we can write down the mean value of the angular power spectrum of $\Theta(N_p)$. This is given by
\begin{equation}
    \langle C_\ell^\Theta \rangle = \frac{\Omega^2}{(2\ell+1)}h(\lambda) \sum_m \sum_{p,q} Y_{\ell m}^*(n_p)Y_{\ell m}(n_q) + \frac{\Omega \, g(\lambda)}{\lambda}.
    \label{eq:cltheta}
\end{equation}
Note that we have not assumed the continuum limit here. For $\ell \leq \ell_{{\rm Ny}}$, the spherical harmonics are orthogonal over the HEALPix grid, and hence the first term in Equation~\eqref{eq:cltheta} becomes simply $4\pi h(\lambda) \delta_{\ell 0}$. For $\ell > \ell_{{\rm Ny}}$ we lose this orthogonality and acquire aliased power, although the symmetries of the HEALPix grid still guarantee that only $m=0$ modes contribute to the summation, which is zero unless $\ell$ is even. Nevertheless, it is a reasonable approximation to replace the first term in Equation~\eqref{eq:cltheta} by $4\pi h(\lambda) \delta_{\ell 0}$, since high frequency modes are strongly suppressed when integrated over the sphere with uniform weight. Note that the second term in Equation~\eqref{eq:cltheta} is exact for all $\ell$. We thus have
\begin{equation}
     \langle C_\ell^\Theta \rangle \approx  4\pi h(\lambda) \delta_{\ell 0} + \frac{\Omega \, g(\lambda)}{\lambda},
     \label{eq:excltheta}
\end{equation}
This form was used in appendix D of Ref.~\citep{2024arXiv240415402V} with the approximation $C_\ell^\Theta \approx \langle C_\ell^{\Theta}\rangle$, but here we only make use of it to tell us what the variance of the noise term in Equation~\eqref{eq:model_fixedpos} should be.

Expressions for the mixing matrices may be found in, for example, Ref.~\citep{2005MNRAS.360.1262B}. Substituting in Equation~\eqref{eq:excltheta}, their expected values are
\begin{align}
    \langle M_{\ell \ell'}^+ \rangle &\approx h(\lambda)\delta_{\ell \ell'} + \frac{\Omega \, g(\lambda)}{2\lambda} \frac{2\ell'+1}{4\pi} \\
    \langle M_{\ell \ell'}^- \rangle &\approx \frac{\Omega \, g(\lambda)}{2\lambda} \frac{2\ell'+1}{4\pi}.
\end{align}
The mode-mixed power spectra following from the ansatz Equation~\eqref{eq:model_fixedpos} is therefore
\begin{align}
    \langle \tilde{C}_\ell^{E} \rangle &\approx h(\lambda)W_\ell^2C_\ell^{E} + \frac{g(\lambda)}{2\bar{n}} \sum_{\ell'} \frac{(2\ell'+1)}{4\pi} C_{\ell'}^E W_{\ell'}^2 + \langle N_\ell \rangle\\
    \langle \tilde{C}_\ell^{B} \rangle &\approx  \frac{g(\lambda)}{2\bar{n}} \sum_{\ell'} \frac{(2\ell'+1)}{4\pi} C_{\ell'}^E W_{\ell'}^2 + \langle N_\ell \rangle,
\end{align}
where $N_\ell$ is the power spectrum of the noise term in Equation~\eqref{eq:model_fixedpos}, and the angle brackets refer to the average over Poisson sources on the full sky. Comparison with Equation~\eqref{eq:EEBB_full_hp} suggests that the residual noise power spectrum should be (including shape noise)
\begin{equation}
   \langle  N_\ell \rangle \approx \frac{f(\lambda)}{2\bar{n}} \left[\sigma_e^2 + \sigma_\gamma^2 - \sum_{\ell'} \frac{(2\ell'+1)}{4\pi} C_{\ell'}^E W_{\ell'}^2 \right]
\end{equation}
Recalling the definition of $f(\lambda)$, this suggests that the \emph{source realization-specific} noise bias is given by
\begin{equation}
    N_\ell \approx \frac{\Omega^2}{8\pi} \left[\sigma_e^2 + \sigma_\gamma^2 - \sum_{\ell'} \frac{(2\ell'+1)}{4\pi} C_{\ell'}^E W_{\ell'}^2 \right] \sum_p \frac{\Theta(N_p)}{N_p^{{\rm eff}}} .
    \label{eq:nlrealspec}
\end{equation}
In other words, the ansatz Equation~\eqref{eq:model_fixedpos} says that the shear map should be modelled as the true shear field convolved with the HEALPix window function, sampled at the centers of non-empty pixels, with empty pixels incorporated into the mask, and an effective noise bias term that includes only sub-pixel variance. Following the lines of discussion in section~\ref{subsec:results_with_sn}, we can remove shape noise bias by subtracting from the power spectrum estimator terms like Equation~\eqref{eq:nest1} or Equation~\eqref{eq:nest2}. These equations remain valid in the situation where the source positions are considered fixed.

In the case of the `unnormalized' estimator (or equivalently inverse-variance pixel weighting) discussed in section~\ref{sec:nn}, the equations above remain valid with the redefinitions $h(\lambda)=g(\lambda)=\lambda^2$. In this case, the noise bias that replaces Equation~\eqref{eq:nlrealspec} is
\begin{align}
    N_\ell^{\mathrm{nn}} &= \frac{\Omega^2}{8\pi} \left[\sigma_e^2 + \sigma_\gamma^2 - \sum_{\ell'} \frac{(2\ell'+1)}{4\pi} C_{\ell'}^E W_{\ell'}^2 \right] \sum_p N_p w_{(2)} \nonumber \\
    &= \frac{\Omega^2}{8\pi} \left[\sigma_e^2 + \sigma_\gamma^2 - \sum_{\ell'} \frac{(2\ell'+1)}{4\pi} C_{\ell'}^E W_{\ell'}^2 \right] \sum_{i=1}^{N_{\mathrm{gal}}} w_i^2 ,
    \label{eq:nlrealspec_nn}
\end{align}
where the final sum runs over all galaxies in the catalogue. Appropriate noise bias estimators to subtract from the power spectra, corresponding to Equations~\eqref{eq:nest1} and Equation~\eqref{eq:nest2}, become
\begin{equation}
    \hat{N}_\ell^{\mathrm{nn}, (1)} = \frac{\Omega^2}{8\pi} \sum_p \left \lvert \sum_{i\in p} w_i \hat{\gamma}_i \right \rvert^2,
    \label{eq:nest1nn}
\end{equation}
and
\begin{equation}
    \hat{N}_\ell^{\mathrm{nn}, (2)} = \frac{\Omega^2}{8\pi} \sum_p  \sum_{i\in p} \lvert w_i \hat{\gamma}_i \rvert^2 = \frac{\Omega^2}{8\pi}  \sum_{i=1}^{N_{\mathrm{gal}}}\lvert w_i \hat{\gamma}_i \rvert^2 .
    \label{eq:nest2nn}
\end{equation}
As in section~\ref{subsec:results_with_sn}, these estimators will both remove the contribution to shape noise variance, but Equation~\eqref{eq:nest1nn} will also subtract additional white noise coming from shear correlations. Note that for Poisson distributed sources, the shear correlation contributions cancel out, c.f.~Equation~\eqref{eq:nn_expected}, so these two estimators have the same expectation value; consider setting $f(\lambda)=g(\lambda)=\lambda^2$ in Equation~\eqref{eq:nhat1}. Practically speaking, the shape noise variance dominates over the shear variance, so either estimator may be used. Ref.~\citep{PKWL} uses the analogue of $\hat{N}_\ell^{\mathrm{nn}, (2)}$ for the `global normalization' estimator Equation~\eqref{eq:gamma_pkwl}, which simply divides Equation~\eqref{eq:nest2nn} by $\bar{w}^2$, where $\bar{w}$ is given in Equation~\eqref{eq:wbar}. The correction due to shear correlations is then absorbed into a redefinition of the mixing matrix by de-biasing the weight map power spectrum; we refer to Ref.~\citep{PKWL} for further details. This approach has been independently proposed by Ref.~\citep{2024arXiv240721013W}.

\subsection{Tests against simulations}
\label{subsec:sims_prepos}

To test the performance of our statistical model for the power spectrum in the fixed-source-position case, we run simulations of full sky, unit shear weight, Gaussian shear catalogues using the same specifications as section~\ref{sec:simulations}, but keeping the realization of the source position field fixed.

\begin{figure}
\centering
    \includegraphics[width=0.45\columnwidth]{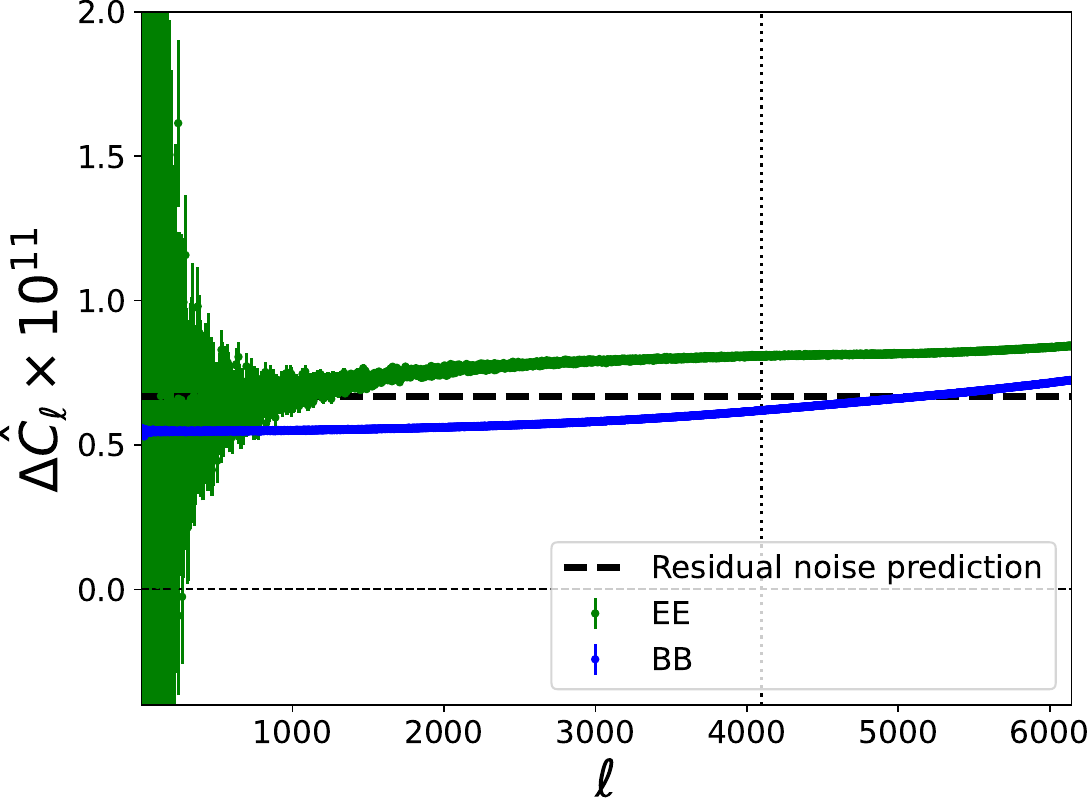}
    \includegraphics[width=0.47\columnwidth]{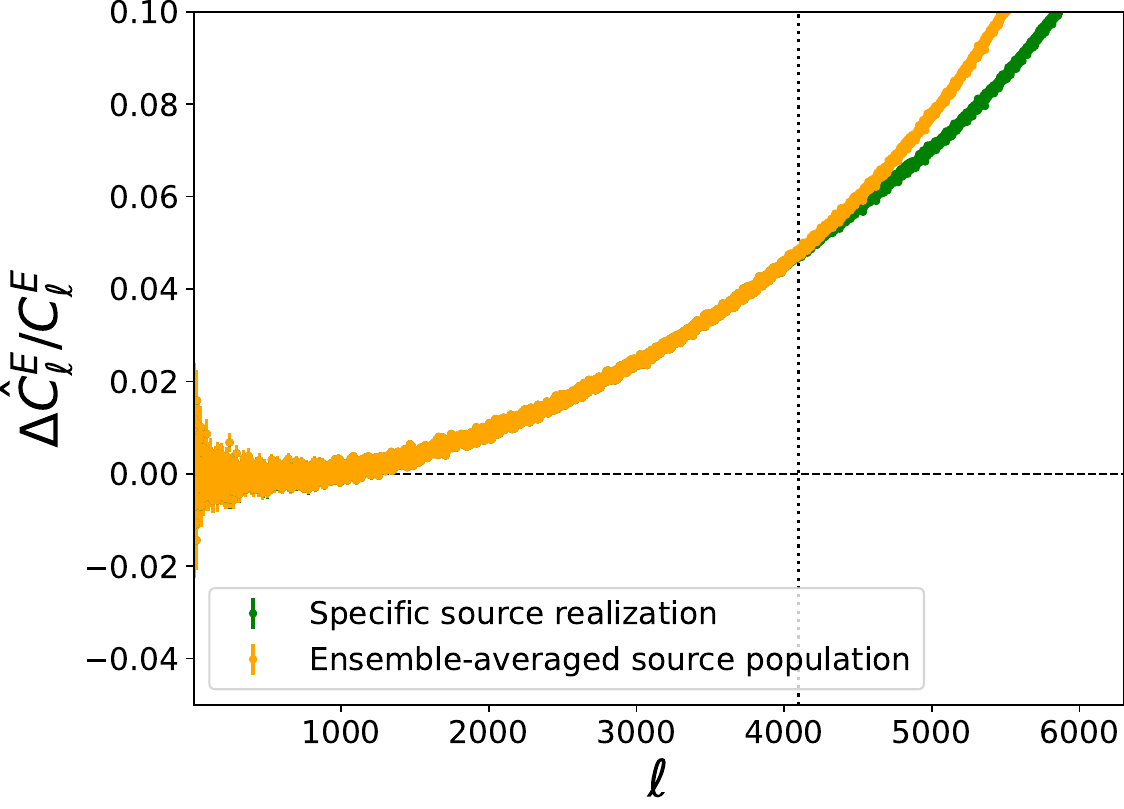}
    \caption{\emph{Left panel}: Difference between measured and modelled $EE$ and $BB$ spectra from simulations with fixed source positions. The shear map is made on a grid of resolution $N_{\mathrm{side}}=2048$. The model uses a mixing matrix derived from the standard estimator Equation~\eqref{eq:correct}, using the binary footprint map as the effective survey mask. The predicted residual noise term from sub-pixel shear variance is shown as the horizontal dashed line, and the vertical line shows the scale at which $\ell = 2N_{\mathrm{side}}$. \emph{Right panel}: Fractional difference between the measured and modelled $EE$ spectra. In green is shown the fractional difference from a model that uses a mixing matrix based on the actual observed sources, and in orange we show a model that uses the ensemble average prediction, equivalent to that used in section~\ref{sec:simulations}} 
    \label{fig:EEBB_pp}
\end{figure}

In the left panel of Figure~\ref{fig:EEBB_pp} we show residual differences between the mean power spectra and the model of Equation~\eqref{eq:model_fixedpos}, along with the noise bias prediction Equation~\eqref{eq:nlrealspec}, for a shear map with on average $\lambda=2$ galaxies per pixel. The performance of the estimator is excellent, with performance very similar to the ensemble-averaged results shown in Figure~\ref{fig:residuals}. In the right panel of Figure~\ref{fig:EEBB_pp} we show the fractional residual bias relative to the signal for both the realization-dependent model Equation~\eqref{eq:model_fixedpos} and the ensemble-averaged model Equation~\eqref{eq:EEBB_full_hp}. The residual bias, mostly due to aliasing, is very similar between these two methods, with the realization-dependent model performing slightly better, as expected.

\begin{figure}
\centering
    \includegraphics[width=0.45\columnwidth]{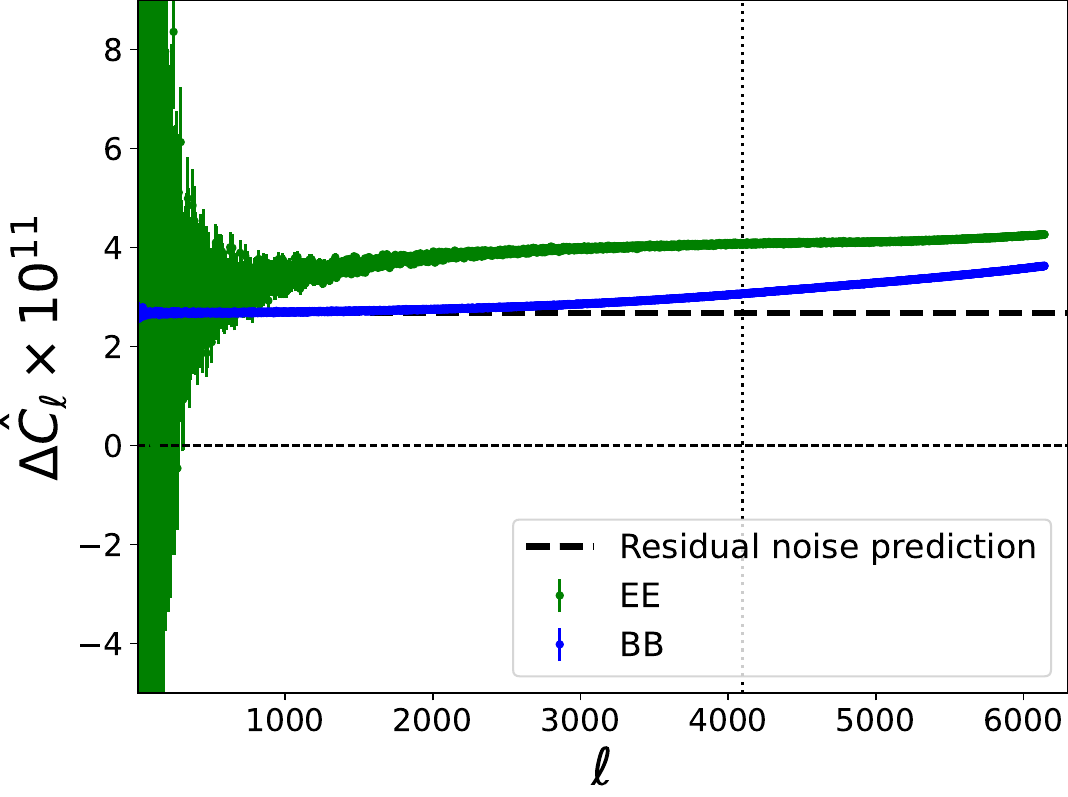}
    \includegraphics[width=0.47\columnwidth]{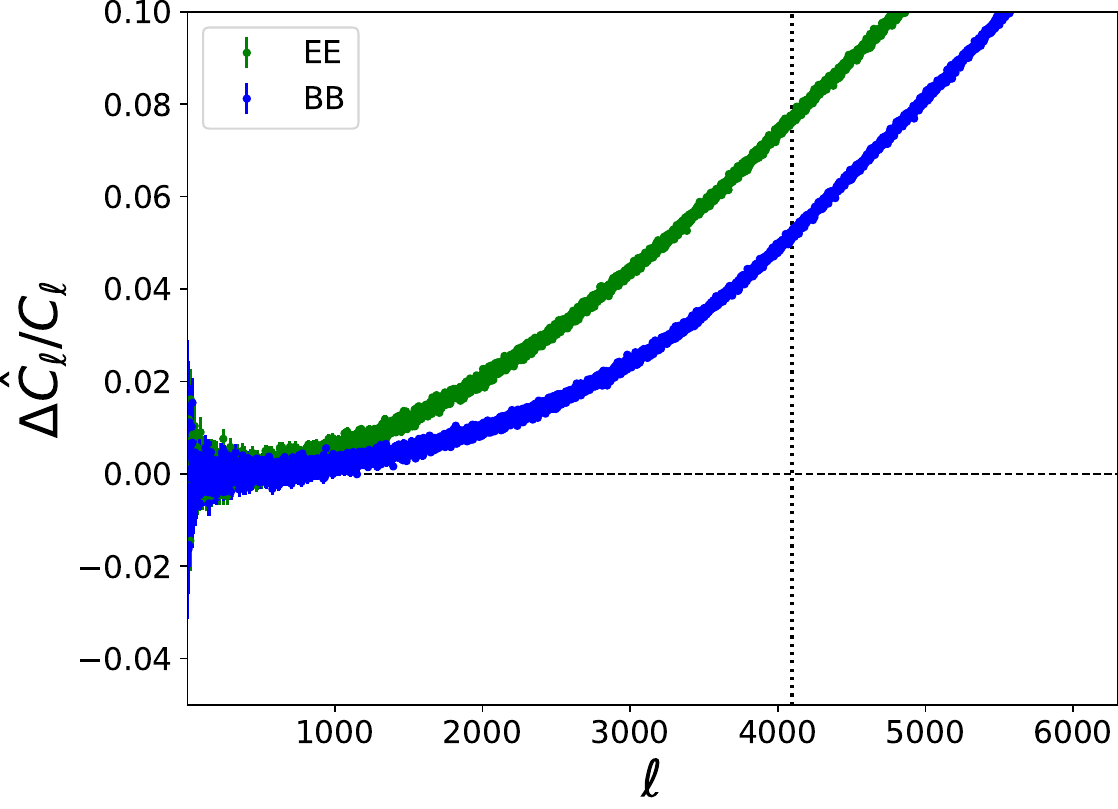}
    \caption{As Figure~\ref{fig:EEBB_pp} but using a shear map with inverse-variance pixel weighting, i.e.~an unnormalized map.} 
    \label{fig:EEBB_nn_pp}
\end{figure}

In Figure~\ref{fig:EEBB_nn_pp} we show results for the unnormalized estimator Equation~\eqref{eq:nn}. Encouragingly, the small residual bias in the $BB$ spectrum has disappeared, with biases in both $EE$ and $BB$ now at the 2\% and 1\% level at $\ell=N_{\mathrm{side}}$ for $EE$ and $BB$ respectively. This is likely due to the suppressed dependency of the noise bias on the pixel window function that occurs for Poisson sources and the estimator Equation~\eqref{eq:nn}, as shown explicitly in Equation~\eqref{eq:nn_expected}; this noise bias is the sole source of $BB$ power in our simulations. The bias in $EE$ at $\ell = 2N_{\mathrm{side}}$ is however higher than in the uniform pixel weighting case at around 7\% compared with 5\%.

To summarize the results of this section, we have shown that our models for the expected shear power spectra for fixed source positions work very well, and correctly capture the residual white noise power due to shear correlations. Our simulations have been simplistic, but sufficient to demonstrate the key properties of standard pseudo-$C_\ell$ algorithms.


\section{Interlacing HEALPix grids to mitigate aliasing}
\label{sec:interlacing}

Cosmological fields are not band-limited, and hence angular power spectra estimated via the creation of maps are vulnerable to biases from aliasing. This is demonstrated explicitly in Figure~\ref{fig:pows}, where the impact of aliasing is shown to be strongest for input spectra that have significant power on small scales. The biases seen in Figure~\ref{fig:pows} closely resemble the biases seen in all numerical tests described in the previous sections.

In this section, we explore a method for mitigating aliasing inspired by the `interlacing' technique presented in Refs.~\cite{1988csup.book.....H, 2016MNRAS.460.3624S} in the context of estimating the three-dimensional power spectrum of points using a Cartesian grid. The interlacing method works by constructing the Fourier modes of the density field as the average of two separate sets of modes built from Cartesian grids that are translated with respect to each other in every direction by half the side-length of a grid cell. This results in a significant cancellation of aliased images, and is now a standard method for estimating spectra from $N$-body simulation particle data. We refer to Ref.~\cite{2016MNRAS.460.3624S} for further details of this technique.

For the HEALPix grid, there is no direct analogue of interlacing because of the non-regularity of pixel shapes that break translational (or more correctly, rotational) invariance. However, in the equatorial region, pixels within a HEALPix ring are separated by a fixed azimuthal angle of $\pi/2N_{\mathrm{side}}$. Therefore, in analogy to the Cartesian interlacing method, we first generate a map from a set of points and then a second map rotated by half this distance, i.e.~a map that has been rotated about the $z$-axis by an angle $\Delta \phi = \pi/4N_{\mathrm{side}}$. This scheme will not be perfect, because in the polar region the pixels occupy a greater range of azimuthal angles and hence the rotation will have less of an effect on map values, but it is the simplest application of interlacing to the spherical setting.

\begin{figure}
\centering
    \includegraphics[width=0.38\columnwidth]{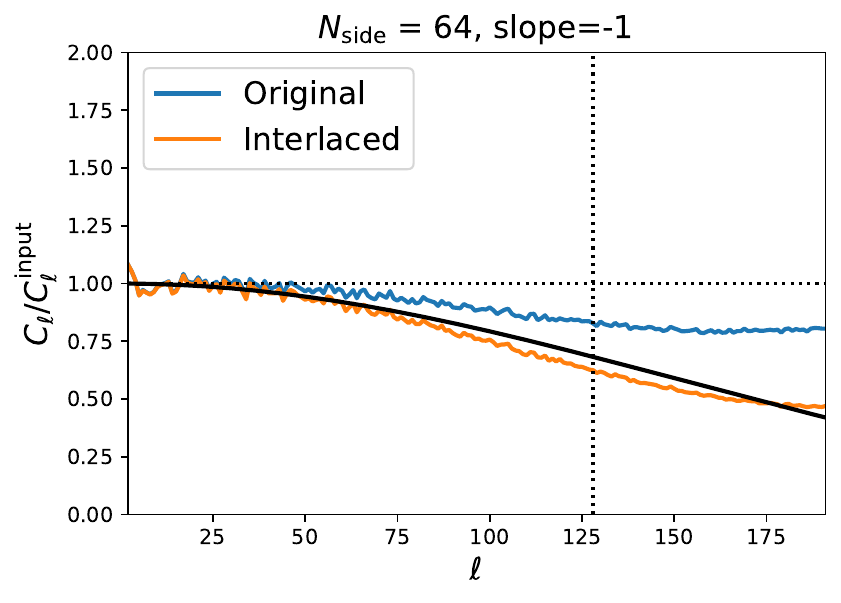}
    \includegraphics[width=0.38\columnwidth]{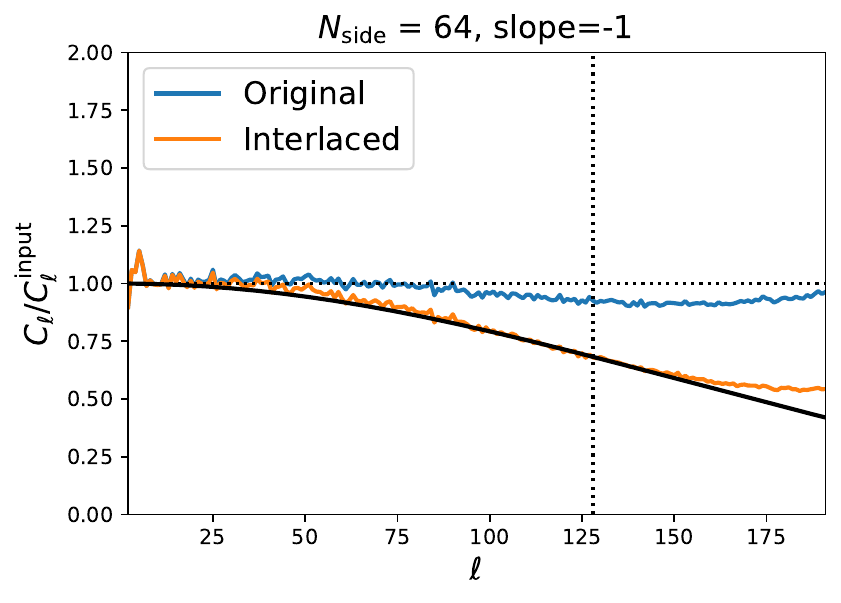}
    \includegraphics[width=0.38\columnwidth]{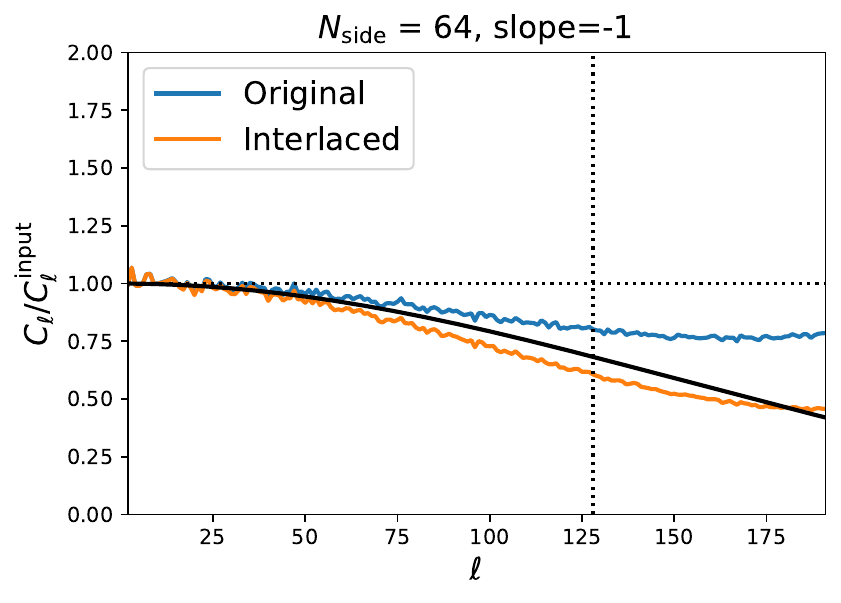}
    \includegraphics[width=0.38\columnwidth]{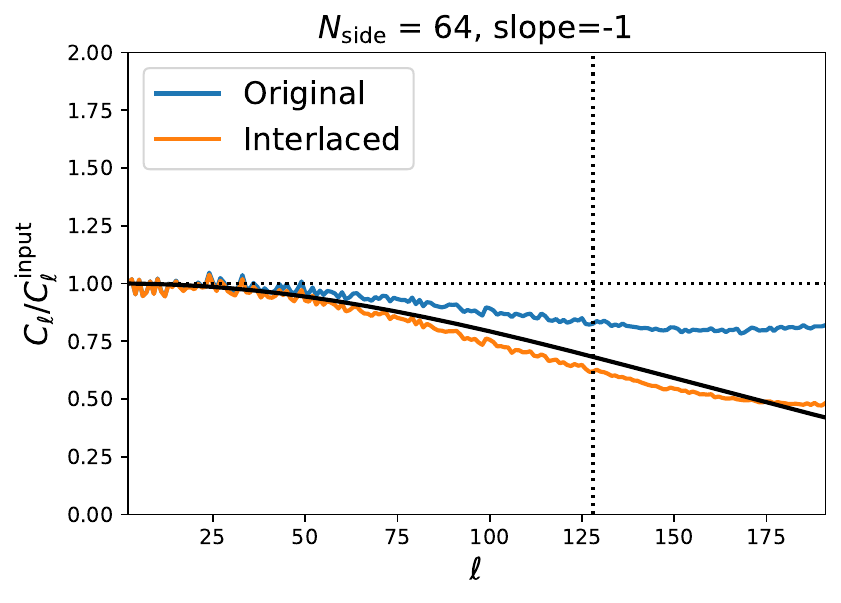}
    
    \caption{Ratio of recovered power spectra from an $N_{{\rm side}} = 64$ grid to the input power spectrum for a power law spectrum having $C_\ell \propto \ell^{-1}$. Blue curves show recovered spectra using the standard algorithm based on maps, orange curves show the results of using interlaced maps. The top row shows results from catalogues with field values assigned from an $N_{{\rm side}} = 256$ grid, the bottom row uses an $N_{{\rm side}} = 1024$ grid. The left column uses a catalogue density such that there are $2$ galaxies per high-resolution pixel, the right column uses $1$ galaxy per high-resolution pixel. Spectra have been corrected for shot noise, and the black curve is the square of the HEALPix pixel window function.}
    \label{fig:il_slopem1}
\end{figure}

In Figure~\ref{fig:il_slopem1} we show the recovered power spectra using interlaced maps from an input power-law spectrum $C_\ell \propto \ell^{-1}$. In these examples, we draw spin-0 Gaussian realizations of a high-resolution map having either $N_{{\rm side}} = 256$ (top row) or $N_{{\rm side}} = 1024$ (bottom row). The resulting map values are assigned to $N_{\mathrm{gal}}$ points placed uniformly over the sphere, with a density drawn from a Poisson distribution such that on average $2$ (left column) or $1$ (right column) galaxies reside in each high-resolution pixel. This catalogue is then randomly rotated to erase memory of the grid from which field values were assigned, and then a low-resolution $N_{{\rm side}} = 64$ map is formed by simple averaging. 

Figure~\ref{fig:il_slopem1} shows the resulting power spectrum with and without the interlacing method (after shot noise subtraction) alongside the prediction based on the HEALPix window function (black curves). In these examples, the interlacing method successfully mitigates bias due to aliasing, and the resulting power spectrum is better described by the input spectrum smoothed with the HEALPix pixel window function. The performance of the method is sensitive to the details of our numerical experiment however, and depends on the resolution of the underlying high-resolution map and the density of the catalogue.

\begin{figure}
\centering
    \includegraphics[width=0.8
    \columnwidth]{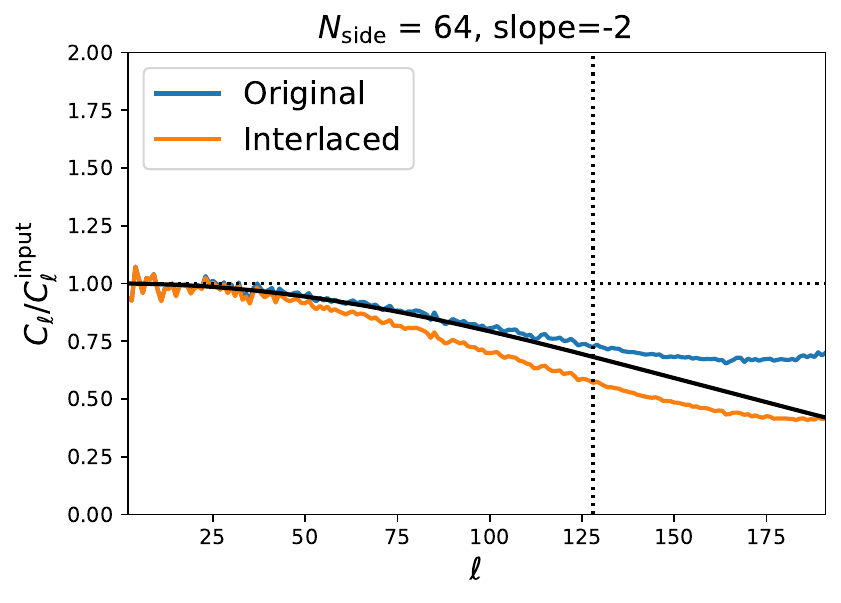}

    \caption{Same as Figure~\ref{fig:il_slopem1} but for an input spectrum $C_\ell \propto \ell^{-2}$. Catalogues were created from high-resolution maps at resolution $N_{{\rm side}}=256$ with on average two galaxies per pixel, and then low-resolution maps were created at $N_{{\rm side}}=64$, i.e.~settings corresponding to the top-left panel of Figure~\ref{fig:il_slopem1}. Alternative setups corresponding to the other panels in Figure~\ref{fig:il_slopem1} show no visual difference.}
    \label{fig:il_slopem2}
\end{figure}

In Figure~\ref{fig:il_slopem2} we show results with the same catalogue settings but with an input spectrum $C_\ell \propto \ell^{-2}$, i.e.~a spectrum having less aliased power. The interlacing method is much less effective here, and exhibits too much smoothing. The default method suffers from relatively little aliasing for $\ell \lesssim 2N_{{\rm side}}$ in this example, and the additional smoothing imposed by interlacing actually biases the power spectrum low. The sensitivity of the method to the details of the catalogue generation depends on the input spectrum itself; Figure~\ref{fig:il_slopem2} shows no visual differences when we vary the simulation parameters as done across the four panels in Figure~\ref{fig:il_slopem1}.

In conclusion, the interlacing method is a promising technique for mitigating aliasing on the sphere, but its efficacy depends on the precise input spectrum and details of the catalogue. A detailed study is beyond the scope of this paper, but we can speculate that survey footprints that can be mostly captured by equatorial pixels in a HEALPix grid may benefit from the interlacing method, since the grid structure is close to the Cartesian case where interlacing has already been shown to be effective.

\section{Pixelization on the masked sky}
\label{sec:masked}

In section~\ref{sec:aliasing} we suggested that the HEALPix pixel window function approximation may be improved by restricting its computation to the observed survey footprint. To see this, recall the definition of the HEALPix pixel window function:
\begin{equation}
    W_\ell^2 \equiv \frac{1}{N_{\mathrm{pix}}} \sum_{p=1}^{N_{\mathrm{pix}}} (W_\ell^p)^2 ,
    \label{eq:wlp}
\end{equation}
where $(W_\ell^p)^2$ is (up to a numerical factor) the angular power spectrum of pixel $p$. Figure~\ref{fig:wlp} shows a map of this quantity for $\ell=128$ and $N_{\mathrm{side}}=64$, demonstrating that the pixel window function can vary by 10\% over the sky, particularly in the polar regions (this was explored briefly in Ref.~\cite{2003ApJ...599..786R}). This suggests that an improved pixel window function can be constructed by replacing the average in Equation~\eqref{eq:wlp} with an average over pixels in the \emph{observed} region.

To test this idea, we take the shear catalogues described in section~\ref{sec:simulations} and, prior to computing the power spectra, apply a survey footprint on an $N_{\mathrm{side}}=64$ grid to each map consisting of only equatorial pixels (having $|{\cos\theta}| < 2/3$); these have particularly stable pixel shapes and low variation in the pixel power spectra, which suggests that the HEALPix approximation of a pixel-independent window function should work better. We also create a second set of power spectra using only the polar pixels, i.e.~the complement of the equatorial footprint.


\begin{figure}
\centering
    \includegraphics[width=0.45\columnwidth]{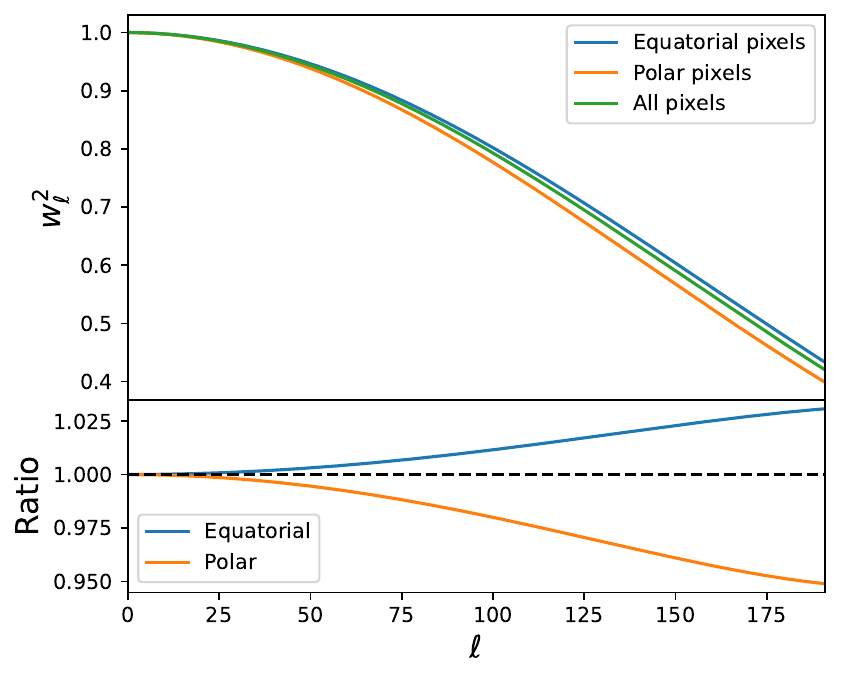}
    \includegraphics[width=0.45\columnwidth]{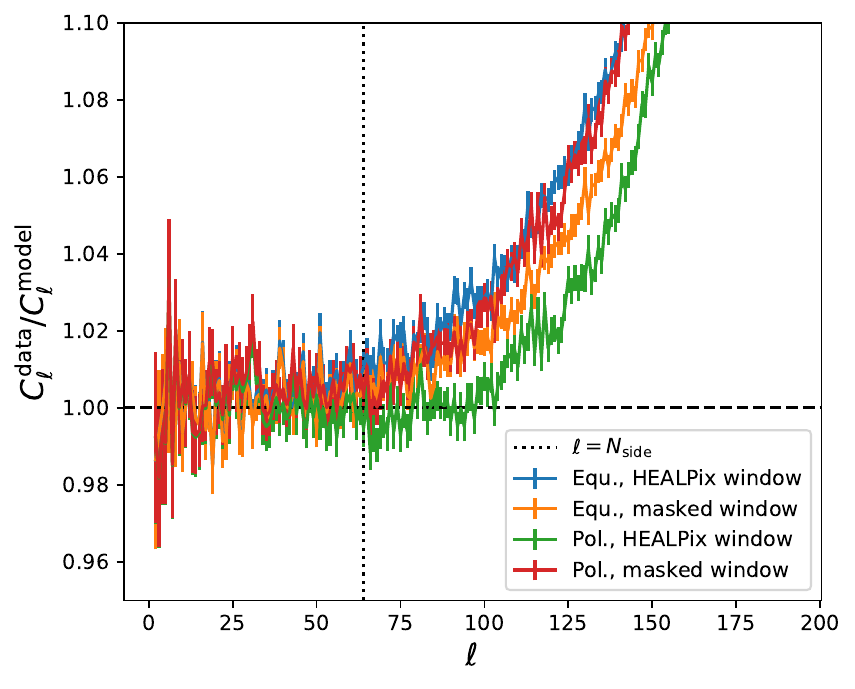}
    \caption{\emph{Left panel}: The average squared pixel window function over the whole sky (orange), along with the average over only equatorial pixels (blue) and polar pixels (green), for $N_{\mathrm{side}}=64$, along with their ratio (lower panel). \emph{Right panel}: Ratio of the measured power spectrum in equatorial pixels to a model using the global pixel window function (blue) and the equatorial-averaged pixel window (orange), alongside the ratio of the measured power spectrum in polar pixels to a model using the global pixel window function (green) and the polar-averaged pixel window (red).} 
    \label{fig:wlp_masked}
\end{figure}

Figure~\ref{fig:wlp_masked} shows the results of this exercise. In the left panel we show the average squared pixel window function over the whole sky, along with the average over only equatorial pixels and polar pixels, for $N_{\mathrm{side}}=64$. Restricting the average to the observed equatorial region boosts the pixel window by up to 3\%, with largest differences on small scales. Larger differences are seen for polar pixels, with the pixel window function suppressed by up to 5\%. The corresponding impact on the power spectrum is shown in the right panel of Figure~\ref{fig:wlp_masked}. Here, the theory power spectrum is multiplied by our bespoke pixel window and then convolved with the appropriate mixing matrix. In the case of the equatorial mask, use of a bespoke pixel window function seemingly improves the bias (the orange curve is below the blue curve), although both are strongly biased by aliasing on small scales. In the case of a polar mask, a bespoke pixel window function actually increases the bias (red curve lies above the green curve). This suggests that aliasing is dominating the bias, with deficiencies in the precise form of the pixel window function subdominant. This is consistent with Figure~\ref{fig:pows}, where all recovered spectra are biased high by aliasing, such that multiplying the theory spectrum by a smaller pixel window function (as is the case for polar pixels) enhances the bias.

\begin{figure}
\centering
    \includegraphics[width=0.8\columnwidth]{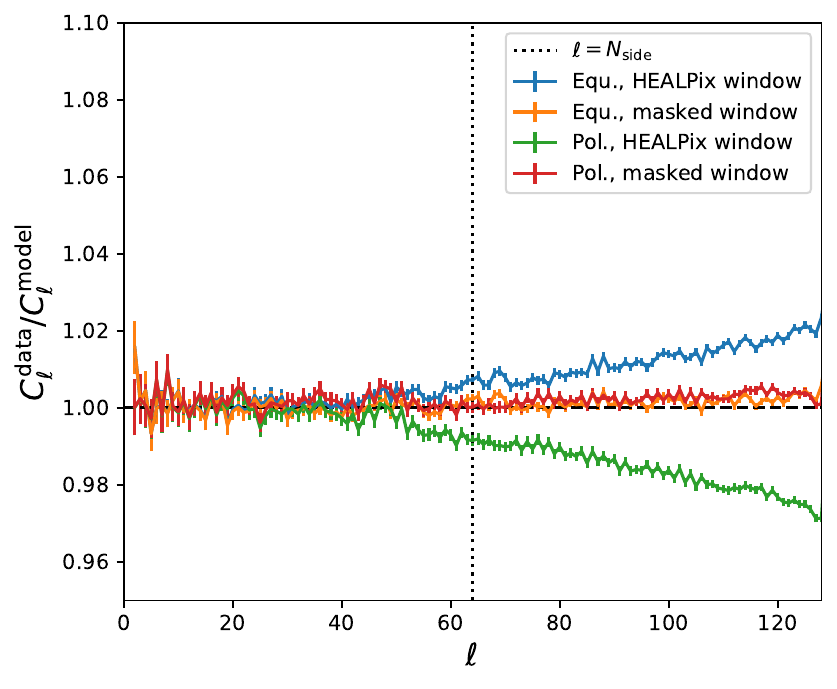}
    \caption{Ratio of the measured power spectrum in equatorial pixels to a model using the global pixel window function (blue) and the equatorial-averaged pixel window (orange), alongside the ratio of the measured power spectrum in polar pixels to a model using the global pixel window function (green) and the polar-averaged pixel window (red), for band-limited input spectra.} 
    \label{fig:wlp_masked_bl}
\end{figure}

Given that aliasing obscures the efficacy of our bespoke pixel window functions, we repeat the above experiment with band-limited input spectra having no power above twice the $N_{{\rm side}}$ of the shear map. For simplicity, in this experiment we simulate high-resolution $N_{{\rm side}}=1024$ spin-0 maps and then degrade them to $N_{{\rm side}}=64$, rather than using our simulated shear catalogues. Figure~\ref{fig:wlp_masked_bl} shows the resulting power spectra compared with models that use the bespoke pixel window functions constructed from only equatorial or polar pixels. The performance of the window functions is now much better; use of the full-sky HEALPix pixel window function induces biases in the spectra at the few percent level which are brought almost to zero by use of bespoke pixel window functions adapted to the survey footprint.

Although the results shown in Figure~\ref{fig:wlp_masked_bl} are below cosmic variance at these scales, we would expect this behaviour to be repeated at higher $\ell$ if a finer grid were used, because the pixel window function varies on roughly the same large scales (shown in Figure~\ref{fig:wlp}) for all map resolutions.

The results suggest that improving the pixel window function will only be useful if aliasing can be mitigated. This is challenging for cosmic shear, which inevitably has significant power on small scales. One interesting direction would be to combine the interlacing method presented in section~\ref{sec:interlacing} with the bespoke pixel window functions presented in this section. We defer this study to a future work. Direct evaluation of the spherical harmonic transforms, avoiding maps entirely, would partially avoid aliasing, although the number density of the source galaxies will always result in a sparse sampling of the underlying shear field and hence aliasing.


\section{Conclusions}
\label{sec:finalconc}

We have conducted a thorough study into a number of biases that show up on small scales in the standard pseudo-$C_\ell$ estimator applied to cosmic shear catalogues. These effects are all related to the pixelization of the shear catalogue, a step which usually directly precedes power spectrum estimation from the resulting shear map. We have built models for these biases and suggested a number of improvements to the standard estimator in various analysis regimes, demonstrating good agreement of our models with simulations. These improvements are motivated by upcoming cosmic shear surveys that will make precise measurements of the shear power spectra on small scales ($\ell_{\mathrm{max}} \approx 5000$). While theoretical modelling uncertainty on these scales due to non-linear clustering remains a significant challenge, we have made headway in understanding several outstanding issues on the measurement side. The main conclusions of this work are as follows:

\begin{itemize}
    
    \item We wrote down exact expressions for the action of pixelization on the full-sky shear two-point functions, and along the way we derived a spin-2 version of the convolution theorem. We elucidated the approximations that lead to the HEALPix window function in the spin-2 case, which have been missing from the literature.

    
    \item We showed how restricting power spectrum estimation to pixels that contain galaxies leads to multiplicative biases that scale with the expected pixel occupancy. We derived simple approximations to these biases that are exact in the limit of constant shear weights and Poisson sources, and found good agreement with mock shear maps. These biases can be highly significant for typical shear map resolutions and the source number densities expected from a $\mathcal{O}(10)$ redshift bin survey at a nominal Euclid-like depth. The models we developed account for most of this bias to high precision. Residual biases exceed 1\% above a multipole $\ell_* \approx N_{\mathrm{side}}$, and are mostly due to aliasing (see Figure~\ref{fig:fracdiffs}). These biases can always be safely pushed to below the pixel scale by using a sufficiently high-resolution HEALPix grid; we have shown how to deal with the resultant scenario of empty pixels.
    
    \item We showed that biases from averaging over pixel occupancy are substantially easier to compute when the shear map is not normalized by the total weight in each pixel, or equivalently by applying an inverse-variance pixel weighting. In addition, choosing not to normalize makes the statistics of the weights easier to model. We derived expressions for the expectation values of these new estimators, including noise bias subtraction. 
    
    \item We showed that further improvements to the estimators can be made when the source galaxy positions and weights are considered fixed. We wrote down a modified estimator that incorporates source density and weight fluctuations into the mixing matrix, Equation~\eqref{eq:model_fixedpos}, and extended this to the case of unnormalized shear maps. 

    \item We have shown (in appendix~\ref{sec:pt}) how the standard shear map estimator neglects a phase factor necessary for consistent averaging of the spin-2 shear field. Neglecting this factor can lead to order unity multiplicative biases in pixels near the poles, but for typical pixels the bias is much smaller, with a mean value of less than $10^{-8}$. The r.m.s.\ value of the multiplicative bias is around $10^{-4}$ for a nominal Euclid-like depth and scales as $\bar{n}^{-1/2}$ where $\bar{n}$ is the source number density. Spurious $E$ and $B$ power (i.e.~additive bias) is generated along with multiplicative bias. For map resolutions with $N_{\mathrm{side}} \geq 128$ both these biases are within Stage-IV requirements. We confirmed our predictions for the size of these biases to within an order of magnitude using mock catalogues.
    
    \item Our recommended estimator uses fixed sources and weights, globally normalized shear maps, and a catalogue-based noise bias subtraction. This estimator is thoroughly tested in Ref.~\citep{PKWL} and is implemented in the Euclid analysis pipeline.

    \item We explored an approximate interlacing technique applied to HEALPix maps as a method to mitigate aliasing. This technique works reasonably well, but shows sensitivity to the precise form of the input spectrum and the details of the catalogue construction. If this technique is to be applied on real data, a careful performance study will be required to avoid accidentally increasing power spectrum biases.

    \item We suggested a further modification to the standard pseudo-$C_\ell$ pipeline by showing how to construct bespoke pixel window functions based on the observed survey footprint. For band-limited spectra, this can reduce percent-level biases in the power spectra to zero. However, the performance of this approach is degraded for non-band-limited spectra, such as those expected from cosmic shear.
    
\end{itemize}

Residual biases at the percent level around the pixel scale remain in the power spectra due to aliasing and the variation of the HEALPix pixels over the sky. To eliminate these, there are several directions that one can take. Firstly, one can choose to replace the model for the power spectra with simulations that have the pixelization effect in them by construction. This `brute force' approach is guaranteed to be unbiased, at the cost of a potentially significant computation cost. Alternatively, one can choose to bypass maps altogether, as in Ref.~\citep{2023arXiv231212285B, 2024arXiv240721013W, PKWL}, and directly compute the spherical harmonics at the galaxy positions. This method is highly promising, although further work is needed to reduce the run-time to acceptable levels at the required resolution. Finally, a novel idea is presented in Ref.~\citep{PKWL}; instead of averaging in HEALPix pixels, one can average in circular disks tied to the centers of HEALPix pixels. In this setup, pixels necessarily overlap, but the operation is mathematically a convolution and hence the pixel window function can be exactly accounted for if the convolution kernel is chosen appropriately. This method has a free parameter (the kernel width), but should eliminate biases. Further work is required to thoroughly test this method in the context of upcoming cosmic shear surveys.

\acknowledgments
We thank Anton Baleato Lizancos and Eric Hivon for useful correspondence and discussion, and the anonymous referee for a highly constructive review. We acknowledge the extensive use of the \texttt{healpy}~\citep{Zonca2019} Python package. AH is supported by a Royal Society University Research Fellowship. For the purpose of open access, the author has applied a Creative Commons Attribution (CC BY) licence to any Author Accepted Manuscript version arising from this submission.

\bibliographystyle{JHEP}
\bibliography{biblio.bib}

\appendix

\renewcommand\thefigure{\thesection.\arabic{figure}}    

\section{Bias due to neglecting parallel transport of shear}
\setcounter{figure}{0}

\label{sec:pt}

In this section we discuss in detail the bias in angular power spectra due to neglecting the phase factor required to correctly form a shear map.

First, let us re-examine the defining equation for the shear map in the standard approach,
\begin{equation}
\label{eq:wrongeq}
    \hat{\gamma}_p = \frac{\sum_{i \in p} w_i \, \hat{\gamma}_i}{\sum_{i\in p}w_i}.
\end{equation}
There is a problem with Equation~\eqref{eq:wrongeq}. The shear field is the projection of a two-dimensional tensor field $P_{ab}$ defined on the sphere, with $\gamma_1 \pm \mathrm{i}\gamma_2 = \mathrm{e}_{\pm}^a \mathrm{e}_{\pm}^b P_{ab}$ for helicity basis vectors $\mathbf{e}_{\pm}$. We must account for the fact that adding tensors on a curved manifold only makes sense at the same point in the manifold. Furthermore, the basis vectors defining the shear depend on position, and this dependence must be accounted for in the averaging procedure. 

\subsection{Parallel transport of shear}
\label{subsec:ptofshear}

\begin{figure}
\centering
    \includegraphics[width=0.6\columnwidth]{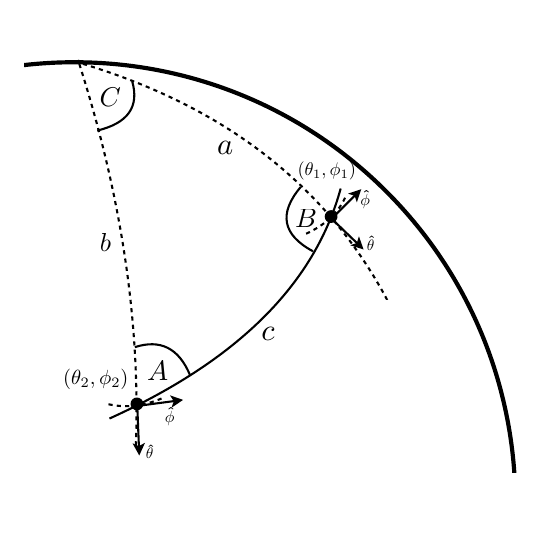}
    \caption{Spherical geometry for parallel transport.}
    \label{fig:sphere}
\end{figure}

To construct an averaging procedure that makes sense for shear, we must \emph{parallel transport} the shear of each galaxy to the centre of its parent pixel. We will first assume that the `polarization tensor' $P_{ab}$ is constant over the pixel area (i.e.,~has vanishing covariant derivatives), such that it is sufficient to consider just the action of parallel transport on the local $(\hat{\mathbf{x}}, \hat{\mathbf{y}})$ basis defining the shear. Consider a galaxy located at coordinates $(\theta_2, \phi_2)$. We define the shear with respect to a local $x$-$y$ basis given by the normalized spherical basis vectors $(\hat{\boldsymbol{\theta}}, \hat{\boldsymbol{\phi}})$.\footnote{Note that this potentially differs to other conventions, with some authors defining the shear on a left-handed basis set with the local $z$-axis parallel to the outward normal to the sphere. In this case the shear is spin $-2$ and the final rotation we derive here should have an extra minus sign. Our convention matches the HEALPix and CAMB conventions, but differs from the IAU convention.} First consider parallel transporting these basis vectors up to the North Pole and then down the meridian to the pixel centre at $(\theta_1, \phi_1)$. By considering Figure~\ref{fig:sphere}, it is clear that the basis rotates by an angle $C$ in a left-handed sense about the outward normal $\hat{\mathbf{n}}_1$ relative to the $(\hat{\boldsymbol{\theta}}, \hat{\boldsymbol{\phi}})$ basis at $(\theta_1, \phi_1)$. Parallel transporting back to the galaxy's position along the geodesic results in a net rotation of $A+B+C-\pi$, the excess area of the spherical triangle. This rotation is in a left-handed sense about $\hat{\mathbf{n}}_2$. This implies that the effect of parallel transporting from $(\theta_1, \phi_1)$ to $(\theta_2, \phi_2)$ is to rotate the vectors by $A+B-\pi$. Therefore, parallel transporting in the opposite direction, from $(\theta_2, \phi_2)$ to $(\theta_1, \phi_1)$ must result in a rotation of $\beta \equiv \pi-A-B$ in a left-handed sense about $\hat{\mathbf{n}}_1$ relative to the basis there. This rotation transforms the helicity basis vectors as $\mathbf{e}_{\pm} \rightarrow e^{\pm \mathrm{i} \beta}\mathbf{e}_{\pm}$. 

The shear therefore transforms under parallel transport as
\begin{equation}
    (\gamma_1 \pm \mathrm{i} \gamma_2)(\hat{\mathbf{n}}_2) \rightarrow (\gamma_1 \pm  \mathrm{i} \gamma_2)(\hat{\mathbf{n}}_1)e^{\pm 2\mathrm{i} \beta}.
\end{equation}
The physical quantity we wish to average in the pixel is the shear tensor at each galaxy's position parallel transported to the pixel centre $\hat{\mathbf{n}}_p$. Evaluating this average on the helicity basis at $\hat{\mathbf{n}}_p$ gives $\hat{\gamma}_p$. The helicity basis at $\hat{\mathbf{n}}_p$ can be expressed in terms of the parallel-transported basis by \emph{undoing} the rotation by $\beta$. The correct averaging procedure is therefore given by
\begin{equation}
    \hat{\gamma}_p = \frac{\sum_{i \in p} w_i \, \hat{\gamma}_i \, e^{-2 \mathrm{i}\beta_i}}{\sum_{i\in p}w_i},
    \label{eq:correct}
\end{equation}
where $\hat{\gamma}_i$ is measured in the standard helicity basis at $\hat{\mathbf{n}}_i$. Note that this estimator is equivalent to first expressing the galaxy shear on a basis aligned with the geodesic - this is achieved by rotating the basis at $\hat{\mathbf{n}}_i$ by $A$ in a left-handed sense about $\hat{\mathbf{n}}_i$. Such a rotation is equivalent to a rotation of the global (lab frame) coordinate system by an angle $\phi_1$ in a right-handed sense about the global $Z$ axis, followed by a rotation by $\theta_1$ in a right-handed sense about the new $Y$ axis. The geodesic is now at constant longitude, such that parallel transport of the shear to $\hat{\mathbf{n}}_p$ gives no additional rotation. Subsequently performing the reverse rotation of the global axes then amounts to rotating the local basis at $\hat{\mathbf{n}}_p$ by an angle $B-\pi$ in a left-handed sense about the outward normal there. The new shear at $\hat{\mathbf{n}}_p$ is hence $\hat{\gamma}_i\, e^{-2 \mathrm{i}\beta_i}$, and is expressed in the correct global helicity basis.

The angle $\beta$ can be derived using the laws of spherical trigonometry. For example, we have
\begin{align}
    &\cos c = \cos \theta_1 \cos \theta_2 + \sin \theta_1 \sin \theta_2 \cos (\phi_2 - \phi_1)\\
    &\sin A = \frac{-\sin \theta_1 \sin (\phi_2 - \phi_1)}{\sin c}\\
    &\cos A = \frac{\cos \theta_1 \sin \theta_2 - \sin \theta_1 \cos \theta_2 \cos (\phi_2 - \phi_1)}{\sin c}\\
    &\sin B = \frac{-\sin \theta_2 \sin (\phi_2 - \phi_1)}{\sin c}\\
    &\cos B = \frac{\cos \theta_2 \sin \theta_1 - \sin \theta_2 \cos \theta_1 \cos (\phi_2 - \phi_1)}{\sin c}
\end{align}

For small pixels with $\theta_2 = \theta_1 + \delta \theta$ and $\phi_2 = \phi_1 + \delta \phi$ with $\delta \theta, \delta \phi \ll 1$, the angle $\beta$ has the expansion
\begin{equation}
\label{eq:blim}
    \beta = \delta \phi\left(-\cos{\theta_1} + \frac12 \delta \theta  \, \sin \theta_1 + \dots\right).
\end{equation}
The rotation angle is thus largest near the poles and is at least quadratic around the equator. Under repeated draws from a uniform galaxy position field the average angle is zero to at least quadratic order, since $\langle\delta \theta \sin \theta \rangle = -\langle \delta \cos \theta \rangle =0$. Note that Equation~\eqref{eq:blim} is also valid for any $\delta \phi$ for small pixels around the poles, which is important because those pixels wrap around and $|\delta \phi| \lesssim \pi/2$.

The formula Equation~\eqref{eq:blim} is consistent with the numerical results of Ref.~\citep{PKWL}. In particular, we have $\beta=0$ along lines of longitude, consistent with the fact that parallel transport along a geodesic gives no effect. Similarly, around the equator where $\theta_1 = \pi/2$ and $\delta \theta = 0$, the effect is also zero.

\subsection{Bias in shear maps incurred by ignoring the phase factor}
\label{sec:bias}

\begin{figure}
\centering
    \includegraphics[width=0.8\columnwidth]{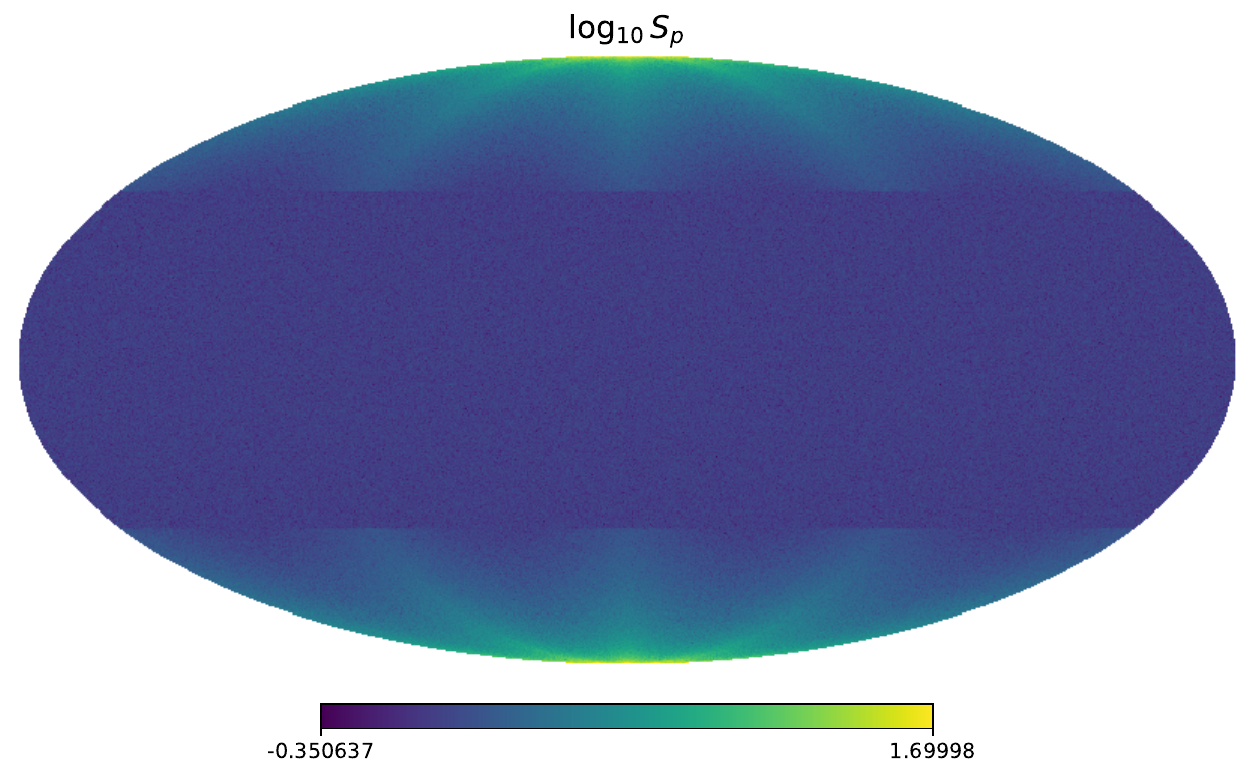}
    \caption{Map of the normalized azimuth variance within pixels, $S_p$, defined as $\langle \delta \phi^2 \rangle = S_p^2 \Omega/12$, for a map with resolution $N_{\mathrm{side}}=512$. Note that there is Monte Carlo noise of around 15\% in every pixel.}
    \label{fig:Sp}
\end{figure}

Neglect of the individual phase factors results in a multiplicative bias on each shear in the pixelized map. Continuing to assume that the polarization tensor is constant over the pixel, the incorrect expression Equation~\eqref{eq:wrongeq} evaluates as
\begin{equation}
    \hat{\gamma}_p = \gamma_p\frac{\sum_{i \in p} w_i e^{2 \mathrm{i}\beta_i}}{\sum_{i \in p} w_i},
\end{equation}
where $\gamma_p$ is the shear at the centre of the pixel. For small pixels such that $\beta \ll 1$ we can use Equation~\eqref{eq:blim} to write $\hat{\gamma}_p \approx \gamma_p(1+m_p)$, where the multiplicative bias is 
\begin{equation}
    m_p = -2 \mathrm{i} \cos\theta_p \frac{\sum_{i\in p}w_i \delta \phi_i}{\sum_{i \in p}w_i} -  \mathrm{i} \frac{\sum_{i\in p}w_i \delta \phi_i \, \delta \mu_i}{\sum_{i \in p}w_i} - 2\cos^2\theta_p\frac{\sum_{i\in p}w_i \delta \phi_i^2 }{\sum_{i \in p}w_i} + \mathcal{O}\left(\epsilon^3\right), 
\end{equation}
where we defined $\mu = \cos \theta$, and $\epsilon$ is an order-counting parameter of order $\delta \phi$. The linear bias is zero on average since galaxies are equally likely to be on the `left' side of a pixel as the `right'. The bias is greatest near the poles of the map, as expected, and is zero around the equator. At leading order, we have
\begin{equation}
    \langle m_p \rangle = -2 \cos^2 \theta_p \langle \delta \phi^2 \rangle + \mathcal{O}\left(\epsilon^3\right).
\end{equation}
We will set $\langle \delta \phi^2 \rangle = S_p^2 \Omega/12$ where $S_p$ is a shape factor of order unity that depends on the shape of the given pixel. For square pixels on the equator we have $S_p=1$, which is a good approximation for most of HEALPix pixels which lie in the equatorial region. For circular pixels on the equator, we have $S_p = \sqrt{3/\pi} \approx 1$. In the polar regions we expect $S_p \gg 1$. In Figure~\ref{fig:Sp} we plot $S_p$ for a map with resolution $N_{\mathrm{side}}=512$. The large-scale features of the HEALPix grid are clearly visible, and although $S_p \approx 1$ for most of the pixels it can be several orders of magnitude lager than that near the poles. We find $\langle S_p \rangle \approx 1.5$ and $\langle S_p^2 \rangle \approx 8$, which is only weakly dependent on the resolution; for $N_\mathrm{side} = 64$ we find $\langle S_p^2 \rangle \approx 6$, for example. We find the maximal value of $S_p$ to be $S_p^{\mathrm{max}} \approx 1.3 N_\mathrm{side}$, reflecting the fact that there are always 4 pixels around the poles for all resolutions, meaning $\langle \delta \phi^2 \rangle$ is roughly constant for these pixels and so $S_p \propto 1/\sqrt{\Omega} \propto N_{\mathrm{side}}$ for the most polar pixels; these pixels are also where $S_p$ is maximal.

The average bias in a pixel has a maximum value (near the poles) of $\langle m_p \rangle \approx -0.3$. Most pixels have $\langle m_p \rangle$ much less than this however; for $N_\mathrm{side} = 4096$ we have $|\langle m_p \rangle| \lesssim 10^{-8}$ for example. For the typical grid resolutions used in weak lensing surveys, the \emph{average} bias across the map is therefore negligible, except for the most northerly and southerly pixels. The maximum bias among all \emph{realizations} of the source position field occurs in the most polar pixels when every galaxy is at the edge of its parent pixel. This bias is roughly $\mathrm{max} \, |m_p| \approx S_p^{\mathrm{max}} \sqrt{2 \Omega} \approx 1.88$, independent of map resolution. These pixels will therefore be strongly biased unless the rotation factor is accounted for.

If there are $N_p$ galaxies in pixel $p$, the r.m.s.\ multiplicative bias in that pixel is
\begin{equation}
    m^{\mathrm{rms}}_p = \mathrm{i} \frac{\cos \theta_p}{\sqrt{3}} \, S_p \left \langle \frac{w_{(2)}(p)}{w_{(1)}^2(p)} \right \rangle^{1/2} n_p^{-1/2} + \mathcal{O}\left(n_p^{-1/2} \Omega\right),
\end{equation}
where $w_{(n)} \equiv \sum_{i \in p} w_i^n /N_p$, and $n_p \equiv N_p/\Omega$ is the number density of galaxies in the pixel. For typical values of $N_{\mathrm{side}}$ the second-order term can be safely neglected. Assuming galaxies are placed into $N_z$ tomographic redshift bins of equal number density, neglecting density fluctuations across the map, and denoting the total number density across all redshift bins by $\bar{n}_{\mathrm{tot}}$, the r.m.s.\ bias in a given bin evaluates as

\begin{equation}
  m^{\mathrm{rms}}_p \approx  \mathrm{i} \cos\theta_p \, S_p \left \langle \frac{w_{(2)}(p)}{w_{(1)}^2(p)} \right \rangle^{1/2} \left(\frac{N_z}{10}\right)^{1/2} \left(\frac{\bar{n}_{\mathrm{tot}}}{30 \, \mathrm{arcmin}^{-2}}\right)^{-1/2} \times 10^{-4}.
  \label{eq:rmsm}
\end{equation}
A typical pixel therefore has a (purely imaginary, at leading order) multiplicative bias of magnitude $\sim 10^{-4}$. Allowing for variations of the galaxy density across the sky, the r.m.s.\ bias can in principle be larger than Equation~\eqref{eq:rmsm} in low density regions; for example if there is only a single galaxy in the pixel then $m_p \approx  \mathrm{i} \cos \theta_p \, S_p \sqrt{\Omega/3}$. For $N_{\mathrm{side}} = 8192$ we get back the same $\sim 10^{-4}$ r.m.s.\ bias. A significantly larger bias would require large empty pixels, which is unlikely given the typical map resolutions used in practice. Note that since $S_p \approx 1.3 N_{\mathrm{side}}$ in the extremal polar caps the r.m.s.\ bias is still significant even at high densities for typical resolutions.

\subsection{Spatial variation of the bias and impact on the power spectra}
\label{eq:pCl}

The multiplicative bias varies across the sky. On angular scales well above the pixel size we can take the dominant spatial dependence of $m_p$ to come from the $\cos \theta_p$ term in Equation~\eqref{eq:blim}. This corresponds to a pure $\ell=1, m=0$ spherical harmonic, coupling each harmonic coefficient of the shear to its neighbour in $\ell$, whilst leaving different $m$-modes uncoupled.

Although neglecting the phase factor gives map-level biases to the shear map, a more pressing question is to what extent the pseudo-$C_\ell$ are impacted by this bias. Treating the source positions as uncorrelated between pixels implies zero correlation between the bias in different pixels. Keeping terms to leading order in $\Omega$ and assuming zero intrinsic $B$-mode power, we can show that the biased power spectra are given, on the full sky and for uniform pixel weighting, by
\begin{align}
    \left \langle \hat{C}_\ell^{EE} \right \rangle &\approx \left(1 - \frac{\langle S_p^2 \rangle \Omega}{9}\right)C_\ell^{EE} + \frac{\langle S_p^2 \rangle \Omega}{18} \frac{\sigma_\gamma^2}{\bar{n}_p} \left \langle \frac{w_{(2)}(p)}{w_{(1)}^2(p)} \right \rangle \\
    \left \langle \hat{C}_\ell^{BB} \right \rangle &\approx \frac{\langle S_p^2 \rangle \Omega}{18} \frac{\sigma_\gamma^2}{\bar{n}_p} \left \langle\frac{w_{(2)}(p)}{w_{(1)}^2(p)} \right \rangle \\
    \left \langle \hat{C}_\ell^{EB} \right \rangle &=0 
\end{align}
where $\sigma_\gamma^2$ is the total shear variance. If the pixels are small enough that they contain only a few galaxies, which will be the case for, say, nominal Euclid Wide Survey depth with $N_z=10$ and $N_{\mathrm{side}} = 4096$, then we may take galaxies as shot noise distributed within a pixel. To get a rough order of magnitude estimate of the bias we will set $n_p = \bar{n}_{\mathrm{tot}}/N_z$ and neglect the variation of $S_p$ over the survey, setting $S_p=1$. For complete sky coverage, this gives a white noise additive bias in both $EE$ and $BB$ of amplitude

%
%

%
\begin{equation}
    \frac{\langle S_p^2 \rangle \Omega}{18} \frac{\sigma_\gamma^2}{\bar{n}_p} \left \langle\frac{w_{(2)}(p)}{w_{(1)}^2(p)} \right \rangle  \approx (2.2 \times 10^{-19}) \left(\frac{\sigma_\gamma}{0.017}\right)^2 \left(\frac{4096}{N_{\mathrm{side}}}\right)^2  \left(\frac{\langle S_p^2 \rangle}{8}\right) \left(\frac{N_z}{10}\right) \ \left(\frac{\bar{n}_{\mathrm{tot}}}{30 \, \mathrm{arcmin}^{-2}}\right)^{-1}.
    \label{eq:addbias}
\end{equation}
For nominal Euclid-like survey values and taking $\sigma_\gamma = 0.017$ (its value for $\ell_\mathrm{max} = 13000$ and a fiducial Planck cosmology), this bias is well within the total requirement on a constant additive power spectrum bias recommended by~\citep{2013MNRAS.429..661M} in order to keep the bias on dark energy parameters to below $0.1 \sigma$ (see their Figure 2). We note that the bias is effectively a small correction to the shape noise by a factor $\mathcal{O}(10^{-8})$ for $N_{\mathrm{side}} = 4096$. Even if additive bias purely from power spectrum estimation is allocated 10\% of the budget identified by~\citep{2013MNRAS.429..661M}, requirements are satisfied as long as $N_{\mathrm{side}} \geq 16$, which is trivially satisfied for a high-resolution shear map.


In the case of Poisson-distributed sources, the bias is of a white noise form, and hence could be removed exactly if the usual pseudo-$C_\ell$ noise bias is removed with Monte Carlo simulations. If instead the noise bias is estimated from the data itself using the methods of~\citep{2021JCAP...03..067N} then the additive bias will remain.

The multiplicative bias is negative and given by
\begin{equation}
- \frac{\langle S_p^2 \rangle \Omega}{9} \approx -(5.5 \times 10^{-8}) \left(\frac{\langle S_p^2 \rangle}{8}\right)   \left (\frac{4096}{N_{\mathrm{side}}}\right)^2.
\label{eq:multibias}
\end{equation}
The total requirement on the multiplicative bias for a Euclid-like survey is $4 \times 10^{-3}$~\citep{2013MNRAS.429..661M}. If we allocate 5\% of this budget to the biases from power spectrum estimation, the requirement is satisfied for $N_{\mathrm{side}} \geq 128$. This is slightly stricter than the additive bias requirement, but still trivially satisfied for a high-resolution map.

In reality galaxies are clustered, and this introduces pixel-to-pixel correlations in the shear bias. The only relevant modes of the density field in this case are those with wavelengths below the pixel scale. Consider two galaxies with azimuthal offsets $\delta \phi_i$ and $\delta \phi_j$ belonging to pixels $p$ and $p'$ with $p \ne p'$ and inter-pixel separation $r_{pp'}$. One can show that, in the flat-sky approximation and when $r_{pp'} \gg \sqrt{\Omega}$, the bias covariance is given by
\begin{equation}
    \langle m_p m_{p'}^* \rangle \approx - \frac{4}{27}\cos \theta_p \cos \theta_{p'} \xi'(r_{pp'})\frac{S_p S_{p'} \Omega}{r_{pp'}},
\end{equation}
where $\xi(r)$ is the galaxy angular correlation function and a prime denotes differentiation. Unless sources are strongly clustered on sub-pixel scales, these spatial correlations are suppressed by two factors of $\sqrt{\Omega}/r_{pp'}$, here assumed to be $\ll 1$. There will be larger effects for pixel separations comparable to the pixel size, e.g.~for neighbouring pixels, but these must be studied numerically. On angular scales well above the pixel scale however, we expect the non-whiteness of the multiplicative bias to be negligible, such that $m_p$ can essentially be taken as uncorrelated between pixels.

\subsection{Tests on simulations}
\label{subsec:pixrot_sims}

\begin{figure}
\centering
    \includegraphics[width=\columnwidth]{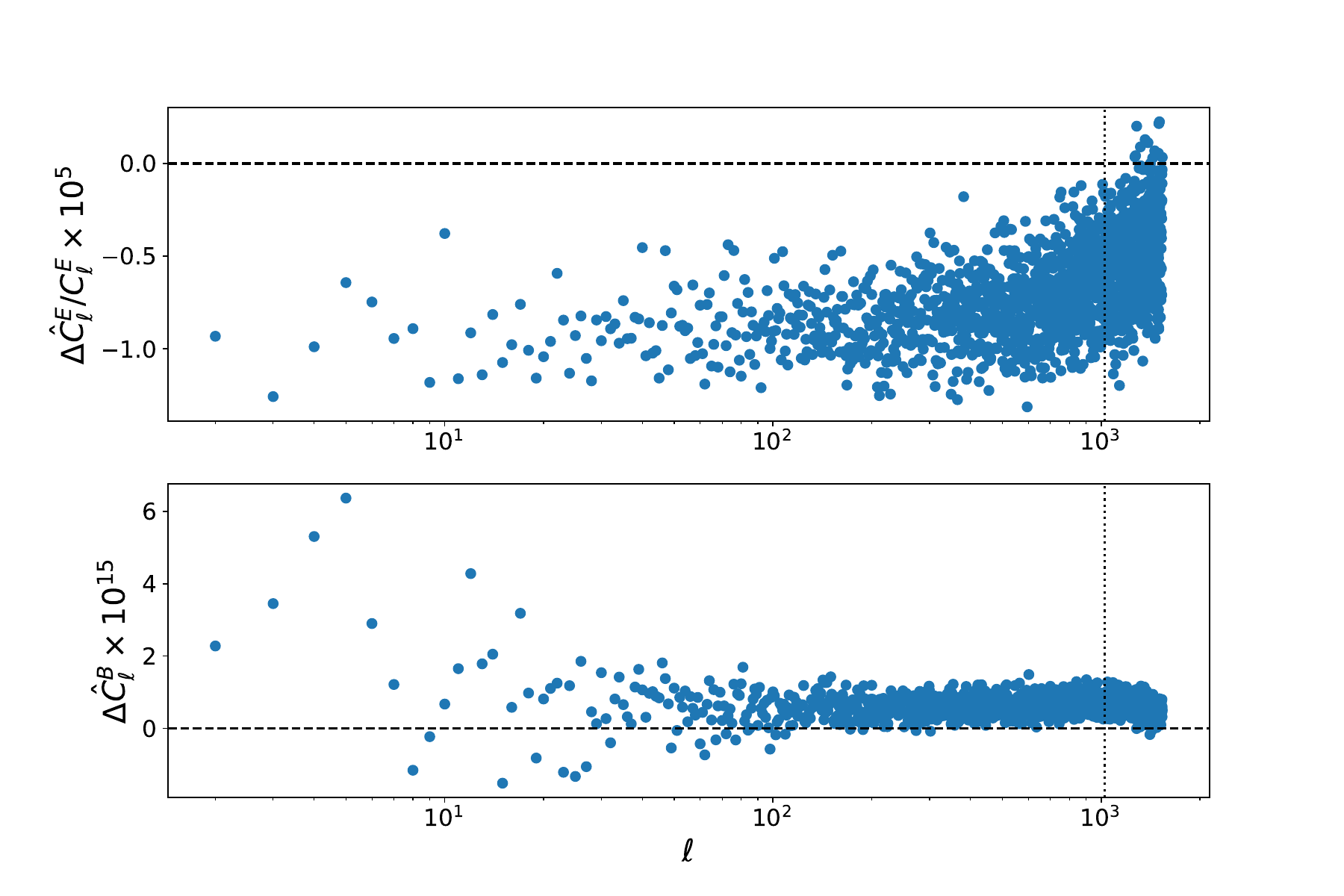}
    \caption{\emph{Top panel}: Fractional bias to the recovered $E$-mode power spectrum from neglecting phase rotation for a shear map with $\lambda=32$ expected pixel occupancy and resolution $N_{\mathrm{side}}=512$. \emph{Bottom panel}: Absolute bias to the $B$-mode power spectrum from neglecting phase rotation. The vertical dashed line indicates the scale given by $\ell = 2N_{\mathrm{side}}$.}
    \label{fig:pixrotbias}
\end{figure}

Using the mock catalogues described in section~\ref{sec:simulations}, we can test the impact of neglecting the phase rotation factor on the power spectra. In Figure~\ref{fig:pixrotbias} we show the fractional correction to the $E$-mode power spectrum, i.e.~the multiplicative bias, and the $B$-mode power spectrum, which is a proxy for the additive bias to the $E$ mode power, for a map with $\lambda=32$, i.e.~$N_{\mathrm{side}}=512$. 

The qualitative picture established in section~\ref{sec:bias} that ignoring phase rotation leads to scale-independent multiplicative and additive biases, the latter sourcing equal amplitudes of $E$ and $B$ modes, appears to be correct. The amplitude of the multiplicative bias is around $-8\times 10^{-6}$, while the additive bias is around $5 \times 10^{-16}$. This is to be compared with Equations~\eqref{eq:multibias} and \eqref{eq:addbias}, which predict biases of around $-4 \times 10^{-6}$ and $6 \times 10^{-17}$. Our prediction thus underestimates the multiplicative bias by factor 2, and the multiplicative bias by a factor 10, but in both cases the biases are well within Stage-IV requirements.

For a shear map with resolution $N_{\mathrm{side}}=2048$ we find a multiplicative bias of order $-10^{-7}$ and no detection of $B$ mode power (i.e.~additive bias), the upper limit being a few $\times 10^{-17}$. These are consistent with our analytic predictions, and confirm that the bias to power spectra from neglecting sub-pixel phase rotation is negligible for typical shear map resolutions.

\subsection{Recommendations for Stage-IV surveys}
\label{subsec:concs}

Neglecting to parallel transport individual galaxies to the centre of their parent pixel prior to averaging to form a shear map results in a multiplicative bias to the value of the shear in each pixel. At leading order this bias is purely imaginary and is negligible across the map except in the extremal polar regions. The bias varies with an r.m.s.\ amplitude of roughly $10^{-4}$ across most of the sky, but can be much larger than this near the poles. This is comparable to the upper limit of requirements on the shear multiplicative bias uncertainty. However, it is approximately uncorrelated between pixels, which leads to multiplicative and additive biases in both $E$ and $B$ that are within Euclid-like requirements as long as the shear map has $N_{\mathrm{side}} \geq 128$ resolution. Furthermore, the extremal polar regions occupy an increasingly negligible fraction of the survey area as the map resolution increases.

If one wants to construct low resolution shear maps, the bias is easy to correct simply by using the correct parallel transport formulae when forming the shear map, given by Equation~\eqref{eq:correct}. This is already implemented in correlation function codes that use large effective pixel areas when correlating shears over large scales, and numerical tips on implementing the expressions can be found in the source code of these software packages. For the typical map resolutions used in practice, it is acceptable to use the series expansion given in Equation~\eqref{eq:blim}, truncated at leading or next-to-leading order.

Our recommendation is that surveys should apply the correct phases factors computed here when using low resolution ($N_{\mathrm{side}} \leq 64$) maps, otherwise the pixel rotation may be neglected or trivially implemented using the first-order expansion.

\section{Effect of pixelization on super-pixel shear modes}
\setcounter{figure}{0}   

\label{subsec:superpix_pert}

We can get a rough idea of the impact of pixelization by considering scales much larger than the pixel scale. In this section we borrow heavily from the formalism of~\citep{2014A&A...571A..17P}, generalizing it to spin-2 fields.

Let us first consider what effect the action of pixelizing the shear catalogue has on the underlying spectrum. We can write the shear map as
\begin{align}
    \hat{\gamma}_p &= \frac{\sum_{i \in p} w_i \, \hat{\gamma}_i \, e^{-2 \mathrm{i}\beta_i}}{\sum_{i\in p}w_i} \nonumber \\
    &= \frac{\sum_{\ell m} (E_{\ell m} +  \mathrm{i}B_{\ell m}) \, \sum_{i \in p} w_i  \, e^{-2\mathrm{i}\beta_i} \, {}_2 Y_{\ell m}(\hat{n}_i)}{\sum_{i\in p}w_i} \nonumber \\
    &\equiv \sum_{\ell m} (E_{\ell m} + \mathrm{i}B_{\ell m}) \Upsilon^p_{\ell m},
    \label{eq:gammap_def}
\end{align}
where $\Upsilon^p_{\ell m}$ is a pixel-dependent filtered harmonic.

If we consider only modes that are far above the pixel scale, we can approximate the spin-weighted spherical harmonic as only slowly varying over the pixel scale and Taylor expand it around the pixel centre at $\hat{n}_p$. To do this, we have to be careful to account for the rotation of the basis vectors defining the shear. Recall that $\hat{\gamma_i}\, e^{-2\mathrm{i}\beta_i}$ is the measured shear of galaxy $i$ parallel transported to the pixel centre. In order to Taylor expand around $\hat{n}_p$, we need to express this in terms of the shear that would be measured at the pixel centre, for which $\beta=0$ by definition. The shear tensor at $\hat{n}_i$ parallel transported to $\hat{n}_p$, $\tilde{P}_{ab}$, is given in terms of the shear tensor at $\hat{n}_p$ by~\citep{2002PhRvD..66l7301C}
\begin{equation}
    \tilde{P}_{ab}(\hat{n}_p) = P_{ab}(\hat{n}_p) + d^c \nabla_c P_{ab}(\hat{n}_p) + \frac12 d^c d^d \nabla_c \nabla_d P_{ab}(\hat{n}_p) + \dots,
\end{equation}
where $\nabla_a$ is a covariant derivative on the sphere, and $d^a$ is the vector on the sphere at $\hat{n}_p$ that lies tangent to the geodesic starting at $\hat{n}_p$ and ending at $\hat{n}_i$. To get the shear at the pixel center we need to evaluate this on the helicity basis at $\hat{n}_p$. This gives~\citep{2002PhRvD..66l7301C}
\begin{equation}
    \hat{\gamma_i}\, e^{-2 \mathrm{i}\beta_i}  = \hat{\gamma} - \frac12 \left({}_1d \, \bar{\eth} + {}_{-1}d \, \eth \right)\hat{\gamma} + \frac18 \left({}_1d \, {}_1d \, \bar{\eth} \, \bar{\eth} + {}_1d \, {}_{-1}d \, \bar{\eth}\, \eth  + {}_{-1}d \, {}_{1}d \, \eth \, \bar{\eth} + {}_{-1}d \, {}_{-1}d \, \eth \, \eth \, \right)\hat{\gamma} + \dots,
\end{equation}
where $\eth$ and $\bar{\eth}$ are spin raising and lowering operators respectively, ${}_{\pm1}d$ are the spin $\pm1$ components of the displacement vector (on the same helicity basis as the shear), and all terms are evaluated at $\hat{n}_p$. Substituting this into the expression for the shear map gives
\begin{align}
    \hat{\gamma}_p =& \hat{\gamma}(\hat{n}_p) - \frac12 \left(\widehat{\langle {}_1d \rangle}_p \, \bar{\eth} + \widehat{\langle{}_{-1}d\rangle}_p \, \eth \right)\hat{\gamma}(\hat{n}_p) \nonumber \\
    & +\frac18 \left(\widehat{\langle{}_1d \, {}_1d \rangle}_p\, \bar{\eth} \, \bar{\eth} + \widehat{\langle{}_1d \, {}_{-1}d \rangle}_p\, \bar{\eth}\, \eth  + \widehat{\langle{}_{-1}d \, {}_{1}d \rangle}_p\, \eth \, \bar{\eth} + \widehat{\langle{}_{-1}d \, {}_{-1}d\rangle}_p \, \eth \, \eth \, \right)\hat{\gamma}(\hat{n}_p) + \dots,
    \label{eq:pixbias}
\end{align}
where $\widehat{\langle \cdot \rangle}_p$ denotes the weighted empirical average over the pixel $p$, an unbiased estimate of $\langle \cdot \rangle_p$. This estimate is for a given realization of the galaxy position field.

Averaging Equation~\eqref{eq:pixbias} over unclustered galaxy positions within a pixel gives a non-zero bias to the pixelized shear. Terms linear in the displacement average to zero by construction.
%
%
%
This gives, neglecting correlations between shear and galaxy weights,
\begin{align}
    \langle {}_{\pm2}\Upsilon_{\ell m}^{p} \rangle &= \left[1 - \frac14(\ell^2 + \ell - 4)\, \sigma_d^2(p) + \dots \right]{}_2Y_{\ell m}(\hat{n}_p) \nonumber \\
    & + \frac{1}{8}\left[\sqrt{(\ell+2)(\ell+1)\ell(\ell-1)} \, {}_2I_p \, Y_{\ell m}(\hat{n}_p) \right. \nonumber \\
    &\left. + \sqrt{(\ell+4)(\ell+3)(\ell-2)(\ell-3)} \, {}_{-2}I_p \, {}_4 Y_{\ell m}(\hat{n}_p) \right] + \dots,
    \label{eq:pixelbias}
\end{align}
where $\sigma_d^2(p)$ is the variance of the displacement within pixel $p$, i.e.~the pixel moment of inertia divided by the pixel area. The mean across all pixels of the term on the first line in square brackets in Equation~\eqref{eq:pixelbias} is the low-$\ell$ expansion of the standard HEALPix window function for polarization. The quantities ${}_{\pm 2}I_p$ are the spin $\pm 2$ components of the moment of inertia tensor, given explicitly by ${}_{\pm2}I_p = e_{\pm}^a e_{\pm}^b I^p_{ab}$ with
\begin{equation}
    I^p_{ab} \equiv \int \frac{\mathrm{d}^2 \hat{n}}{\Omega} W_p(\hat{n}) (\hat{n} - \hat{n}_p)_a (\hat{n} - \hat{n}_p)_b
\end{equation}
Note that $\sigma_d^2(p) = \mathrm{Tr}(I^p)$. Note also that ${}_{\pm 2}I_p = 0$ for pixels having four-fold rotational symmetry. Deviations from this particular symmetry in the pixels couple with the higher spin derivatives of the shear field.

\begin{figure}
\centering
    \includegraphics[width=0.7\columnwidth]{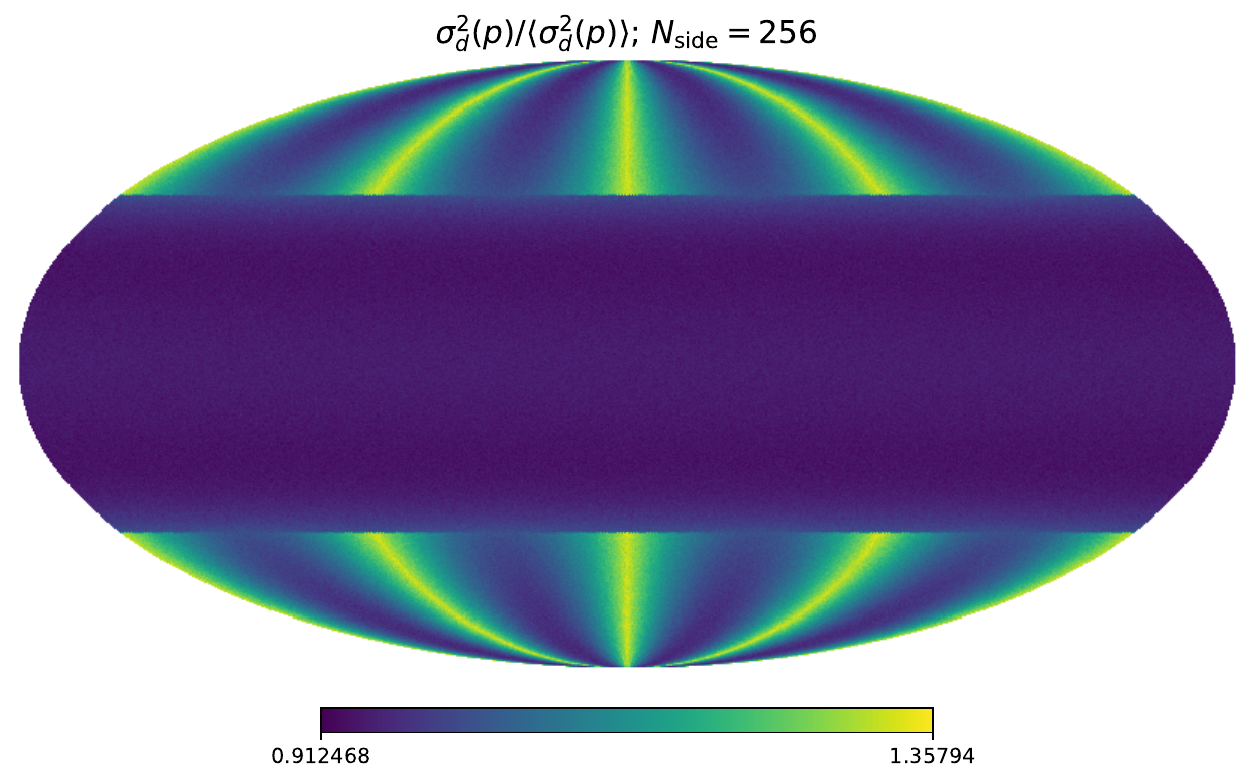}
    \caption{Variation of the pixel moment of inertia with respect to its mean value, for $N_{\mathrm{side}}=256$. Note that there is percent-level Monte Carlo noise in this map, uncorrelated between pixels.}
    \label{fig:varMoI}
\end{figure}

Equation~\eqref{eq:pixelbias} is an `average' convolution kernel that depends on the shape of the given pixel. Since $\sigma^2_d(p)$ and ${}_{\pm 2}I_p$ are both $\mathcal{O}(\Omega)$, the requirement for the mode to be super-pixel is that subsequent terms in the expansion (denoted by ellipses) are suppressed, i.e., $\ell \sqrt{\Omega} \ll 1$. In Figure~\ref{fig:varMoI} we show the variation of $\sigma_d^2(p)$ around its mean value (see~\citep{P15_beams} for a similar plot). This variation, which is especially prominent outside of the equatorial region defined by $\lvert\cos \theta \rvert< 2/3$, is \emph{not} accounted for in the standard HEALPix pixel window function. Likewise, the spin-weighted inertias ${}_{\pm2}I_p$ are not accounted for by the HEALPix pixel window.

\begin{figure}
\centering
    \includegraphics[width=0.7\columnwidth]{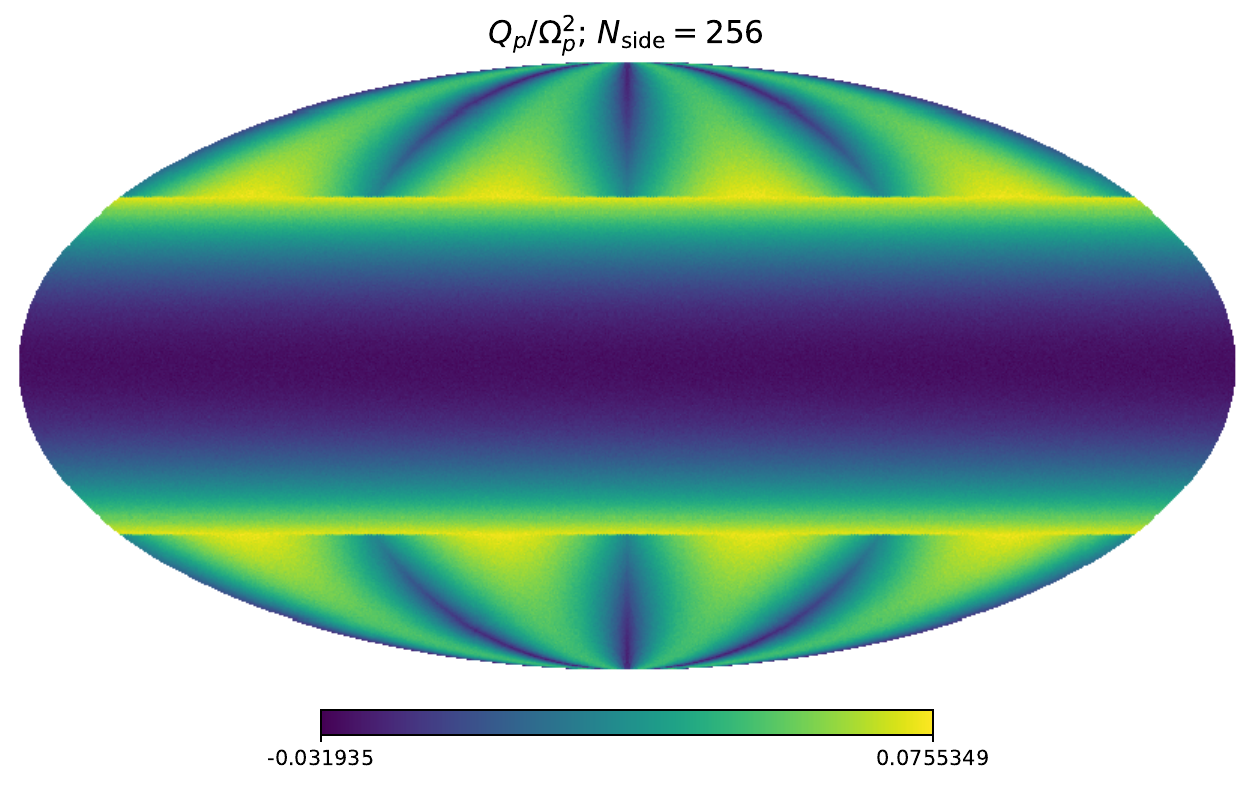}
    \includegraphics[width=0.7\columnwidth]{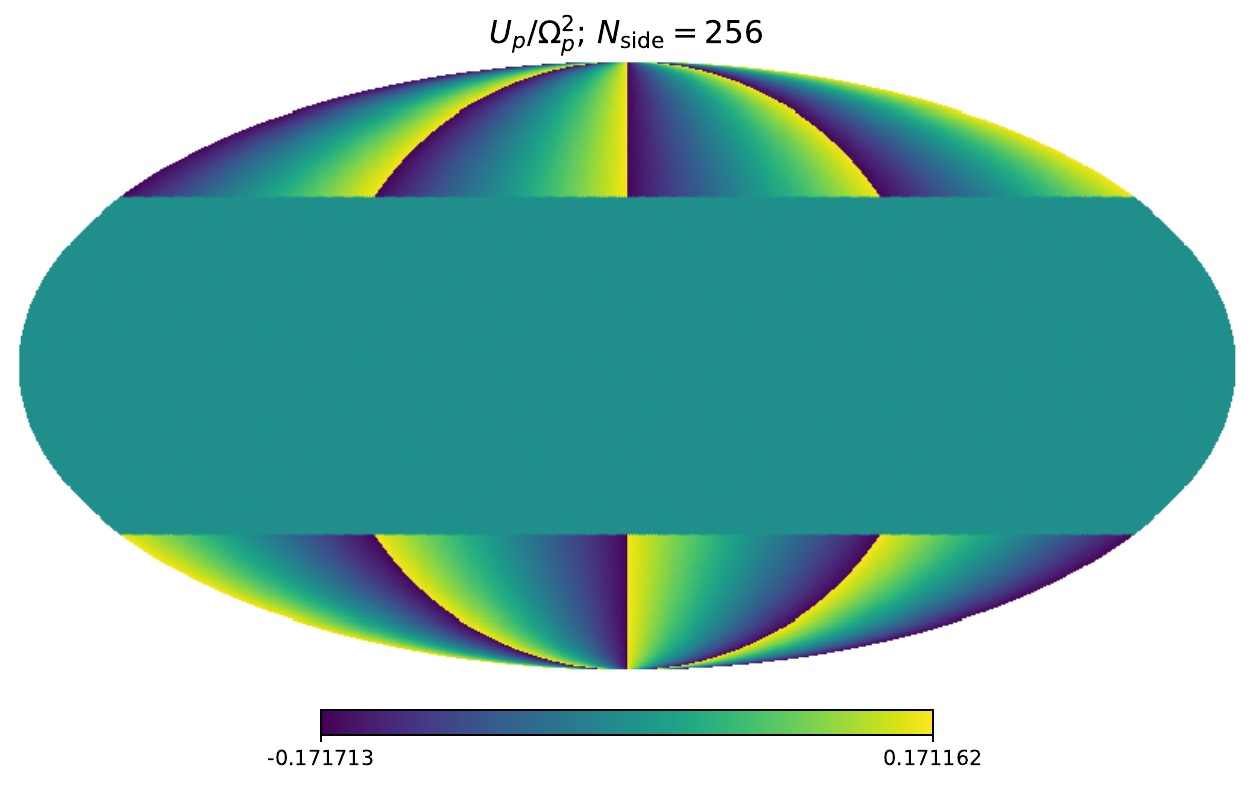}
    \caption{Real (top) and imaginary (bottom) parts of the spin-2 moment of inertia ${}_2I_p$, for $N_{\mathrm{side}}=256$, in units of the pixel area squared.}
    \label{fig:QUMoI}
\end{figure}

In Figure~\ref{fig:QUMoI} we show the two components of the pixel anisotropy ${}_2I_p$, analogous to the $Q$ and $U$ Stokes parameters of polarization. Around the equatorial region the $U$ component is almost zero, with the $Q$ component strongly suppressed, in line with the fact that the pixels are close to square in this region and so possess the four-fold rotational symmetry that gives vanishing ${}_{\pm2}I_p$. In particular, these pixels are oriented approximately North-South and East-West, implying zero $U$ polarization. 

Equation~\eqref{eq:pixelbias} implies that pixelization induces a bias in the shear map due to the suppression of sub-pixel modes. This can be mostly removed by correcting the multipoles with the HEALPix window function, or by incorporating the window into the model. Residuals then come from the variation shown in Figure~\ref{fig:varMoI} over the observed area, the spin-weighted inertia terms, and higher-order terms not accounted for in the perturbative expansion of Equation~\eqref{eq:pixelbias}.

\subsection{Effect of pixelization on super-pixel shear map correlation functions}
\label{subsubsec:realpix}

We can gain further insight into the effects of pixelization by considering the configuration-space correlation function of points on the HEALPix grid. 

The two-point functions of the weighted displacements can be computed if we assume that the weights are uncorrelated with the displacements (we will return to this point later on). These read, for fixed pixel occupancy,
\begin{align}
    \langle \widehat{\langle {}_{\pm1} \bar{d} \rangle}_p  \widehat{\langle {}_{\pm 1} \bar{d} \rangle}^*_{p'} \rangle &= \delta_{pp'} \frac{1}{N_p} \left \langle \frac{w_{(2)}(p)}{w_{(1)}^2(p)} \right \rangle \sigma_d^2(p)\\
    \langle \widehat{\langle {}_{\pm 1} \bar{d} \rangle}_p  \widehat{\langle {}_{\mp 1} \bar{d} \rangle}^*_{p'} \rangle &= \delta_{pp'}  \frac{1}{N_p} \left \langle  \frac{w_{(2)}(p)}{w_{(1)}^2(p)} \right \rangle {}_{\pm 2} I_p,
\end{align}
where $\bar{d}$ is the displacement vector on a spin basis aligned with the geodesic connecting $p$ and $p'$, and, as in appendix~\ref{sec:pt}, $N_p$ is the pixel occupancy and $w_{(n)}(p) \equiv \sum_{i \in p} w_i^n /N_p$. The angle brackets around the shear weight terms here denote averages over the distribution of the shear weights at fixed pixel occupancy. For the purposes of the present discussion we will neglect the correlation between the galaxy positions in the numerator of Equation~\eqref{eq:gammap_def} and those in the normalizing factor, i.e.~in the denominator of Equation~\eqref{eq:gammap_def}. This is equivalent to assuming no correlation between source position and shear weight. In this case the shear correlation functions on the grid, averaged over the source displacements, are
\begin{align}
    \hat{\xi}_{+}(p,p') &\approx \hat{\xi}_{+}(\hat{n}_p, \hat{n}_{p'}; W_p, W_{p'}) \nonumber \\
 &+ \frac{1}{4}\delta_{p,p'}\frac{1}{N_p} \left \langle 
 \frac{w_{(2)}(p)}{w_{(1)}^2(p)} \right \rangle \left [ \sigma_d^2(p) (\bar{\eth} \hat{\gamma} \, \bar{\eth}^* \hat{\gamma}^* + \eth \hat{\gamma} \, \eth^* \hat{\gamma}^*) + {}_2I_p \, \bar{\eth} \hat{\gamma} \, 
 \eth^* \hat{\gamma}^* + {}_{-2}I_p \, \eth \hat{\gamma} \,  \bar{\eth}^* \hat{\gamma}^*\right ], \\
     \hat{\xi}_{-}(p,p') &\approx \hat{\xi}_{-}(\hat{n}_p, \hat{n}_{p'}; W_p, W_{p'}) \nonumber \\
     &+ \frac{1}{4}\delta_{p,p'}\frac{1}{N_p} \left \langle  \frac{w_{(2)}(p)}{w_{(1)}^2(p)} \right \rangle \left [ \sigma_d^2(p) (\bar{\eth} \hat{\gamma}\, \eth \hat{\gamma}+ \eth \hat{\gamma} \, \bar{\eth} \hat{\gamma}) + {}_2I_p \, \bar{\eth} \hat{\gamma} \, \bar{\eth} \hat{\gamma} + {}_{-2}I_p \, \eth \hat{\gamma} \, 
     \eth \hat{\gamma} \right],
\end{align}
where $\hat{\xi}_{\pm}(\hat{n}_p, \hat{n}_{p'}; W_p, W_{p'})$ are shear correlation function estimates using shears filtered with the window functions in Equation~\eqref{eq:pixelbias}, i.e.~the zeroth order term plus `1-3' terms. Further averaging over galaxy ellipticities gives
\begin{align}
    \xi_{+}(p,p') &\approx \xi_{+}(\hat{n}_p, \hat{n}_{p'}; W_p, W_{p'}) + \delta_{p,p'} \frac{\sigma_d^2(p)}{2N_p} \left \langle \frac{w_{(2)}(p)}{w_{(1)}^2(p)}\right \rangle R_\gamma \\
    \xi_{-}(p,p') &\approx \xi_{-}(\hat{n}_p, \hat{n}_{p'}; W_p, W_{p'})
\end{align}
where $R_\gamma$ is the variance of the shear gradient, given by
\begin{equation}
    R_\gamma = \sum_{\ell \geq 2} \frac{2\ell+1}{4\pi} (C_\ell^E + C_\ell^B)(\ell^2 + \ell - 4).
\end{equation}
%
%
%
Thus, in the limit of purely super-pixel shear modes and for Poisson-distributed galaxies, pixelization induces \emph{white noise} that depends on the gradient power of the shear field. Note that this does not affect the $\xi_-$ correlation function, which vanishes at zero lag when only super-pixel modes are present.

For a single pixel containing only a single galaxy we have $\xi_\pm(p,p) \approx \xi_\pm(\hat{n}_p,\hat{n}_{p};1,1)$, i.e.~all dependence on the pixel shape disappears. This makes sense intuitively, since in this case the process of pixelization is just to translate this galaxy to the pixel centre. Since the underlying field is isotropic the variance is translation invariant, so the final variance estimate is invariant to this translation. When $N_p>1$ the variance estimate accumulates shears from galaxies that are not co-located, which brings in sensitivity to the shear gradient. 

We note that, like shape noise, the additional white noise from pixelization vanishes when cross-correlating shear maps from two different redshift bins. Unlike shape noise however, it is present when correlating a shear map with a galaxy number counts map constructed from the same set of galaxies, due to the non-zero correlation of the galaxy overdensity gradient with the intrinsic alignment gradient. The white noise also depends on the cosmological model, unlike shape or shot noise.

The white noise contributed by the shear gradient has a simple interpretation. Any binning of galaxies in a pixel will result in a shear value that is slightly rotated with respect to the `truth' due to the finite number of galaxies and the fact the shear varies below the pixel scale. This noise is like shape noise, and is suppressed when $N_p \gg 1$. 

The gradient power is formally infinite, although in practice the combined effect of excluding very close galaxies and strongly lensed galaxies will keep it finite. Even so, we have lost control of our perturbative expansion ($\ell\sqrt{\Omega} \ll 1$). The small random rotation, uncorrelated between pixels, should manifest as a roughly white noise power contributing to $E$ and $B$. In the case that $p\neq p'$ we have shown that the effect of pixelization is to smooth the shear field by the window function in Equation~\eqref{eq:pixelbias} and sample at pixel centres.

\section{Conditional Poisson statistics of pixelized density fields}
\label{app:conddensstats}

In section~\ref{sec:aliasing}, we consider the statistics of shear estimators after averaging over \emph{unclustered} source positions at fixed pixel occupancy, and then ultimately over the pixel occupancies themselves. In section~\ref{sec:nn} we consider an alternative approach where the source galaxy positions are held fixed and shears averaged over. In this section, we give some useful results for the statistics of the density field conditioned on a fixed pixel occupancy.

The expectation value of the weighted density, taken over repeated realizations of unclustered galaxy positions, is
\begin{equation}
    \langle n_w(\hat{n}) \rangle = \frac{1}{\Omega}\sum_{i \in p}^{N_p} w_i = \frac{N_p}{\Omega} w_{(1)}(p) ,
    \label{eq:meannw}
\end{equation}
where $w_{(1)}(p) \equiv \sum_{i \in p} w_i/N_p$ for non-empty pixels, $N_p$ is the (unweighted) number of galaxies in the pixel, $p$ refers to the pixel containing position $\hat{n}$, and $\Omega$ is the pixel area. The summation in Equation~\eqref{eq:meannw} is over galaxies in pixel $p$. Note that the individual galaxy weights are also held fixed here. This effectively decouples the shear weights from the source galaxy positions, which is clearly unrealistic, but simplifies the discussion considerably. In reality, the position of a galaxy in the focal plane impacts the shear measurement quality, and hence the weight, due to proximity to bright stars, CCD edge effects, chip defects, foreground contamination, etc. We will assume that all of these effects have been corrected for in the image processing or shear measurement, and assume that the weights are essentially random numbers drawn from a position-independent distribution that are assigned to each galaxy. The approach of section~\ref{sec:fixedpos} avoid this assumption by fixing both the source positions and their weights.

Similarly, the second moment of $n_w$ is
\begin{align}
    \langle n_w(\hat{n}) n_w(\hat{n}') \rangle &= \frac{1}{\Omega^2}\sum_{i \in p}^{N_p} w_i \, \sum_{j \in q}^{N_q} w_j + \delta^D(\hat{n} - \hat{n}') \frac{1}{\Omega} \sum_{i \in p}^{N_p} w_i^2 - \delta_{pq} \frac{1}{\Omega^2}\sum_{i \in p}^{N_p}w_i^2 \nonumber \\
    &= \frac{N_p N_q}{\Omega^2} w_{(1)}(p)w_{(1)}(q) + \delta^D(\hat{n} - \hat{n}') \frac{N_p}{\Omega} w_{(2)}(p) - \delta_{pq} \frac{N_p}{\Omega^2} w_{(2)}(p),
    \label{eq:wncov}
\end{align}
where $p$ and $q$ denote the pixels containing $\hat{n}$ and $\hat{n}'$ respectively, and $w_{(2)}(p) \equiv \sum_{i \in p} w_i^2/N_p$ for non-empty pixels. The second two terms in Equation~\eqref{eq:wncov} are the covariance of the weighted density field. Note that there is an additional term here compared with the usual situation in galaxy clustering statistics; this arises due to conditioning on the pixel occupancy. If there is only a single source galaxy in the sample, then for $\hat{n} \neq \hat{n}'$ and $p=q$ we are averaging over three events: 1) only point $\hat{n}$ is co-located with the source, 2) only point $\hat{n}'$ is co-located with the source, and 3) neither are co-located with the source. Two of these events cancel out as they average over the density at one point fixing the other point to be not co-located with a source, leaving an excess residual probability that one of the pair hits a source and the other does not. In other words, if there is precisely one galaxy in a pixel, the average density around that galaxy is lower than average (i.e.~guaranteed to be empty space).

\end{document}